\documentclass[twocolumn,tighten]{aastex62}
\pdfoutput=1 

\usepackage{amsmath,amssymb,amstext}
\usepackage{apjfonts}
\usepackage[T1]{fontenc}

\newcommand{\avg}[1]{\left\langle{#1}\right\rangle}

\makeatletter
\newcommand{\biggg}{\bBigg@\thr@@}
\newcommand{\Biggg}{\bBigg@{3.5}}
\def\bigggl{\mathopen\biggg}

\def\bigggr{\mathclose\biggg}

\makeatother

\iftrack
\renewcommand{\added}[1]{{\bf #1}}
\renewcommand{\deleted}[1]{\ }
\renewcommand{\replaced}[2]{{\bf #2}}
\else
\renewcommand{\replaced}[2]{#2}
\fi

\graphicspath{{./}{figures/}}

\shorttitle{Carbon Monoxide Line-intensity Cross-correlations}
\shortauthors{Chung et al.}

\begin{document}

\title{Cross-correlating Carbon Monoxide Line-intensity Maps with Spectroscopic and Photometric Galaxy Surveys}


\correspondingauthor{Dongwoo T. Chung}\email{dongwooc@stanford.edu}
\author{Dongwoo T.~Chung}
\affiliation{Kavli Institute for Particle Astrophysics and Cosmology \& Physics Department, Stanford University, Stanford, CA 94305, USA}
\author{Marco P.~Viero}
\affiliation{Kavli Institute for Particle Astrophysics and Cosmology \& Physics Department, Stanford University, Stanford, CA 94305, USA}
\author{Sarah E.~Church}
\affiliation{Kavli Institute for Particle Astrophysics and Cosmology \& Physics Department, Stanford University, Stanford, CA 94305, USA}
\author{Risa H.~Wechsler}
\affiliation{Kavli Institute for Particle Astrophysics and Cosmology \& Physics Department, Stanford University, Stanford, CA 94305, USA}
\affiliation{SLAC National Accelerator Laboratory, Menlo Park, CA 94025, USA}
\author{Marcelo A.~Alvarez}
\affiliation{Berkeley Center for Cosmological Physics, University of California, Berkeley, CA 94720, USA}
\author{J.~Richard Bond}
\affiliation{Canadian Institute for Theoretical Astrophysics, University of Toronto, 60 St.~George Street, Toronto, ON, M5S 3H8, Canada}
\author{Patrick C.~Breysse}
\affiliation{Canadian Institute for Theoretical Astrophysics, University of Toronto, 60 St.~George Street, Toronto, ON, M5S 3H8, Canada}
\author{Kieran A.~Cleary}
\affiliation{California Institute of Technology, Pasadena, CA 91125, USA}
\author{Hans K.~Eriksen}
\affiliation{Institute of Theoretical Astrophysics, University of Oslo, P.O.~Box 1029 Blindern, N-0315 Oslo, Norway}
\author{Marie K.~Foss}
\affiliation{Institute of Theoretical Astrophysics, University of Oslo, P.O.~Box 1029 Blindern, N-0315 Oslo, Norway}
\author{Joshua O.~Gundersen}
\affiliation{Department of Physics, University of Miami, 1320 Campo Sano Avenue, Coral Gables, FL 33146, USA}
\author{Stuart E.~Harper}
\affiliation{Jodrell Bank Centre for Astrophysics, School of Physics and Astronomy, The University of Manchester, Oxford Road, Manchester, M13 9PL, U.K.}
\author{H\aa vard T.~Ihle}
\affiliation{Institute of Theoretical Astrophysics, University of Oslo, P.O.~Box 1029 Blindern, N-0315 Oslo, Norway}
\author{Laura C.~Keating}
\affiliation{Canadian Institute for Theoretical Astrophysics, University of Toronto, 60 St.~George Street, Toronto, ON, M5S 3H8, Canada}
\author{Norman Murray}
\affiliation{Canadian Institute for Theoretical Astrophysics, University of Toronto, 60 St.~George Street, Toronto, ON, M5S 3H8, Canada}
\author{Hamsa Padmanabhan}
\affiliation{Institute for Particle Physics and Astrophysics, ETH Zurich, Wolfgang-Pauli-Strasse 27, CH 8093 Zurich, Switzerland}
\author{George F.~Stein}
\affiliation{Canadian Institute for Theoretical Astrophysics, University of Toronto, 60 St.~George Street, Toronto, ON, M5S 3H8, Canada}
\author{Ingunn K.~Wehus}
\affiliation{Institute of Theoretical Astrophysics, University of Oslo, P.O.~Box 1029 Blindern, N-0315 Oslo, Norway}
\collaboration{(COMAP Collaboration)}

\begin{abstract}
Line-intensity mapping (LIM or IM) is an emerging field of observational work, with strong potential to fit into a larger effort to probe large-scale structure and small-scale astrophysical phenomena using multiple complementary tracers. Taking full advantage of such complementarity means, in part, undertaking line-intensity surveys with galaxy surveys in mind. We consider the potential for detection of a cross-correlation signal between COMAP and blind surveys based on photometric redshifts (as in COSMOS) or based on spectroscopic data (as with the HETDEX survey of Lyman-$\alpha$ emitters). \replaced{For more than a factor-of-2 advantage in detection significance over the CO auto spectrum, we find that obtaining $\sigma_z/(1+z)\lesssim0.003$ accuracy in redshifts and $\gtrsim10^{-4}$ sources per Mpc$^3$ with spectroscopic redshift determination is sufficient}{We find that obtaining $\sigma_z/(1+z)\lesssim0.003$ accuracy in redshifts and $\gtrsim10^{-4}$ sources per Mpc$^3$ with spectroscopic redshift determination should enable a CO--galaxy cross spectrum detection significance at least twice that of the CO auto spectrum}. Either a future targeted spectroscopic survey or a blind survey like HETDEX may be able to meet both of these requirements.
\end{abstract}


\section{Introduction}

The technique of line-intensity mapping or intensity mapping (LIM or IM) images the aggregate emission in specific spectral lines from the galaxy population at large, rather than attempting to resolve individual galaxies. Line-intensity surveys thus trade understanding of individual galaxies for improved statistical insight into global astrophysics and cosmology within a significant survey volume. The 21-cm hydrogen line is one example of a line emitted commonly enough to be viable as a target, but other lines such as carbon monoxide and ionized carbon lines can be tied to molecular gas and star-formation activity. Surveys in each of these lines have the potential to yield, for example, a greatly improved understanding of cosmic star formation and ionization histories. (See~\citealt{Kovetz17} for a general overview of the theoretical and experimental landscape.)

While line-intensity mapping is relatively new, with 21-cm detections at $z\lesssim1$ only arising within the past decade~\citep{Chang10,Switzer13,Anderson18}, searching for individual galaxies is a tried and true method of mapping the luminous matter beyond our own galaxy. Current and future galaxy surveys are massive undertakings in collecting and processing high-resolution optical and infrared (IR) imagery, extracting galaxy catalogues from this imagery, and calculating redshifts and other galaxy properties for each object. The resulting data represent a wealth of astrophysical and cosmological information serving as important tests of our models of the early Universe.

However, optical and infrared surveys cannot detect indefinitely faint galaxies. One of the deepest surveys currently public is the Hawk-I UDS and GOODS Survey (HUGS;~\citealt{HUGS}), which reaches AB magnitude limits of $K\simeq26$--28 ($5\sigma$ limit per 0.4 square arcseconds) throughout the 10 arcminute wide GOODS-South~\citep{GOODS} and 20 arcminute wide UKIDSS~\citep{UKIDSS} Ultra Deep Survey fields. The depth of this imaging has allowed studies of $z\gtrsim4$ galaxies with stellar mass functions measured down to as low as $10^9\,M_\odot$~\citep{Grazian15}. While impressive, the scientific output of HUGS and other ultra-deep surveys are ultimately limited by their field size and therefore sample variance. Looking at shallower but wider fields, the COSMOS2015 catalogue~\citep{Laigle16} is complete down to $K_{\rm s}=24.0$ (AB magnitude, $3\sigma$, 3$''$ aperture) over a square-degree-scale intersection of the COSMOS~\citep{COSMOS} and UltraVISTA~\citep{UltraVISTA} fields, corresponding to a 90\% stellar mass completeness limit of $10^{10}\,M_\odot$. These data cover an area several orders of magnitude beyond typical ultra-deep fields, but the correspondingly reduced depth and mass completeness may lead to missing a majority of the total cosmic star-formation activity, happening in galaxies below the COSMOS2015 catalogue's stellar mass limit. (See~\citealt{Juneau05} and~\citealt{Sobral14} for studies at $z\lesssim2$ of contributions of galaxies of different stellar mass ranges to the global star formation rate.)

However, resolving and cataloguing individual galaxies over COSMOS-scale fields with HUGS-level depth is challenging with the cameras currently online. Considering the 10--30 hour exposure times per 70 square arcminute pointing used in HUGS, covering the 1.58 square degree (or 5688 square arcminute) area of the COSMOS2015 catalogue with the same camera (Hawk-I, the High Acuity Wide-field $K$-band Imager, at the ESO VLT) to the same depth as the HUGS data would require $\sim10^3$ hours. A project requiring this amount of time is difficult to run on community instruments, and would only output near-IR imagery with further follow-up requiring more time on other instruments.

Such is the niche that line-intensity surveys aim to fill, by operating dedicated instruments to map line emission over galaxy survey fields to greater depths than conventional galaxy surveys. As previously mentioned, however, the increased depth is not necessarily accompanied by an understanding of each individual object emitting in the observed line---only a statistical understanding of the whole emitter population---and additionally requires removal of significant foregrounds and systematics to meaningfully achieve.

Overall, the range of different trade-offs, systematics, advantages, and challenges in galaxy and line-intensity surveys means that the two techniques provide complementary views into the early Universe, and could be even more powerful in coordination. This will only become truer with further developments in line-intensity mapping, and in near-IR imaging technology and analysis. Work is already progressing on how to exploit cross-correlations both within line-intensity mapping~\citep[as in][]{BreysseRahman17} and between line-intensity and galaxy surveys~\citep[as in][]{Wolz17} to provide novel insights into star formation and galaxy evolution.

This leads into our own interest in prospects for cross-correlation between galaxy surveys and line-intensity surveys, which is specifically in the context of the Carbon monOxide Mapping Array Pathfinder (COMAP, as explored in~\citealt{Li16}). The initial phase of COMAP targets the CO(1-0) line (rest frequency 115.27 GHz) at redshifts 2.4--3.4 over square degree scale patches. The patch size and redshift range are well-matched to a galaxy catalogue like the COSMOS2015 catalogue, leading to the question of whether a potential COMAP detection of CO could be augmented by cross-correlation with the COSMOS2015 data, or even potentially an independent spectroscopic follow-up.

We bring up the idea of spectroscopic \emph{follow-up} specifically because galaxy surveys typically undertake wide-field photometric imaging followed by deeper, targeted spectroscopy of objects selected from the former. However, surveys operating outside of this paradigm are to come online in the near-future. One example is the Hobby--Eberly Telescope Dark Energy Experiment~\citep[HETDEX;][]{HETDEX}, a wide-field, blind spectroscopic survey and a possible platform for Lyman-$\alpha$ line-intensity mapping~\citep[e.g.~as considered in][]{Fonseca17}. While the main product of the HETDEX survey will be a catalogue of $\sim10^6$ Lyman-$\alpha$ emitters (LAEs) over $\sim400$ square degrees of sky, the locations of these LAEs are not predetermined. Rather, the survey footprint is blindly and sparsely sampled (with a fill factor of $1/4.5$---see~\autoref{sec:hetdex} for details) with the VIRUS spectrograph~\citep{VIRUS}, with individual LAEs extracted from the resulting spectra. This places HETDEX at the intersection of conventional catalogue-oriented surveys and blind line-intensity surveys, and potentially allows for generation of both LAE catalogues and Lyman-$\alpha$ intensity cubes from the same data. The redshift coverage of HETDEX ($z=1.9$--3.5) is well-matched to that of COMAP, which naturally then leads also to the question of how detectable a COMAP--HETDEX cross-correlation would be, and how it would compare to a COMAP--COSMOS cross-correlation---not only using HETDEX as a conventional cataloguing machine, but also as a line-intensity mapper.

We aim to answer the following questions.
\begin{itemize}
\item What number of sources do we need for significant cross-correlation, in the case of a hypothetical spectroscopic follow-up to complement COMAP?
\item What redshift accuracy must the reference galaxy catalogue achieve to enable significant cross-correlation?
\item What would be the detection significance of the various cross-power spectra under consideration?
\end{itemize}

The paper is structured as follows: in~\autoref{sec:expmeth} we outline the different experimental methods that COMAP, COSMOS, and HETDEX use to survey galaxies, then introduce our methods for simulating CO, galaxy, and Lyman-$\alpha$ observations in~\autoref{sec:simmeth}. We present expected cross-correlation results in~\autoref{sec:results}. After some discussion of these results and their implications for COMAP in~\autoref{sec:discussion}, we present our conclusions in~\autoref{sec:conclusions}.

Where necessary, we assume base-10 logarithms, and a $\Lambda$CDM cosmology with parameters $\Omega_m = 0.286$, $\Omega_\Lambda = 0.714$, $\Omega_b =0.047$, $H_0=100h$\,km\,s$^{-1}$\,Mpc$^{-1}$ with $h=0.7$, $\sigma_8 =0.82$, and $n_s =0.96$, broadly consistent with nine-year \emph{WMAP} results~\citep{WMAP9}. Distances carry an implicit $h^{-1}$ dependence throughout, which propagates through masses (all based on virial halo masses, proportional to $h^{-1}$) and volume densities ($\propto h^3$).

\section{Context: Experimental Methods}
\label{sec:expmeth}

\subsection{CO Line-intensity Mapping: COMAP}
\label{sec:comap}

\begin{deluxetable}{cc}[t!]
\tabletypesize{\footnotesize}
\tablecaption{\label{tab:comapparams}
COMAP instrumental and survey parameters assumed for this work.}
\tablehead{
\colhead{Parameter} & \colhead{Value}}
\startdata
System temperature & 40 K\\
Angular resolution & $4'$\\
Frequency resolution & 15.625 MHz\\
Observed frequencies & 26--30 GHz; 30--34 GHz\\
Number of feeds & 19\\
Survey area per patch & $\sim2.5$ deg$^2$\\
On-sky time per patch & 1500 hours\tabularnewline
\hline
\enddata
\tablecomments{Feeds are single-polarization. The survey observes frequencies of 26--34 GHz with two separate backend systems each covering a 4 GHz band in that range. The angular resolution above is the full width at half maximum of the Gaussian beam profile, for the receiver's central pixel. We simulate only one patch, though we expect to observe more than one at least for CO autocorrelation. A patch with 8.6\% observing efficiency could expect 1500 hours of integration time in two years, compared to typical values of $\sim10\%$ for fields close to the celestial equator (conditioning observability from the COMAP site on solar altitudes below $-10^\circ$, field altitudes above $30^\circ$, and elongations greater than $30^\circ$ from the Moon).}
\end{deluxetable}

\autoref{tab:comapparams} describes the current anticipated parameters for the initial phase (or Phase I) of COMAP. The receiver is currently undergoing commissioning at the Owens Valley Radio Observatory (OVRO) in California, where we expect the Phase I instrument to undertake a two-year observing campaign.

While a wide range of predictions exist for the CO power spectrum at $z\sim3$~\citep{Righi08,VL10,Pullen13,Breysse14,Li16,Padmanabhan18}, the sensitivity calculations made in~\cite{Li16}---given their fiducial model---place COMAP Phase I squarely in a regime where instrumental noise dominates over sample variance, which is still true after various changes to COMAP parameters made since the writing of~\cite{Li16}. This dictates the optimal observing strategy to some extent, pushing COMAP towards surveying at most several small fields (as close as possible to the $\sim1$ deg$^2$ field of view) with maximum observing efficiency. In a 2D analysis assuming a total on-sky time of one year ($\sim9000$ hours) split across four patches (for $\sim2200$ hours per patch),~\cite{Breysse14} found that a survey footprint of four patches with almost 4 deg$^2$ per patch would maximize total signal-to-noise. If the optimal area scales linearly with on-sky time, the fiducial area per patch of 2.5 deg$^2$ is close to ideal for a survey time of 1500 hours per patch as assumed in this work.

\subsection{Conventional Galaxy Survey: COSMOS2015}
\label{sec:cosmos2015}

Conventional galaxy surveys are a natural target for cross-correlation with line-intensity mapping, and the successful detection of 21-cm line emission from galaxies at $z\sim1$ comes from cross-correlation with spectroscopic galaxy surveys~\citep{Chang10,Switzer13}. However, spectroscopic data are currently limited in depth and abundance at $z\sim3$, so we look to existing public photometric datasets.

The COSMOS2015 catalogue contains half a million galaxies observed in $1<z<6$ across 1.58 square degrees of sky near the celestial equator. The catalogue is $K_s$-selected (the $K_s$ band being at 2.2 mm), and as mentioned previously the completeness limit is $K_s=24.0$. The $K_s$ magnitude correlates well with stellar mass up to $z\sim4$ (and magnitudes in longer-wavelength bands may be used at higher redshifts; see e.g.~\citealt{Davidzon17}). The redshift distribution skews largely towards lower redshift, but the source abundances are still relatively high for the redshift range relevant to COMAP, within which we find just under 20000 sources over 1.58 square degrees (of which 0.2 square degrees are masked due to saturated pixels) with $K_s\leq24$.

The critical limiting factor of the COSMOS2015 catalogue for studies of 3D large-scale structure is the redshift accuracy. \cite{Laigle16} quote photometric redshift errors at $3<z<6$ to be $\sigma_z=0.021(1+z)$, with some fraction of catastrophic failures; certain subsets even reach $\sigma_z\lesssim0.01(1+z)$. However,~\cite{Davidzon17} suggest that the error is higher for $z\gtrsim3$ galaxies, and is closer to $\sigma_z=0.03(1+z)$.

Deep low- to medium-resolution spectroscopic follow-up exists in the COSMOS field, but the surveys either do not satisfactorily cover $z>2$ or are limited in area. A recent catalogue of ten thousand objects selected across the COSMOS field~\citep{DEIMOS10k} only contains $\sim10^2$ objects in the redshift range of interest to COMAP, with the majority of spectroscopic redshifts well below (or above) that range. Meanwhile, the VIMOS Ultra-Deep Survey~\citep[VUDS;][]{VUDS} reports one of the largest $z>2$ emission-line galaxy samples, with $\sim2800$ spectroscopic redshifts at $z=2.5$--3.5 down to $i_\text{AB}\simeq25$ over a square degree of sky. However, the initial data release~\citep[the only public data release, at time of writing;][]{VUDSDR1} covers less than 10\% of this area, and even the full square degree is split across three patches covering $\sim10^3$ square arcminutes each, including half a square degree of the COSMOS field, or around a third of the COSMOS2015 coverage (and a fifth of the expected area per COMAP patch). Such surveys are well-suited for measuring stellar mass functions and average spectral properties, but the areas covered are less than ideal for cross-correlation against line-intensity maps. Uncertainty in the resulting cross spectra is roughly proportional to the inverse square root of the survey volume, so factors of 3--5 in sky area can noticeably affect detection significance.

\subsection{Blind Spectroscopic Survey: HETDEX}
\label{sec:hetdex}

\begin{deluxetable}{cc}[t!]
\tabletypesize{\footnotesize}
\tablecaption{\label{tab:hetdexparams}
HETDEX instrumental and survey parameters assumed for this work.}
\tablehead{
\colhead{Parameter} & \colhead{Value}}
\startdata
On-sky area per fibre & $1.8$ arcsec$^2$\\
Resolving power & 700\\
Observed wavelengths & 350--550 nm\\(or frequencies)&(857--545 THz)\\
Fill factor & $1/4$ (1 in SHELA field)\\
Line sensitivity & $4\times10^{-17}$ erg s$^{-1}$ cm$^{-2}$\\
\hline
\enddata
\tablecomments{Per~\cite{HETDEX2016}, the line sensitivity estimate is based on integrating 20 minutes per shot with three dithered exposures of 180 s, and the actual fill factor outside the SHELA field is closer to $1/4.5$ in reality.}
\end{deluxetable}

The central stated goal of the HETDEX survey is constraining the expansion history of the universe, and specifically detecting dark energy at $3\sigma$ significance, by identifying $\sim10^6$ LAEs through a wide-field spectroscopic survey~\citep{HETDEX,HETDEX2016}. However, we note a few interesting differences between HETDEX and previous conventional spectroscopic galaxy surveys measuring dark energy, e.g.~the Dark Energy Spectroscopic Instrument~\citep[DESI;][]{DESI}, the SDSS-IV Extended Baryon Oscillation Spectroscopic Survey~\citep[eBOSS;][]{eBOSS}, and WiggleZ~\citep{WiggleZ}.
\begin{itemize}
\item HETDEX targets a redshift range of $z=1.9$--3.5\footnote{A 350 nm minimum wavelength (below which strong ozone absorption features exist; see~\citealt{Schachter91}) is typical in ground-based optical spectrographs, setting the minimum redshift for ground-based LAE surveys beyond $z\sim1$ by necessity.}, well beyond the typical redshifts of $z\sim1$ of other dark energy-centric optical and NIR surveys. (eBOSS and DESI will target quasars and thus the Ly-$\alpha$ forest at $z\gtrsim2$, but emission line galaxies only up to $z\sim2$.)
\item HETDEX does not target specific points on the sky based on prior imaging, but rather samples its survey footprint with integral field spectroscopy, integrating for $\sim20$ minutes at each spot in the sky, and picks sources out from the noisy spectra.
\end{itemize}
The first point is of interest to us because unlike many other dark energy-centric surveys, future HETDEX detections will fall squarely in a redshift range relevant to COMAP Phase I. The second point is of interest since in principle, the noisy spectra uniformly sampled across the survey footprint could be processed and analysed as a line-intensity cube.

Due to the survey and instrument design of HETDEX, the survey footprint will not be completely filled in with data. Rather, the integral field unit (IFU) arrangement of the VIRUS instrument covers only $1/4.5$ of the area of each $20'$-diameter `shot', and while some dithering (in three exposures) will fill in areas between fibres, no attempt will be made to fill in the IFU spacing of $100''$, given a greater need for survey volume than for capture of small scales. The patch used for the Spitzer/HETDEX Exploratory Large Area (SHELA) survey~\citep{SHELA} is an exception, and the area between IFUs will be filled in for this field only~\citep{HETDEX2016}. At $13\times2$ square degrees, the SHELA patch could entirely contain a single (appropriately oriented) COMAP patch.

HETDEX is subject to interloper emission, detecting $\sim10^6$ galaxies from $z<0.5$ emitting in the [\ion{O}{2}] doublet~\citep{HETDEX}. When extracting individual emitters from LAE survey spectra, imposing a minimum equivalent width cutoff removes the low-$z$ emitters~\citep[][]{CowieHu98,HETDEXPilot}, and more sophisticated classification using Bayesian methods may also recover more of the underlying LAE sample~\citep{Leung17}. While we find no literature discussing foreground removal strategies in the context of line-intensity mapping with HETDEX, such literature does exist in the context of [\ion{C}{2}] observations~\citep{Cheng16,Lidz16,Sun18} and some strategies may be applicable beyond their original context. Furthermore, the low-$z$ [\ion{O}{2}] emission will have no corresponding component in COMAP data, potentially leading to its amelioration in COMAP--HETDEX line-intensity cross-correlation.

\begin{deluxetable*}{cccccc}[t!]
\tabletypesize{\footnotesize}
\tablecaption{\label{tab:allparams}
A summary of survey coverage and redshift precision for all experiments considered above.}
\tablehead{
\colhead{Experiment} & \colhead{Field size} & \colhead{Range of $z$} & \colhead{$\sigma_z/(1+z)$} & \colhead{Source selection} & \colhead{Source count}\\\colhead{or catalogue}&\colhead{(deg$^2$)}&&&&\colhead{(per deg$^2$ per $\Delta z=1$)}}
\startdata
COMAP&2.5&2.4--3.4&$\sim1/2000$&none (surveys aggregate CO emission)&\dots\\
COSMOS2015&1.58&1--6&$\sim0.02$&$K_s$-band magnitude $\lesssim24.0$&$\sim13000$\\
HETDEX&$300+150$&1.9--3.5&$\sim1/700$&Lyman-$\alpha$ luminosity $\gtrsim3\times10^{42}$ erg s$^{-1}$&$\sim1400$\\
(in SHELA)&(26)&\dots&\dots&\dots&$\sim6000$\\
\hline
\enddata
\tablecomments{The COMAP survey footprint will comprise two or more patches of 2.5 square degrees each; the HETDEX survey footprint includes a 300 deg$^2$ `Spring' field and a 150 deg$^2$ `Fall' field, with possible 50--60\% extensions to each. The `HETDEX in SHELA' row describes HETDEX full-fill coverage of the $13\times2$ deg$^2$ SHELA field.}
\end{deluxetable*}

To end this section, we show in~\autoref{tab:allparams} a summary of the coverage and redshift precision of all surveys discussed above. For simulation purposes, we will expand or truncate coverage as necessary to match the COMAP coverage, as explained in~\autoref{sec:mock}.

\section{Simulation Methods}
\label{sec:simmeth}

We simulate all surveys under consideration using halo catalogues derived from a dark matter simulation. We describe the dark matter simulation in~\autoref{sec:dmsim}, the models of various halo properties in~\autoref{sec:hmll}, and then the mocks of CO, Lyman-$\alpha$, and conventional galaxy survey data using these properties in~\autoref{sec:mock}. Finally, we outline the calculation of auto and cross power spectra from these data in~\autoref{sec:pspec_theory} and the calculation of sensitivity estimates with respect to those spectra in~\autoref{sec:sensest}.
\subsection{Dark Matter Simulation}
\label{sec:dmsim}

We use a cosmological N-body simulation as the basis for our simulations. In particular, we use the {\tt c400-2048} box, which is part of the Chinchilla suite of dark-matter-only simulations. \cite{Li16}, who used the same simulation, provide implementation details of the simulation and subsequent halo identification. The simulation spans $400h^{-1}$ Mpc on each side, and has a dark matter particle mass of $5.9\times10^8h^{-1}\,M_\odot$; we include dark matter halos more massive than $M_\mathrm{vir} = 10^{10}\,M_\odot$ in our analysis\added{, meaning that we assume halos with lower virial halo mass are not massive enough to host galaxies with substantial star-formation activity}.\added{ (\cite{Li16} justify the same choice of cutoff mass for CO simulations in their Appendix A; we consider its effect on Lyman-$\alpha$ simulations in our~\autoref{sec:lyamodel} alongside other details of Lyman-$\alpha$ modelling.)}

To simulate galaxies in our field of observation, we use dark matter halos identified in ``lightcone'' volumes, enclosing all halos within a given sky area and redshift range, with each lightcone based on arbitrary choices of observer origin and direction within the cosmological simulation. We use 100 lightcones spanning $z=1.5$--3.5 and a flat-sky area of $100'\times100'$, each populated with $\sim10^6$--$10^7$ halos. These lightcones form the basis for our simulated observations in $z=2.4$--3.4. Note that this redshift range spans approximately 1 Gpc, so some line-of-sight repetition of the N-body data will occur\replaced{ but should not impact our conclusions}{, as we exploit the periodic boundary conditions of the simulation box---which is only $400h^{-1}\approx570$ Mpc along each side---to extend the lightcone beyond the actual simulated comoving volume. However, the lightcone extents are not so much greater than the simulation volume that we expect this periodicity to impact the results of our study}.

\begin{figure}[t]
\centering\includegraphics[width=0.96\linewidth]{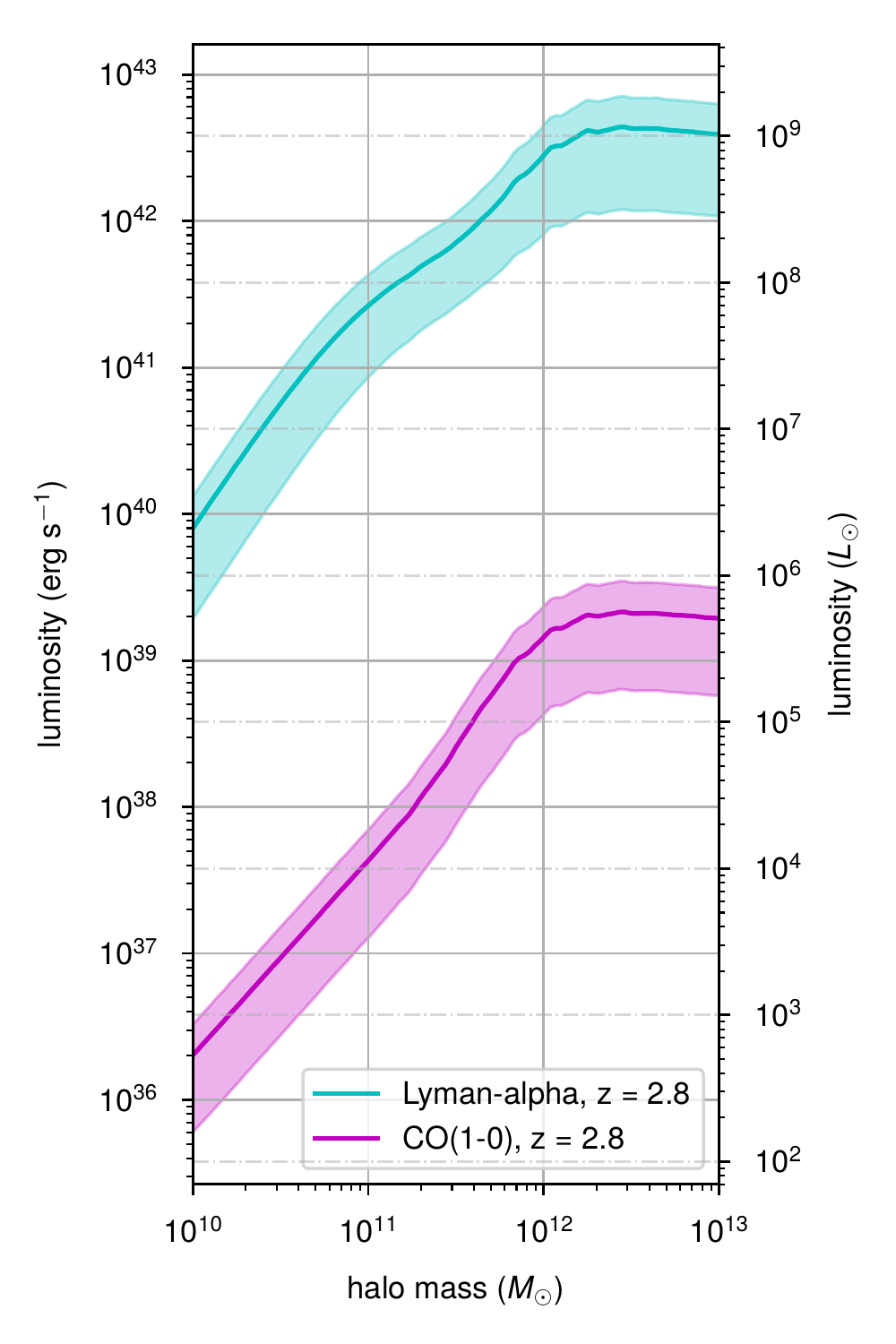}

\caption{Mean relation at redshift 2.8 between halo mass and line luminosity for CO(1-0) and Lyman-$\alpha$ emission. The shaded area around each mean curve indicates the $1\sigma$ log-scatter of the log-normal distribution at each halo mass.}
\label{fig:hmll}
\end{figure}

\subsection{Deriving Halo Properties}
\label{sec:hmll}
We derive CO and Lyman-$\alpha$ line luminosities for each of the halos (above the $10^{10}\,M_\odot$ cutoff mass) from their virial masses and redshifts, with~\autoref{fig:hmll} showing the average halo mass--line luminosity relations at $z=2.8$ (the COMAP mid-band redshift). In addition, we calculate stellar masses (for the galaxy survey selection) and star-formation rates (as an intermediate property for the line luminosities) for each halo. Below we explain the derivation of each of these properties.

\paragraph{Stellar mass} We assign a stellar mass $M_*$ to each halo using the best-fit stellar mass--halo mass relation from~\cite{Behroozi13a,Behroozi13b}. We apply the mean relation and the redshift-dependent scatter from the model, which is $\approx0.23$ dex at the redshifts considered here. The stellar mass is a property in itself and, unlike the star-formation rate, does not influence other properties.

\paragraph{Star-formation rate} We convert halo masses to star-formation rates (SFR) for each halo via interpolation of data from~\cite{Behroozi13a,Behroozi13b}. The main focus of these papers is to constrain the stellar mass--halo mass relation and derived quantities by comparing simulation data with observational constraints, and the resulting data include the average SFR in a halo given its mass and redshift.

We approximate halo-to-halo scatter in SFR by adding 0.3 dex log-normal scatter to the SFR obtained above, preserving the linear mean. The assumption of 0.3 dex scatter is reasonable given 0.2--0.4 true or intrinsic scatter observed in the SFR--stellar mass relation~\citep{Speagle14,Salmon15}, combined with the tight $\sim0.2$ dex scatter in the stellar mass--halo mass relation of~\cite{Behroozi13a}.

We also re-express SFR as infrared (IR) luminosity, using a known tight correlation:
\begin{equation}\frac{{\rm SFR}}{M_\odot\,\mathrm{yr}^{-1}} = 10^{-10}\left(\frac{L_\mathrm{IR}}{L_\odot}\right).\label{eq:SFR_LIR}\end{equation}
As in~\cite{Behroozi13a} and~\cite{Li16}, we assume a Chabrier initial mass function (IMF;~\citealt{Chabrier03}).
\paragraph{CO luminosity} We convert between IR luminosity and observed CO luminosity through power-law fits to observed data, commonly given in the literature:
\begin{equation}\log{\left(\frac{L_\mathrm{IR}}{L_\odot}\right)}=\alpha\log{\left(\frac{L'_\mathrm{CO}}{\text{K\,km\,s}^{-1}\text{\,pc}^2}\right)}+\beta,\end{equation}
where for our fiducial model, we take $\alpha=1.37$ and $\beta=-1.74$ from a fit to high-redshift galaxy data ($z\gtrsim1$) given in~\cite{CW13}, following~\cite{Li16}.

$L'_\mathrm{CO}$ (or indeed any $L'_\mathrm{line}$) is the observed luminosity (or velocity- and area-integrated brightness temperature) of the halo, which we convert into an intrinsic luminosity for each halo, as in~\cite{Li16}:
\begin{equation}\frac{L_\mathrm{line}}{L_\odot}=4.9\times10^{-5}\left(\frac{\nu_\mathrm{line,rest}}{115.27\text{ GHz}}\right)^3\frac{L'_\mathrm{line}}{\text{K\,km\,s}^{-1}\text{\,pc}^2}.\end{equation}

We also add 0.3 dex log-normal scatter in CO luminosity, again preserving the linear mean. We model this scatter as completely independent of the scatter in SFR. If we were simulating CO emission alone, as in~\cite{Li16}, we could end up with the same log-normal distribution width with a total scatter of $\sigma_\text{tot}=(\sigma_\text{SFR}^2/\alpha^2+\sigma_{L_\text{CO}}^2)^{1/2}=0.37$ \replaced{dex}{(with all $\sigma$ in units of dex)} on top of the mean SFR--$L_\text{CO}$ relation. However, unlike in~\cite{Li16}, the scattered SFR for a given halo informs both CO and Lyman-$\alpha$ line luminosities. Furthermore, \replaced{in this work, we preserve the linear mean SFR given halo mass and the linear mean $L_\text{CO}$ (or $L_\text{Ly$\alpha$}$, for that matter) given SFR. This approach is different from preserving the linear mean $L_\text{CO}$ given halo mass (which is what is done in~\citealt{Li16}), and since the $M_\text{vir}$--SFR and $L_\text{IR}$--$L_\text{CO}$ scaling relations are what is obtained from the literature, our approach is justified.}{our approach to scatter is subtly different from adding log-normal scatter to CO luminosity for a given halo mass and redshift; see~\autoref{sec:scatterorder} for details.}

Note that while the scatter exhibited is representative of the amount of scatter seen in the high-redshift galaxy data obtained via~\cite{CW13}, it may not be representative of the galaxy population at large that COMAP will study, and larger scatter in CO luminosity than we have assumed here may reduce our ability to cross-correlate CO against galaxy catalogues. However, this means that were COMAP to indeed confidently detect CO while finding little in the way of CO--galaxy cross-correlation, that in itself would lead to interesting insights about how stochastic CO is in these high-redshift galaxies (or at least those galaxies with sufficiently high mass and luminosity to be catalogued in the cross-correlation sample).

\paragraph{Lyman-$\alpha$ luminosity} Before dust absorption and other attenuation mechanisms, the Lyman-$\alpha$ line is 8.7 times stronger (for case B recombination) than H$\alpha$, a frequently chosen emission-line tracer of star-formation activity~\citep{KE12}. We use an intrinsic Lyman-$\alpha$--SFR calibration (via H$\alpha$) with an escape fraction encapsulating possible attenuation of the intrinsic Lyman-$\alpha$ luminosity. We outline the specifics behind our model in~\autoref{sec:lyamodel}, the end result of which is that for a given halo, we calculate
\begin{equation}L_{\text{Ly}\alpha}=1.6\times10^{42}\left(\frac{\text{SFR}}{M_\odot\text{ yr}^{-1}}\right)f_\mathrm{esc}(\text{SFR},z)\text{ erg s}^{-1},\label{eq:Lyamodel}\end{equation}
with
\begin{align}f_\mathrm{esc}(\text{SFR},z)&=\bigl(1+e^{-1.6z+5}\bigr)^{-1/2}\nonumber\\&\quad\times\bigggl[0.18+\frac{0.82}{1+0.8\bigl(\frac{\text{SFR}}{M_\odot\text{ yr}^{-1}}\bigr)^{0.875}}\bigggr]^2.\label{eq:fescmodel}\end{align}
The escape fraction $f_\text{esc}$ increases with lower SFR and higher redshift so as to allow this model to match observed LAE luminosity functions (including from~\citealt{Sobral17} and~\citealt{Gronwall07}; see~\autoref{sec:lyamodel} for more references) in the redshift range of interest ($z\sim2$--4).

As with the CO luminosity, we add 0.3 dex log-normal scatter in Lyman-$\alpha$ luminosity. Due to the non--power-law nature of the SFR--$L_\text{Ly$\alpha$}$ relation, the net scatter in $L_\text{Ly$\alpha$}$ varies non-monotonically with halo mass (between 0.31 and 0.42 dex in the mean relation shown in~\autoref{fig:hmll} at $z=2.8$).

Note that our model is tuned to LAE observations and assigns luminosities only at the centres of dark matter halos, and thus does not account for the finer details of Lyman-$\alpha$ radiative transfer beyond each LAE \emph{per se}, which would result for example in diffuse Lyman-$\alpha$ halos or blobs~\citep{Steidel11}. Incorporating such details would introduce additional components to the Lyman-$\alpha$ emission, which we discuss in more detail in~\autoref{sec:Lyamodelbad}. The signal as simulated here could be modified by these components in ways that may be detectable in HETDEX data through line-intensity mapping, but not necessarily through LAE identification. All of this suggests the need to study implications of diffuse Lyman-$\alpha$ emission for HETDEX and cross-correlation with COMAP, but we leave this for future work.

\subsection{Mock Surveys}
\label{sec:mock}
After the processing outlined in the last section, each halo has a sky position, redshift (excluding peculiar velocities, which have minimal effects on the results of this work\footnote{\added{While peculiar velocities do alter $P(k)$ at the scales studied, the effect is at a factor of order unity (boosting the CO $P(k)$ by at most 30\% given the mass-averaged bias expected of CO emission), and is unlikely \emph{per se} to weaken how well our different tracers cross-correlate with each other. Furthermore, we wish to obtain an even comparison to previous studies like \cite{Li16} that also neglect peculiar velocities.}}), virial halo mass, stellar mass, CO luminosity, and Lyman-$\alpha$ luminosity. We now use these properties to simulate survey data for COMAP Phase I, a COSMOS2015-like mass-selected galaxy catalogue, and a HETDEX-like Lyman-$\alpha$ survey.

For ease of analysis, all survey cubes are generated with the same grid of voxels, based on the COMAP observation. The angular extent of each voxel is $\delta_x=\delta_y=0.4'$ or $1.16\times10^{-4}$ rad in each direction (oversampling the COMAP beam width by a factor of 10 and the on-sky VIRUS IFU width by a factor of 2.1), and each voxel spans $\delta_\nu=15.625$ MHz in COMAP frequency (equivalent to 335 GHz in HETDEX frequency) unless otherwise specified. The simulated cubes span $100'\times100'$ and a continuous 26--34 GHz band in COMAP frequency (557--729 THz in HETDEX frequency), enclosing a total comoving volume of $190\times190\times1000=3.6\times10^7$ Mpc$^3$.
\subsubsection{CO \replaced{i}{I}ntensity \replaced{d}{D}ata}
We follow~\cite{Li16} again in generating a temperature cube, taking the same set of steps:
\begin{itemize}
\item Bin the halo luminosities into resolution elements in frequency and angular position, resulting in a certain luminosity $L_\mathrm{line,vox}$ \replaced{associated with each}{for each voxel that is simply the cumulative line luminosity of all halos in that} voxel.
\item Convert these luminosities into surface brightness (apparent spectral intensity, in units of luminosity per unit area, per unit frequency, per unit solid angle):
\begin{equation}I_{\nu,\mathrm{obs}} = \frac{L_\mathrm{line,vox}}{4\pi D_L^2}\frac{1}{\delta_x\delta_y\delta_\nu},\end{equation}
where $D_L$ is the luminosity distance to that voxel.
\item Convert to the expected brightness temperature contribution from each voxel. The Rayleigh--Jeans brightness temperature for a given surface brightness is
\begin{equation}T=\frac{c^2I_{\nu,\mathrm{obs}}}{2k_B\nu_\mathrm{obs}^2},\end{equation}
from which we obtain our temperature $T_\text{CO}(\mathbf{x})$ at each voxel position $\mathbf{x}$ in the data cube.
\end{itemize}

\subsubsection{Galaxy \replaced{o}{O}verdensity \replaced{f}{F}ield}
\label{sec:mockCOSMOS}
We devise an ideal NIR-selected galaxy survey tracing galaxies down to a certain stellar mass limit. We claim that the galaxy--halo connection allows us to model this, starting with a catalogue of halos and imposing stellar mass cuts corresponding to realistic magnitude limits.

\paragraph{Mass-completeness}
To crudely simulate the $K_s$ magnitude cut used by catalogues like COSMOS2015, we assume the $K_s$ magnitude correlates reliably with the stellar mass in our redshift range, an assumption that~\cite{Laigle16} support at least in relating completeness limits for the two quantities. Each step down in magnitude is a factor of $10^{0.4}$ up in brightness, so a constant mass-to-light ratio would result in the same factor up in stellar mass. 
\cite{Laigle16} find their $K_{s,\mathrm{lim}}=24.0$ limit to be equivalent to a stellar mass completeness limit of $M_{*,\mathrm{lim}}=10^{10}\,M_\odot$ in our redshift range. We extrapolate this to different completeness limits with the following relation:
\begin{equation}
\log{M_{*,\mathrm{lim}}} = 10.0 - 0.4(K_{s,\mathrm{lim}}-24.0)
\end{equation}
Then limits of $K_{s,\mathrm{lim}}=(25.0,24.0,23.0,22.0)$ are equal to $\log{(M_*/M_\odot)}=(9.6,10.0,10.4,10.8)$. We use these stellar mass cuts to select our mock galaxy survey sample in each lightcone. Of these, the $\log{(M_*/M_\odot)}>10.0$ cut matches the COSMOS2015 source abundance within a factor of order unity, so we take this as our fiducial $M_*$ cut. 

\paragraph{Redshift accuracy}
To simulate uncertainty in redshifts derived from imagery, we apply different levels of scatter in observed redshift relative to the true cosmological redshift. While we do simulate cross-correlations against a survey with perfect galaxy redshift knowledge, not even spectroscopic surveys have such information. Therefore, we simulate normal scatter of redshifts with $\sigma_z/(1+z)=0.0007$, 0.003, 0.01, 0.02, and 0.03. The first scenario meets the minimum redshift accuracy required for cosmological applications of the Subaru Prime Focus Spectrograph~\citep[PFS; see][]{Takada14}. The second scenario corresponds to lower-resolution spectroscopy, as seen in HST grism surveys like 3D-HST~\citep[$\sigma_z/(1+z)=0.003$;][]{Momcheva16}\replaced{ or}{,} prism surveys like PRIMUS~\citep[$\sigma_z/(1+z)=0.005$;][]{PRIMUS1,PRIMUS2}\added{, or even narrow-band photometric surveys like the PAU Survey~\citep[$\sigma_z/(1+z)=0.0037$;][]{Eriksen18}}. The last three scenarios represent optimistic, fiducial, and pessimistic expectations for photometric redshift accuracy, based on the discussion in~\autoref{sec:cosmos2015}.

After applying the stellar mass cut and redshift scatter (if applicable), we calculate the galaxy overdensity across the voxel grid. We count the number of galaxies $N_\mathrm{gal,vox}(\mathbf{x})$ in each voxel, divide by the comoving volume of the voxel to get the number density $n_\mathrm{gal,vox}=N_\mathrm{gal,vox}/V_\mathrm{vox}$. The quantity we deal with then is normalized by the average number density $\bar{n}_\mathrm{gal}$ across all voxels observed:
\begin{equation}\delta_\text{gal,vox} = \frac{n_\mathrm{gal,vox}(\mathbf{x})}{\bar{n}_\mathrm{gal}}-1.\end{equation}

\subsubsection{Lyman-$\alpha$ \replaced{s}{S}urvey \replaced{s}{S}imulation}
We simulate two data products for the Lyman-$\alpha$ survey: a LAE overdensity cube, and a Lyman-$\alpha$ line-intensity cube. This is in view of our earlier statement in~\autoref{sec:hetdex} that while the primary data product from HETDEX will be a catalogue of high-redshift LAEs, the collection of spectra across the survey footprint could be treated and analysed as a Lyman-$\alpha$ line-intensity data cube.

We calculate the relative LAE overdensity $\delta_\text{LAE,vox}$ for each voxel in much the same way as in the galaxy survey cubes, except our selection criterion is now the Lyman-$\alpha$ luminosity of each halo rather than the stellar mass. The HETDEX Pilot Survey~\citep{HETDEXPilot,Blanc11} reported luminosity limits of 3--$6\times10^{42}$ erg s$^{-1}$ with $5\sigma$ line flux sensitivities of $5\times10^{-17}$ erg s$^{-1}$ cm$^{-2}$---not far from the goal for the final survey shown in~\autoref{tab:hetdexparams}---so we set luminosity cuts at $(3\times10^{42},6\times10^{42})$ erg s$^{-1}$.

The Lyman-$\alpha$ line-intensity cube is generated in much the same way as the CO temperature cube, but rather than converting the observed intensity to a brightness temperature (which is no longer applicable for observations in optical bands), we work with the intensity per unit log-frequency interval $\nu_\mathrm{Ly\alpha}I_{\nu,\mathrm{Ly\alpha}}$, in units of erg s$^{-1}$ cm$^{-2}$.

VIRUS is expected to have a resolving power of $R\sim700$~\citep{VIRUS}, and the Lyman-$\alpha$ emission in LAEs from the HETDEX Pilot Survey has been observed with velocity offsets of several hundred km s$^{-1}$ relative to the galaxy systemic redshifts~\citep{Chonis13}. We translate all of this to an expectation of redshift precision of $\sigma_z/(1+z)\approx0.0015$, mostly based on $1/R\approx0.0014$ but adding on possible velocity offsets of the Lyman-$\alpha$ line relative to the rest of the galaxy (which will result in residuals even after the subtraction of an average velocity offset). We simulate normal scatter with this error in the LAE redshifts when calculating the LAE overdensity. When simulating the Lyman-$\alpha$ intensity cube, we do not apply this scatter; however, we do account for the attenuation from spectral resolution when calculating the cross spectrum and its detection significance.

\paragraph{Sparse sampling}
As previously mentioned, the great majority of the HETDEX survey footprint will be sampled sparsely. 
To emulate this in our simulations, we leave regions of 2 pixels by 2 pixels unmasked, each separated by 2 masked pixels. This results in a fill factor of $1/4$ with a regular pattern of $48''\times48''$ squares with centres spaced apart by $96''$, approximating both the fill factor and the IFU on-sky spacing that will show up in HETDEX data. While the actual IFU arrangement and shot tiling is more complex and results in additional biasing of the $P(k)$ measurement, this serves as a first pass at simulating the effect of sparse sampling on both auto and cross spectra, at a level sufficient for this work. A detailed analysis from~\cite{Chiang13} shows that resulting measurement biases for more complex sparse sampling scenarios (versus perfect tiling as simulated in this work) are within 10\% up to scales of $0.5h$ Mpc$^{-1}$.
\subsection{Simulated Auto and Cross Spectra}
\label{sec:pspec_theory}
We have now defined a grid of voxels and four quantities associated with each voxel: the CO brightness temperature $T_\text{CO}$, the mass-selected galaxy overdensity $\delta_\mathrm{gal}$ (for four different mass cuts), the Lyman-$\alpha$ spectral intensity per log-frequency interval $\nu_\mathrm{Ly\alpha}I_{\nu,\mathrm{Ly\alpha}}$, and the luminosity-selected LAE overdensity $\delta_\mathrm{LAE}$ (for two different luminosity cuts). Following previous works~\citep{VL10,Li16}, we use Fourier estimators of the auto and cross power spectra. If $\tilde{A}(\mathbf{k})$ and $\tilde{B}(\mathbf{k})$ are the Fourier transforms of the fields $A(\mathbf{x})$ and $B(\mathbf{x})$, then the full 3D auto spectra are
\begin{equation}P_\text{A}(\mathbf{k}) = V_\text{surv}^{-1}|\tilde{A}(\mathbf{k})|^2,\quad P_\text{B}(\mathbf{k}) = V_\text{surv}^{-1}|\tilde{B}(\mathbf{k})|^2;\end{equation}
the full 3D cross spectrum is
\begin{equation}P_{\text{A}\times\text{B}}(\mathbf{k}) = V_\text{surv}^{-1}\operatorname{Re}{[\tilde{A}(\mathbf{k})\tilde{B}^*(\mathbf{k})]}.\end{equation}
\added{Since the fields are defined on a grid of voxels at discrete values of $\mathbf{x}$, the Fourier transforms and full 3D spectra are also defined at discrete $\mathbf{k}$.}

With the assumption of isotropy, we then spherically average \replaced{these}{the power spectra} in shells of $k=|\mathbf{k}|$\added{ of width $\Delta k=0.035$ Mpc$^{-1}$}, each containing some number of modes $N_\text{modes}(k)$\added{ for which $k-\Delta k/2<|\mathbf{k}|<k+\Delta k/2$}, to obtain the spherically averaged 3D power spectra $P_\text{A}(k)$, $P_\text{B}(k)$, and $P_{\text{A}\times\text{B}}(k)$. Here we take $A(\mathbf{x})=T_\text{CO}$, while $B(\mathbf{x})$ can be any of the other three fields defined above.

Broadly speaking, we can consider each of the auto and cross $P(k)$ to be the sum of a clustering component that dominates at low $k$ and a constant shot-noise term that dominates at high $k$. The clustering component traces the matter power spectrum with some bias associated with the quantity being observed, while the shot-noise component arises from Poisson fluctuations. Cross shot noise can have interesting interpretations explored in previous work: \cite{Wolz17} show that HI--galaxy cross shot noise may be used to infer the HI content of the cross-correlated galaxies, and~\cite{BreysseRahman17} show that $^{12}$CO--$^{13}$CO cross shot noise within COMAP may be used to learn about $^{12}$CO--$^{13}$CO isotopologue ratios and $^{12}$CO saturation. One could extend the latter idea to cross-correlate between two separate line-intensity surveys, e.g.~between COMAP and HETDEX to learn about the molecular fraction of Lyman-$\alpha$ emitters (using CO as a proxy for molecular gas). However, as we will see, such astrophysical interpretation of the cross shot noise must take into account attenuation both from sparse sampling as seen in HETDEX and from redshift errors in all galaxy samples. We leave for future work a detailed investigation into effects of such attenuation on astrophysical inferences.

\subsection{Sensitivity Estimates}
\label{sec:sensest}
For our purposes, we take the sources of uncertainty to be sample variance, thermal noise in the CO temperature or Lyman-$\alpha$ intensity field, and shot noise in the galaxy density field\footnote{Unlike thermal noise in the line-intensity maps, the shot noise in the galaxy density field emerges naturally from the simulation procedure outlined above, and may be considered a component of the observed/simulated galaxy power spectrum (as is shot noise in the line-intensity power spectra).}. From~\cite{VL10},
\begin{equation}\sigma^2_{P_{\text{A}\times\text{B}}}(k)=\frac{P_\text{A,total}(k)P_\text{B,total}(k)+P^2_{\text{A}\times\text{B}}(k)}{2N_{\rm modes}(k)},\label{eq:VL10sig}\end{equation}
where the `total' power spectra include noise, interloper emission, and other components not necessarily correlated to the tracer. In this work, we ignore these components apart from instrumental noise:
\begin{align}P_\text{CO,total}(k)&=P_\text{CO}(k)+P_{n,\text{COMAP}};\\
P_\text{Ly$\alpha$,total}(k)&=P_\text{Ly$\alpha$}(k)+P_{n,\text{HETDEX}}.
\end{align}
Therefore, it is best to treat the signal-to-noise estimates given in this work as upper bounds on what the actual surveys may ultimately achieve.

\added{In general, the instrumental (thermal) noise power spectrum is given by the root-mean-square temperature or intensity fluctuation per voxel $\sigma_n$ and the comoving voxel volume $V_\text{vox}$, and is assumed to be pure white noise and thus constant across all $k$~\citep{Lidz11}:
\begin{equation}
P_n = \sigma_n^2V_\text{vox}.
\end{equation}
(This is analogous to the inverse of the weight per solid angle $w=(\sigma_\text{pix}^2\Omega_\text{pix})^{-1}$ in the calculation of uncertainties in 2D $C_\ell$ analysis from~\citealt{Knox1995}.) We then only need calculate $\sigma_{n,\text{COMAP}}$ and $\sigma_{n,\text{HETDEX}}$ based on the expected instrumental and survey parameters.
}

Calculation of the COMAP instrumental noise follows the same procedure outlined in\added{ Appendix C of}~\cite{Li16}\added{,} using the parameters in~\autoref{tab:comapparams}. \added{Specifically, $\sigma_{n,\text{COMAP}}$ derives from the system temperature $T_\text{sys}=40$ K, the number of feeds $N_\text{feeds}=19$, the frequency resolution $\delta_\nu=15.625$ MHz, and the survey time per pixel $\tau_\text{pix}=(1500\text{ hr})\cdot\delta_x\delta_y/(2.5\text{ deg}^2)$ (being simply the total survey time per patch divided by the number of pixels per patch):

\begin{equation}
\sigma_{n,\text{COMAP}}=\frac{T_\text{sys}}{\sqrt{N_\text{feeds}\delta_\nu\tau_\text{pix}}}.
\end{equation}}

We simulate and assume only one patch of 2.5 deg$^2$, so the signal-to-noise estimates are also given per patch. If we fix the on-sky time per patch and the solid angle per patch, uncertainties from COMAP instrumental noise (which we expect to dominate total uncertainties) will decrease as the square root of the number of patches (from the linear increase in the number of modes averaged to obtain our best $P(k)$ estimate).

For the HETDEX instrumental noise, we refer to the sensitivity metrics given in~\cite{VIRUS}. Each VIRUS fibre covers a solid angle of $1.8$ square arcseconds at one time, and a dither pattern of three exposures allows the area within each IFU to be completely covered. The line sensitivity expected from each 20 minute shot is $\lesssim4\times10^{-17}$ erg s$^{-1}$ cm$^{-2}$, and dividing by the $5.4$ square arcsecond solid angle per dithered fibre gives \replaced{$\sigma_\text{pix,HETDEX}=3.15\times10^{-7}$}{$\sigma_{n,\text{HETDEX}}=3.15\times10^{-7}$} erg s$^{-1}$ cm$^{-2}$ sr$^{-1}$.\replaced{ Then in analogy to the COMAP noise spectrum, $P_{n,\text{HETDEX}}$ is $\sigma_\text{pix,HETDEX}^2$ times the comoving voxel volume.}{ Thus we now have $P_{n,\text{COMAP}}=\sigma_{n,\text{COMAP}}^2V_\text{vox}$ and $P_{n,\text{HETDEX}}=\sigma_{n,\text{HETDEX}}^2V_\text{vox}$. For COMAP, the dependence on the voxel size (simulated or otherwise) cancels out as $\sigma_n^2$ is proportional to the inverse of the frequency bandwidth per channel as well as the inverse of the solid angle per pixel (via $\tau_\text{pix}$). For HETDEX, we choose $V_\text{vox}$ in the context of $P_{n,\text{HETDEX}}$ to correspond to the 5.4 square arcsecond solid angle per dithered fibre (cancelling the same factor we used to convert from line sensitivity to intensity fluctuation per voxel) and the spectral resolution of the instrument (or a redshift interval of $(1+z)/R\sim0.005$), which comes out to $\sim0.03$ Mpc$^{3}$.}

\replaced{T}{As with the cross power spectrum uncertainty in~\autoref{eq:VL10sig}, t}he errors on the individual auto power spectra are also given by dividing the `total' spectra by the square root of $N_\text{modes}(k)$. In the case of galaxy or LAE overdensities, the `total' power spectrum is equal to the simulated power spectrum, as we never subtract the shot noise term of $1/\bar{n}$. In the other cases, we use the `total' spectra from above:
\begin{align}\sigma_{P_{\text{CO}}}(k)&=\frac{P_\text{CO}(k)+P_{n,\text{COMAP}}}{\sqrt{N_{\rm modes}(k)}};\\\sigma_{P_{\text{Ly}\alpha}}(k)&=\frac{P_\text{Ly$\alpha$}(k)+P_{n,\text{HETDEX}}}{\sqrt{N_{\rm modes}(k)}}
.\end{align}

When calculating the signal-to-noise, we also need to account for attenuation in the CO signal due to the beam. As discussed in~\cite{Li16}, the beam resolution limit attenuates the Fourier-transformed CO temperature field $\tilde{T}_\text{CO}(\mathbf{k})$ at each $\mathbf{k}$ by a factor of $\exp{(-k_\perp^2\sigma_\text{beam}^2/2)}$, where $k_\perp$ is the transverse component of $\mathbf{k}$ and $\sigma_\text{beam}$ the width of the Gaussian profile of the beam (projected into the comoving survey volume). However, angular resolution limits differ significantly between the CO temperature field and the field being cross-correlated against, and the latter (which may be due to pixelization, galaxy survey limits, fibre diameters, and so on) will be much finer than the COMAP $\sigma_\text{beam}$. Then while the full 3D CO auto spectrum $P_\text{CO}(\mathbf{k})\propto\tilde{T}_\text{CO}(\mathbf{k})^2$ is attenuated at each $\mathbf{k}$ by $\exp{(-k_\perp^2\sigma_\text{beam}^2)}$, the cross spectra only scale linearly with the attenuated CO signal and thus are attenuated at each $\mathbf{k}$ by approximately $\exp{(-k_\perp^2\sigma_\text{beam}^2/2)}$.

The spherically averaged auto and cross $P(k)$ are correspondingly attenuated by
\begin{equation}W^2(k) = {\bigl\langle\exp{(-k_\perp^2\sigma_\perp^2)}\bigr\rangle}_\mathbf{k},\label{eq:Wbeam}\end{equation}
where the average is over all (discrete) $\mathbf{k}$ in the $P(\mathbf{k})$ averaging that fall within the shell corresponding to $k$, and $\sigma_\perp$ is the applicable resolution limit for the $P(k)$ being calculated\footnote{\added{$W^2(k)$ is what some works---e.g.~\citealt{Li16}---denote as $W(k)$. Here we adopt the convention that $W(\mathbf{k})$ refers to the window function applied to the Fourier-transformed field, not its squared magnitude.}}. For the CO auto spectrum, $\sigma_\perp$ is simply $\sigma_\text{beam}$, and given the large beam size of the COMAP telescope, effectively $\sigma_\perp\approx\sigma_\text{beam}/\sqrt{2}$ for the cross spectra as discussed above.

Recall that towards the end of~\autoref{sec:mock}, we also discussed attenuation in cross spectra between CO and Lyman-$\alpha$ intensity fluctuations due to the limited spectral resolution of HETDEX. This follows a similar average as in~\autoref{eq:Wbeam}\added{ but with $k_\parallel$, the line-of-sight component of $\mathbf{k}$}:
\begin{equation}W^2_z(k) = {\bigl\langle\exp{(-k_\parallel^2\sigma_\parallel^2)}\bigr\rangle}_\mathbf{k}.\end{equation}
Were we dealing with the HETDEX auto spectrum, we would base $\sigma_\parallel=\sigma_{\parallel,\text{HETDEX}}$ on the redshift-space resolution of $0.0015(1+z)$ at the average redshift of the survey volume, again based on the resolving power $R\sim700$ of VIRUS~\citep{VIRUS} and observed Lyman-$\alpha$ component velocity offsets~\citep{Chonis13} as discussed in~\autoref{sec:mock}. However, in cross-correlation with COMAP, which has much higher redshift precision, we can assume that $\sigma_\parallel\approx\sigma_{\parallel,\text{HETDEX}}/\sqrt{2}$ using similar arguments as for $\sigma_\perp$.

\added{In discussing our results, we will often quote the total signal-to-noise ($\mathrm{S/N}$) over `all' scales, meaning all scales $k\in(0.017,4.2)$ Mpc$^{-1}$ (in linear bins of width $0.034$ Mpc$^{-1}$) that our simulations nominally access (with the minimum and maximum $k$ respectively corresponding to the lightcone and voxel angular widths in comoving space). Following~\cite{Li16}, we calculate this total $\mathrm{S/N}$ as
\begin{equation}
\mathrm{\frac{S}{N}} = \left[\sum_k\left(\mathrm{\frac{S}{N}}(k)\right)^2\right]^{1/2} = \left[\sum_k\left(\frac{P_\text{obs}(k)}{\sigma_P(k)}\right)^2\right]^{1/2},
\end{equation}
where $P_\text{obs}(k)=P(k)W^2(k)$ (or $P(k)W^2(k)W^2_z(k)$ in the case of the CO--Lyman-$\alpha$ intensity cross spectrum) and the sum is over all $k$-bins with central values in the previously specified range of $(0.017,4.2)$ Mpc$^{-1}$. Note that $W^2(k)$ and $W^2_z(k)$ also modify the auto and cross spectra (before thermal noise) in the expressions for $\sigma_P(k)$.
}
\section{Results}
\label{sec:results}

We examine how the signal-to-noise and the cross-correlation signal itself vary with survey variables when cross-correlating COMAP against a conventional galaxy survey or a HETDEX-like survey. While we present the signal-to-noise based on the auto or cross spectra in isolation, we will mostly plot the signal in the form of the normalized cross-correlation coefficient between tracers A and B:
\begin{equation}r(k)=\frac{P_{\text{A}\times\text{B}}(k)}{\sqrt{P_\text{A}(k)P_\text{B}(k)}},\end{equation}
the value of which varies from -1 for perfect anti-correlation to +1 for perfect co-correlation.\added{ This allows us to evenly compare different cross-correlation scenarios---for which the cross $P(k)$ otherwise have significantly varying units and amplitudes---and relevant scales over which correlations wax or wane.} We do not account for instrumental noise or beam response in the plotted $r(k)$, although we do in calculating power spectra signal-to-noise.

We present $r(k)$ and overall cross $P(k)$ signal-to-noise for cross-correlation against a conventional galaxy survey in~\autoref{sec:COSMOSres}, and against a HETDEX-like survey in~\autoref{sec:HETDEXres}. We then summarize the relevant auto and cross $P(k)$ for COMAP and sensitivities for all scenarios in~\autoref{sec:allres}.

\subsection{Cross-correlations with Conventional Galaxy Surveys}
\label{sec:COSMOSres}

We explore two different variables in the galaxy survey: (1) the mass-completeness of a perfect redshift survey ($\sigma_z/(1+z)=0$), and (2) the redshift accuracy of a survey with the fiducial mass-completeness cut ($\log{(M_*/M_\odot)}>10.0$).

\subsubsection{Mass-completeness}
\begin{figure}[t]
\centering\includegraphics[width=\linewidth]{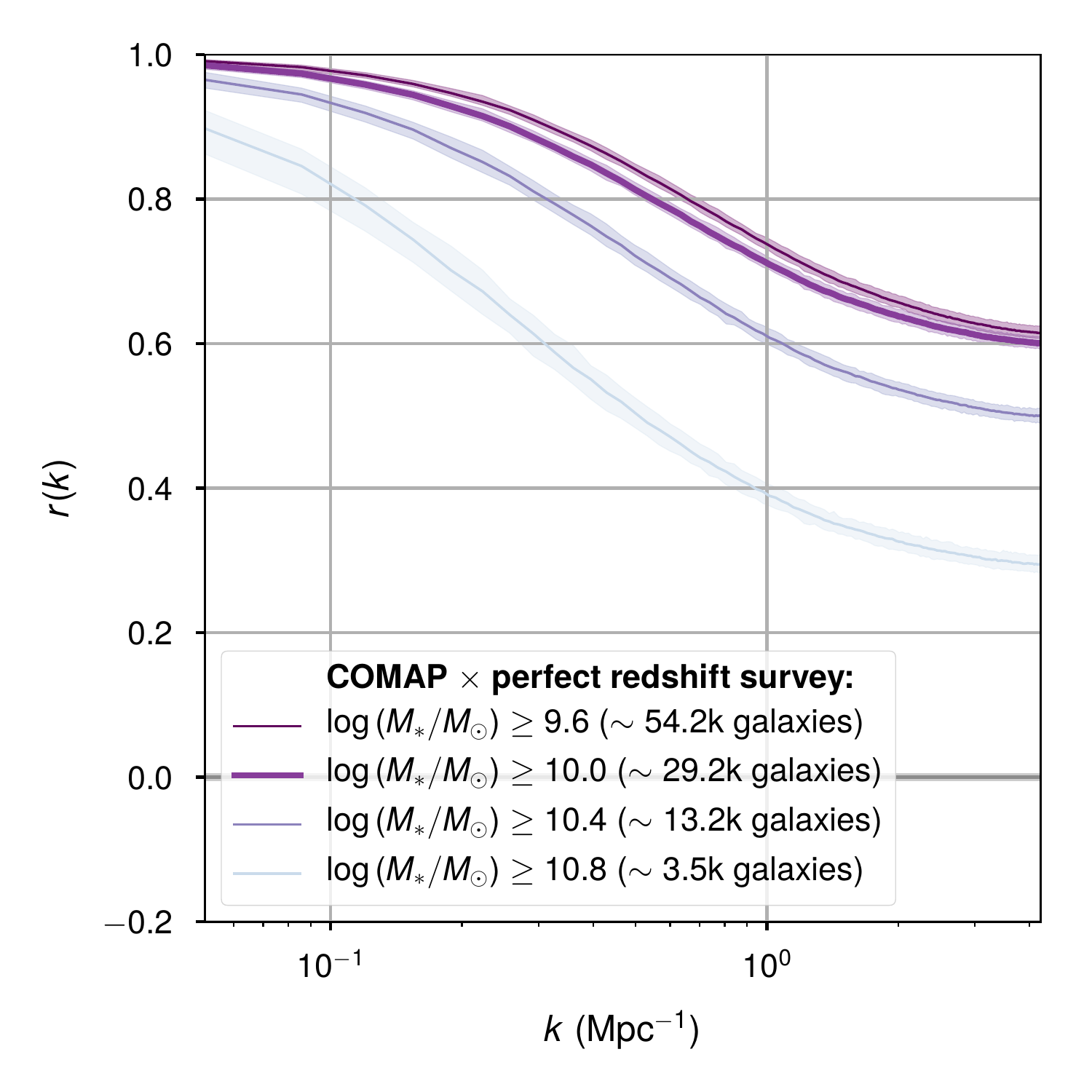}

\caption{Median (curves) and 95\% sample intervals (shaded areas) across 100 lightcones of normalized cross-correlation coefficient $r(k)$ for simulated CO--galaxy cross-correlation, assuming a perfect redshift survey ($\sigma_z/(1+z)$ = 0). The different curves show $r(k)$ for different stellar mass cuts used to select the galaxy sample for cross-correlation, as indicated in the legend. These curves show the underlying $r(k)$ rather than a detectable signal, since all galaxy redshifts are assumed to be perfectly known in these simulations.}
\label{fig:Mhcuts}
\end{figure}

\autoref{fig:Mhcuts} shows how $r(k)$ varies based on mass-completeness of an ideal survey with perfectly determined redshifts. Note that since the CO emission traces faint galaxies well below halo masses of $10^{12}\,M_\odot$ in between the brighter galaxies with $M_\text{vir}\gtrsim10^{12}\,M_\odot$, $r(k)$ falls off with higher $k$ as the CO and galaxy surveys begin to trace less similar fluctuations at smaller comoving scales. The fall-off is greater with less complete surveys, and impacts all scales significantly once we reach $\sim4\times10^3$ galaxies in our survey volume (or densities of $\sim10^3$ galaxies per deg$^2$ per $\Delta z=1$).

We show signal-to-noise ratios for all simulated cross spectra in~\autoref{tab:COSMOS_SNR}. All of these cross spectra---even the one with the lowest assumed density---might be detected with a signal-to-noise ratio above 20, far higher than the 4.6 expected from the CO auto spectrum alone, at least provided that the galaxy sample fully covers the COMAP volume.

\subsubsection{Redshift Accuracy}
\begin{figure}[t]
\centering\includegraphics[width=\linewidth]{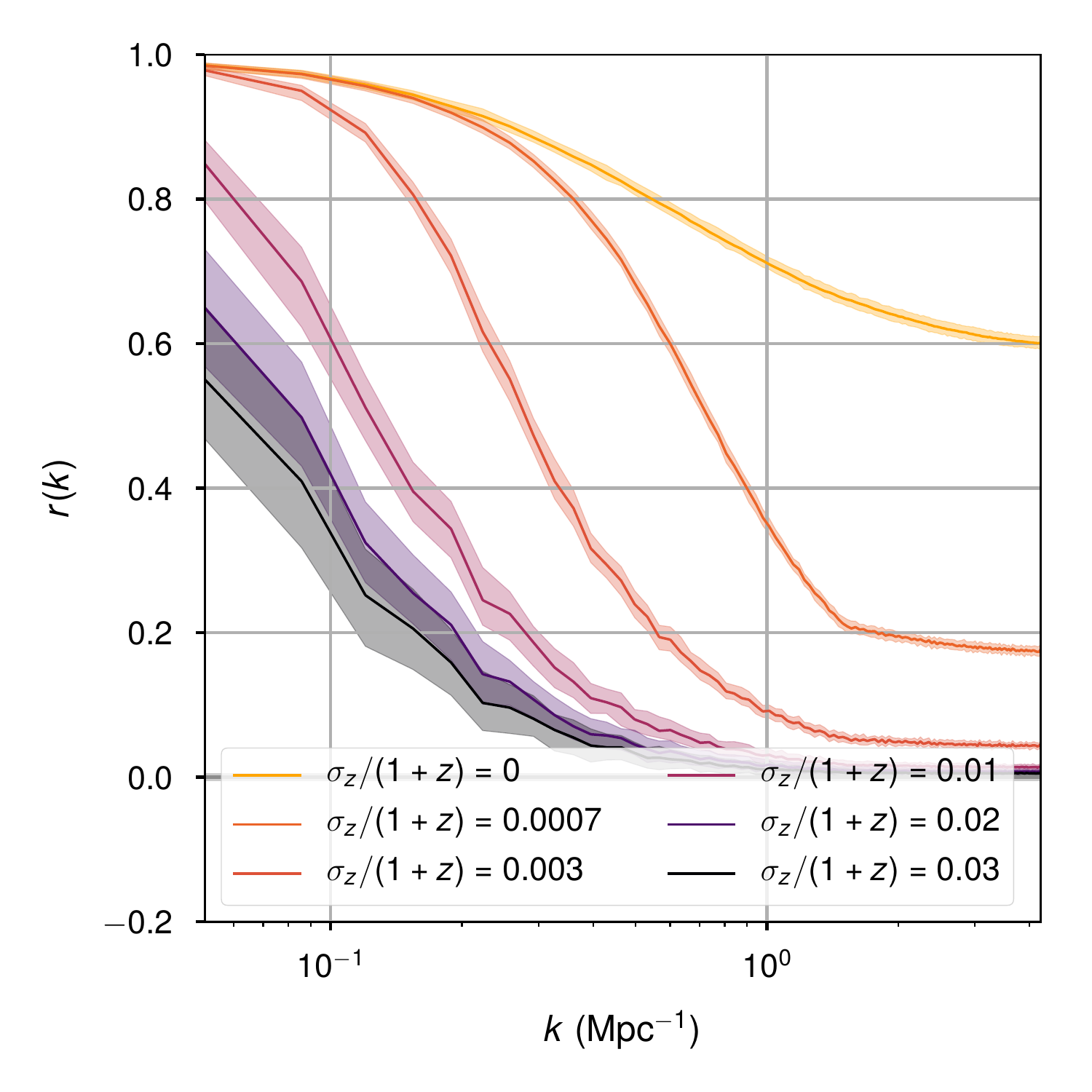}

\caption{Median (curves) and 95\% sample intervals (shaded areas) across 100 lightcones of normalized cross-correlation coefficient $r(k)$ for simulated CO--galaxy cross-correlation. The different curves show $r(k)$ for different redshift errors used to select the galaxy sample used in the cross-correlation exercise, as indicated in the legend. The galaxy samples in these simulations are selected based on a stellar mass cut of $\log{(M_\mathrm{*,min}/M_\odot)}=10.0$.}
\label{fig:pzerr}
\end{figure}

\begin{figure}[t]
\centering\includegraphics[width=\linewidth]{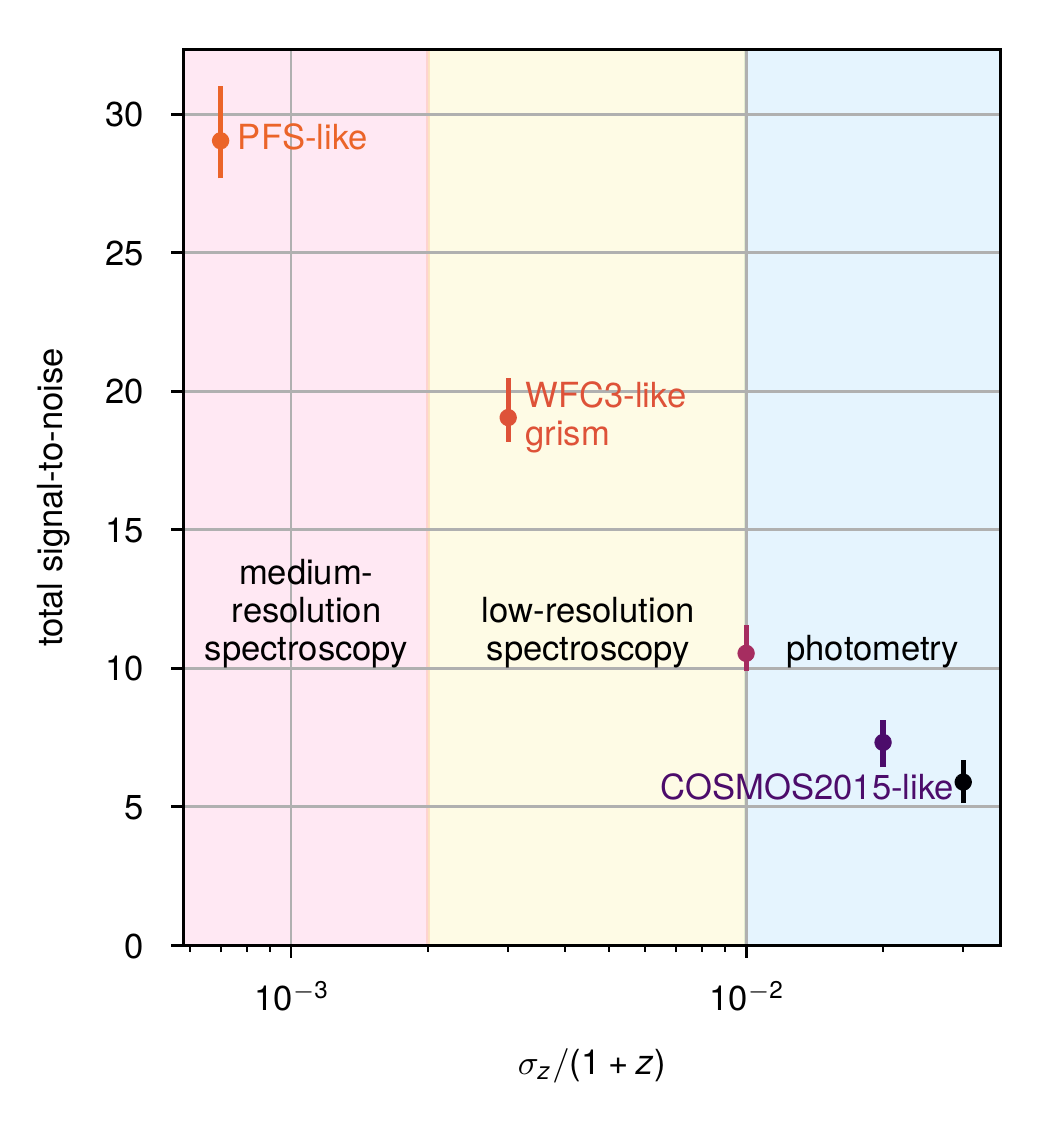}

\caption{Median (circles) and 95\% sample intervals (error bars) across 100 lightcones of total signal-to-noise over all scales $\text{S}/\text{N}=[\sum_k(\text{S}/\text{N})_k^2]^{1/2}$ for simulated CO--galaxy cross spectra for different galaxy $\sigma_z/(1+z)$ values. The annotations indicate what instrument, technique, or catalogue can achieve redshift accuracy broadly similar to each of our simulated scenarios. The galaxy sample is simulated with a minimum stellar mass of $\log{(M_\mathrm{*,min}/M_\odot)}=10.0$. All $\mathrm{S/N}$ are quoted for a single patch of 2.5 deg$^2$ observed for 1500 hours; we may expect up to a factor-of-$\sqrt{2}$ improvement if two equivalent patches are observed for 1500 hours each.}
\label{fig:pzerrSNR}
\end{figure}

Now fixing the mass-completeness cut at the fiducial value of $\log{(M_\mathrm{*,min}/M_\odot)}=10.0$, we vary the redshift accuracy in the survey. \autoref{fig:pzerr} shows the resulting $r(k)$ values for each value of $\sigma_z/(1+z)$ assumed.

Note that even for $\sigma_z/(1+z)=0.0007$, we find significant attenuation of \replaced{the shot noise cross-correlation}{cross-correlation at large $k$, suggesting the effect is particularly great for the cross shot-noise component of the signal (which dominates over the clustering component for $k\gtrsim1$ Mpc$^{-1}$)}.\footnote{While this redshift error seems well-matched to the spectral resolution of the COMAP data ($R=\lambda/\Delta\lambda=\nu_\text{obs}/\delta_\nu\sim2000$ corresponds to $\sigma_z/(1+z)=0.0005$), the COMAP channel width is in reality equivalent to a full width at half maximum, whereas the galaxy redshift errors have been described here as $1\sigma$ errors in each direction.} It is realistic to expect attenuation of this kind given the possible fitting errors in spectroscopic redshift determination, as well as mismatches in line-of-sight pixelization or resolution between the two survey data. This level of precision is also where redshift-space distortions \added{(RSD)} from galaxy peculiar velocities begin to distort line-of-sight structure\replaced{, so there is little scientific motivation for higher-resolution redshift surveys}{. As this precision is thus sufficient for galaxy redshift surveys looking for RSD, they have little motivation to pursue higher spectral resolution---e.g.~\citealt{Gaz12} and~\citealt{Eriksen15} demonstrate that $\sigma_z/(1+z)=0.003$ is adequate for the purposes of the PAU Survey (mentioned above in~\autoref{sec:mockCOSMOS})}.

For both $\sigma_z/(1+z)=0.0007$ and $\sigma_z/(1+z)=0.003$ (the higher- and lower-resolution spectroscopic errors), we see significant attenuation of the \replaced{shot noise cross spectrum}{cross-correlation at large $k$, i.e.~the smallest scales simulated,} but relatively little attenuation at the largest scales probed. This works to our advantage, as our single-dish line-intensity survey aims to detect CO fluctuations at these largest scales rather than the CO shot noise. The signal-to-noise reflects this, falling only from 32.4 to 29.1 if $\sigma_z/(1+z)=0.0007$, and then to 19.1 if we increase $\sigma_z/(1+z)$ to 0.003.

\begin{deluxetable}{cccc}
\tabletypesize{\footnotesize}
\tablewidth{0.9\linewidth}
\tablecaption{\label{tab:COSMOS_SNR}
Mean over all simulated observations of 100 lightcones of total signal-to-noise ratio ($\mathrm{S/N}$) for $P_\text{CO$\times$gal}(k)$ over all modes.}
\tablehead{\\[-1em]$\log{(M_\mathrm{*,min}/M_\odot)}$ & $\sigma_z/(1+z)$ & Median galaxy count & $\mathrm{S/N}$}
\startdata
9.6 &0.    &$5.4\times10^4$&33.2\\
10.0&0.    &$2.9\times10^4$&32.4\\
10.4&0.    &$1.3\times10^4$&29.6\\
10.8&0.    &$3.5\times10^3$&22.7\\\hline
9.6 &0.0007&$5.4\times10^4$&30.1\\
10.0&0.0007&$2.9\times10^4$&29.1\\
10.4&0.0007&$1.3\times10^4$&26.6\\
10.8&0.0007&$3.5\times10^3$&20.4\\\hline
10.0&0.003 &$2.9\times10^4$&19.1\\
10.0&0.01  &$2.8\times10^4$&10.6\\
10.0&0.02  &$2.7\times10^4$&7.32\\
10.0&0.03  &$2.7\times10^4$&5.93\\
\hline
\enddata
\tablecomments{For comparison, the $\mathrm{S/N}$ for $P_\text{CO}(k)$ is 4.6. All signal-to-noise ratios are quoted for a single patch of 2.5 deg$^2$ observed for 1500 hours; we may expect up to a factor-of-$\sqrt{2}$ improvement if two equivalent patches are observed for 1500 hours each.}
\end{deluxetable}

These numbers become more discouraging as we approach errors more typical of wide-band photometric surveys, settling in a range closer to 6--11. We show the effect graphically in~\autoref{fig:pzerrSNR}, and again summarize the signal-to-noise ratios calculated in~\autoref{tab:COSMOS_SNR}. Even taking these ratios at face value, the high $\sigma_z$ significantly dulls the advantage of cross-correlation over auto-correlation in detection significance. We must also carefully consider the integration time of 1500 hours per patch assumed for all scenarios. In the specific case of COMAP, when observing from the site in California, this integration time takes 2--3 times longer to achieve on an equatorial field like COSMOS versus on a field at 50--$70^\circ$ declination. If we had fixed the `real' survey duration for all scenarios instead of the integration time, we would expect to see no advantage in detection significance from cross-correlating against a COSMOS2015-like galaxy catalogue over CO autocorrelation in a field of our choice.

\begin{figure}[t]
\centering\includegraphics[width=\linewidth]{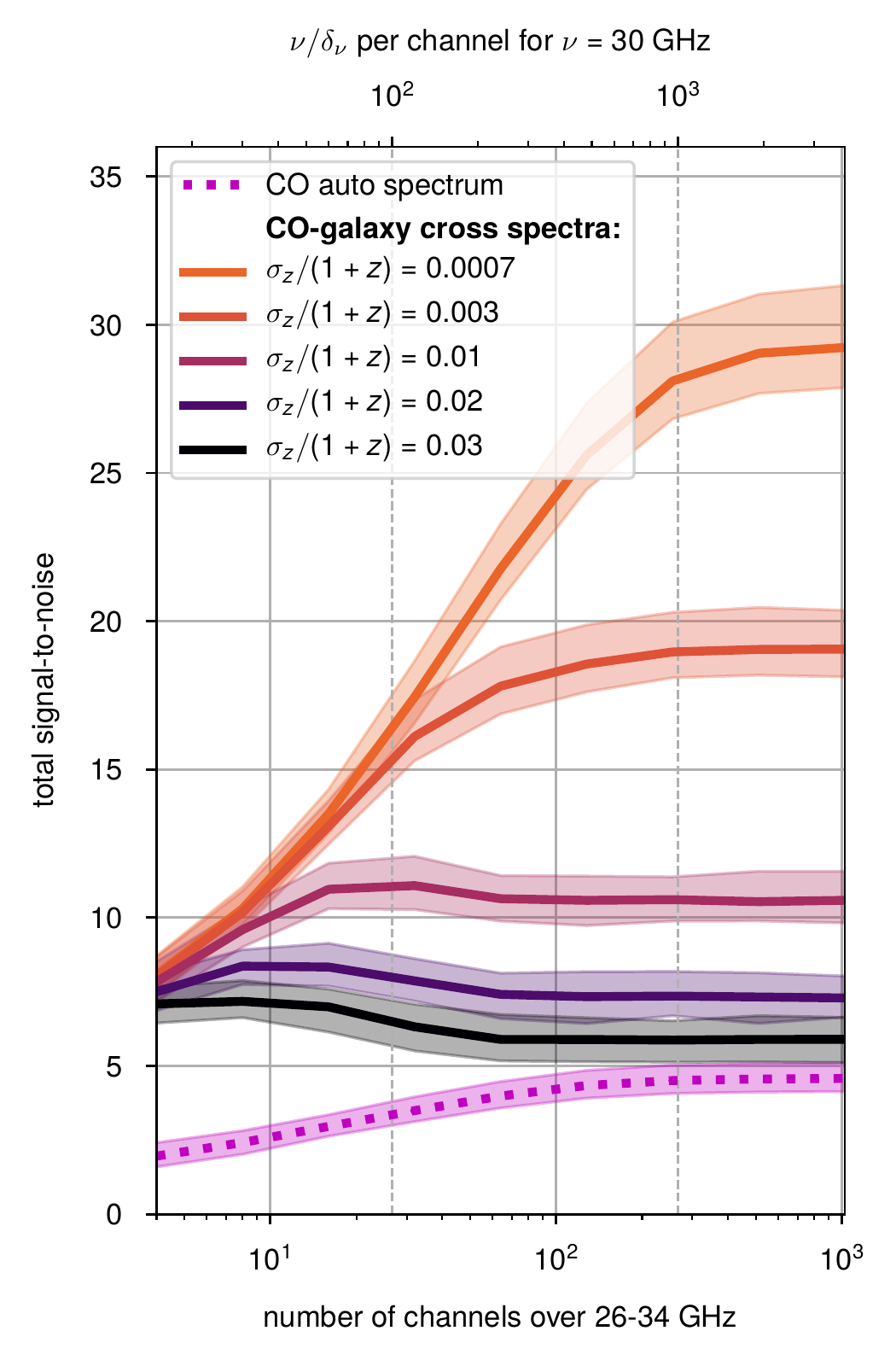}

\caption{A demonstration of the effect of COMAP line-of-sight resolution on the signal-to-noise ratio for auto and cross spectra. We express frequency resolution here as number of channels across the spectrometer bandwidth (also expressed as $\nu/\delta_\nu$ per channel for $\nu = 30$ GHz by multiplying by 3.75), and show how it affects total signal-to-noise over all scales $\text{S}/\text{N}=[\sum_k(\text{S}/\text{N})_k^2]^{1/2}$ for simulated CO auto spectra and CO--galaxy cross spectra for different galaxy $\sigma_z/(1+z)$ values. The thick curves and shaded areas show the median and 95\% interval across 100 lightcones. The simulated galaxy sample is selected with a minimum stellar mass of $\log{(M_\mathrm{*,min}/M_\odot)}=10.0$. All signal-to-noise ratios are quoted for a single patch of 2.5 deg$^2$ observed for 1500 hours; we may expect up to a factor-of-$\sqrt{2}$ improvement if two equivalent patches are observed for 1500 hours each.}
\label{fig:chanSNR}
\end{figure}

One natural step we might take to ameliorate the problem of photometric redshift errors is to coarsen the line-of-sight resolution of the data, so as to make the redshift errors less relevant. While this will result in boosting the signal closer to its true value by essentially removing attenuated line-of-sight modes from consideration, it will also result in increased uncertainties in the end result as the Fourier-space volume and thus the number of modes decreases. The net result is largely a loss in total cross signal-to-noise, as we show in~\autoref{fig:chanSNR}, and only a slight gain for photometric scenarios of $\sigma_z/(1+z)\geq0.01$\added{ (which plateaus when $\sigma_z/(1+z)\approx\delta_\nu/\nu$)}.

\subsection{Cross-correlations with a HETDEX-like Survey}
\label{sec:HETDEXres}

We consider two variations on cross-correlating against a HETDEX-like survey: one in which we cross-correlate against a Lyman-$\alpha$ intensity cube, and one in which we cross-correlate against the LAE overdensity field with different luminosity cuts, as discussed in~\autoref{sec:mock}. We fix $\sigma_z/(1+z)=0.0015$ in all cases, however.

\begin{figure}[t]
\centering\includegraphics[width=\linewidth]{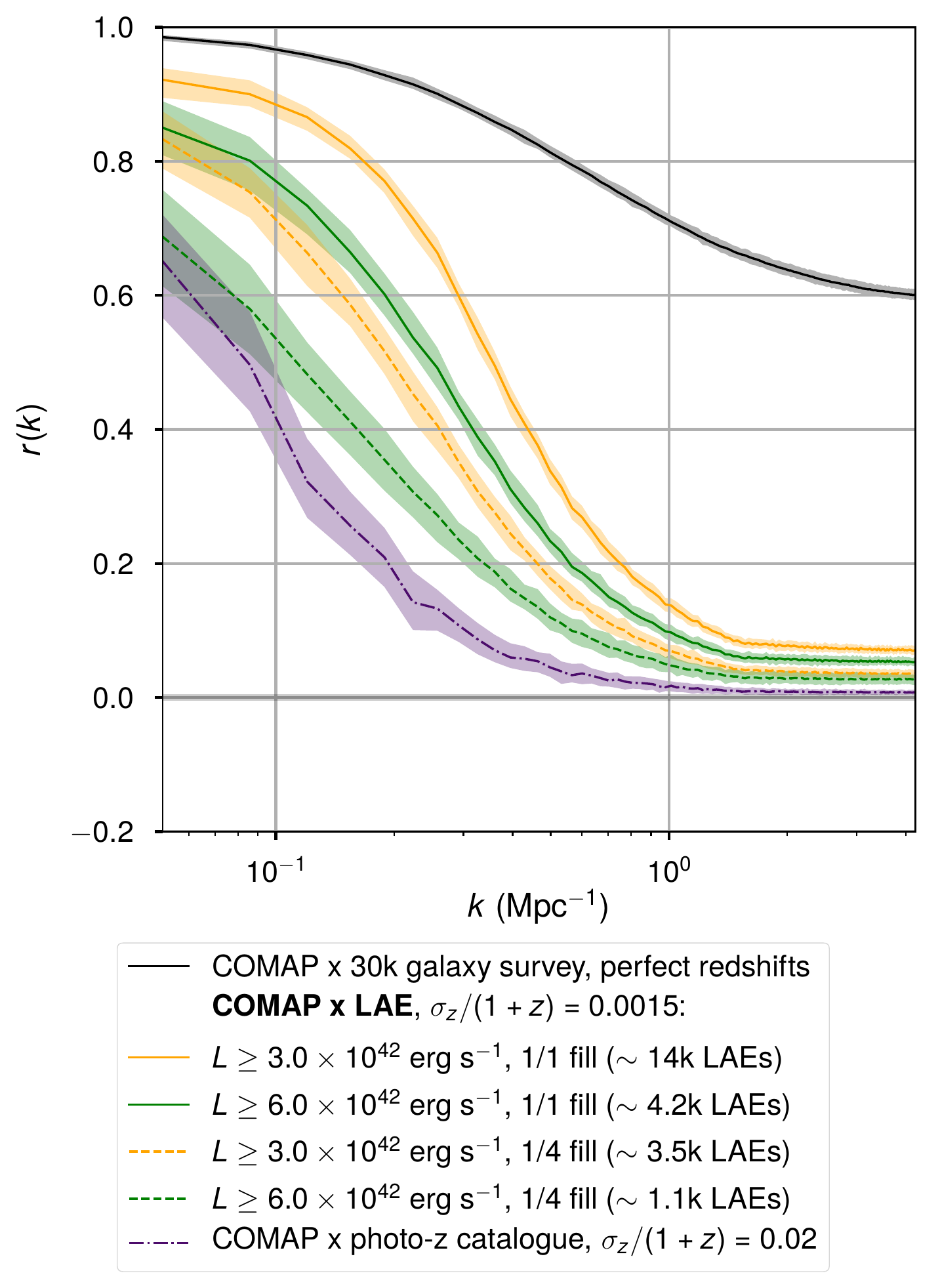}

\caption{Median (curves) and 95\% sample intervals (shaded areas) across 100 lightcones of normalized cross-correlation coefficient $r(k)$ for simulated CO--LAE cross-correlation. The different curves show $r(k)$ for different $L_\text{Ly$\alpha$}$ cuts used to select the LAE sample used in the cross-correlation, both with and without sparse sampling. LAE redshifts are scattered by a normal distribution with $\sigma_z/(1+z)=0.0015$. For comparison, we also show $r(k)$ from cross-correlation of COMAP with a galaxy survey with a minimum stellar mass of $\log{(M_\mathrm{*,min}/M_\odot)}=10.0$, assuming both perfect redshifts (black) and scattered redshifts with $\sigma_z/(1+z)=0.02$ (purple).}
\label{fig:Lyacuts}
\end{figure}

We first consider the COMAP $\times$ LAE scenario, and show the simulated $r(k)$ in~\autoref{fig:Lyacuts}. For luminosity cuts of $(3\times10^{42},6\times10^{42})$ erg s$^{-1}$, we find on average $(1.4\times10^4,4.2\times10^3)$ LAEs in the survey volume, with approximately $1/4$ as many LAEs when the volume is sparsely sampled with a fill factor of $1/4$. The number of LAEs with $L_\text{Ly$\alpha$}>3\times10^{42}$ erg s$^{-1}$ approximately matches the expected source abundance in~\cite{HETDEX} of $8\times10^5$ LAEs across $\sim400$ (sparsely sampled) square degrees in a redshift interval of $\Delta z=1.6$.

If the HETDEX data filled the survey volume completely, as in the SHELA field, the COMAP $\times$ LAE cross spectrum would be detectable with total signal-to-noise as high as $20.7$ for $L_\text{Ly$\alpha$}>3\times10^{42}$ erg s$^{-1}$, even with the LAE redshifts scattered by $\sigma_z/(1+z)=0.0015$ (without which the $\mathrm{S/N}$ might be higher by around 30\%). However, with the $1/4$ fill factor, the $\mathrm{S/N}$ does drop to $14.5$. The numbers are lower by 20--25\% for the more stringent cut of $L_\text{Ly$\alpha$}>6\times10^{42}$ erg s$^{-1}$.

\begin{figure}[t]
\centering\includegraphics[width=\linewidth]{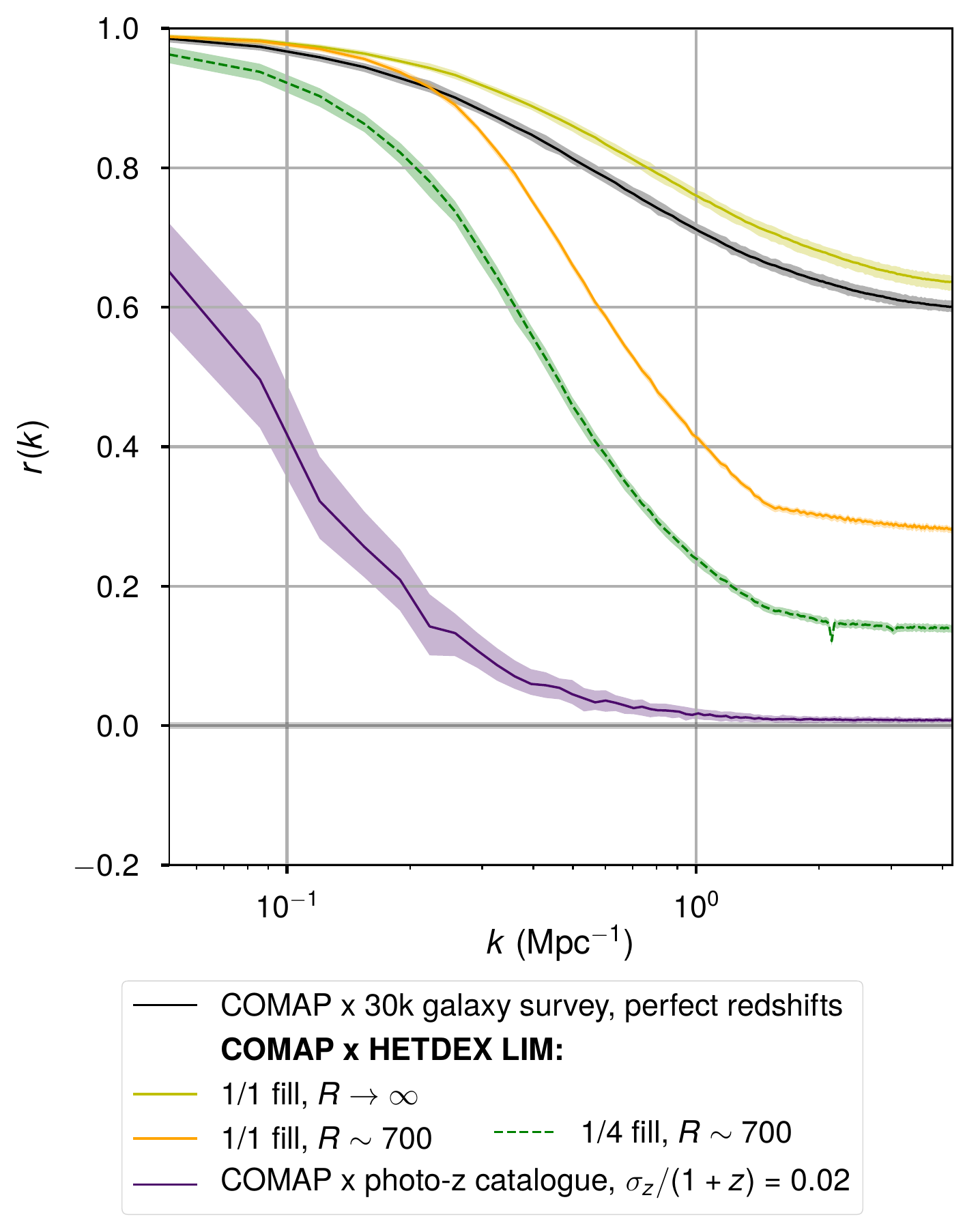}

\caption{Median (curves) and 95\% sample intervals (shaded areas) across 100 lightcones of normalized cross-correlation coefficient $r(k)$ for simulated CO--Ly$\alpha$ intensity cross-correlation. The different curves show $r(k)$ for different assumptions of sparse sampling and HETDEX resolution. Apart from the $r(k)$ curve that we label `$R\to\infty$', the COMAP--HETDEX $r(k)$ curves are attenuated by the amount expected for the VIRUS resolving power of $R\sim700$. We show the same CO--galaxy $r(k)$ curves for comparison as we did in~\autoref{fig:Lyacuts}.}
\label{fig:LyaLIM}
\end{figure}

We now consider cross-correlating against a Lyman-$\alpha$ intensity cube generated from HETDEX, showing $r(k)$ in~\autoref{fig:LyaLIM}. Compared to cross-correlation against LAE overdensity, the roll-off of $r(k)$ with greater $k$ is slower, even with sparse sampling and limited spectral resolution. The signal is potentially detectable at a signal-to-noise of \replaced{26.0}{29.1} without sparse sampling, and a signal-to-noise of \replaced{15.9}{23.3} with sparse sampling. This results in a slight edge versus cross-correlating against individually identified LAEs, although the simulated advantage may change with the Lyman-$\alpha$ model---see~\autoref{sec:LIMvsLAE} (and~\autoref{sec:Lyamodelbad}) for further discussion.

\begin{deluxetable}{ccc}
\tabletypesize{\footnotesize}
\tablewidth{0.9\linewidth}
\tablecaption{\label{tab:HETDEX_SNR}
Mean over all simulated observations of 100 lightcones of total signal-to-noise ratio ($\mathrm{S/N}$) for $P_\text{CO$\times$Ly$\alpha$}(k)$ over all modes.}
\tablehead{\\[-1em]$L_\text{Ly$\alpha$,min}$ (erg s$^{-1}$) & Median LAE count & $\mathrm{S/N}$}
\startdata
none (LIM)&---&\replaced{26.0 (15.9)}{29.1 (23.3)}\\
$3\times10^{42}$&$1.4\times10^4$ ($3.5\times10^3$)&21.2 (14.5)\\
$6\times10^{42}$&$4.2\times10^3$ ($1.1\times10^3$)&16.7 (10.6)\\
\hline
\enddata
\tablecomments{Counts and $\mathrm{S/N}$ in parentheses are for $1/4$-fill sparse sampling; counts and $\mathrm{S/N}$ not in parentheses are for full-fill sampling. We assume $\sigma_z/(1+z)=0.0015$ in all cases. For comparison, the $\mathrm{S/N}$ for $P_\text{CO}(k)$ is 4.6. All $\mathrm{S/N}$ are quoted for a single patch of 2.5 deg$^2$ observed for 1500 hours; we may expect up to a factor-of-$\sqrt{2}$ increase if two equivalent patches are observed for 1500 hours each and a further roughly linear increase with more integration time.}
\end{deluxetable}

\begin{figure}[t]
\centering\includegraphics[width=\linewidth]{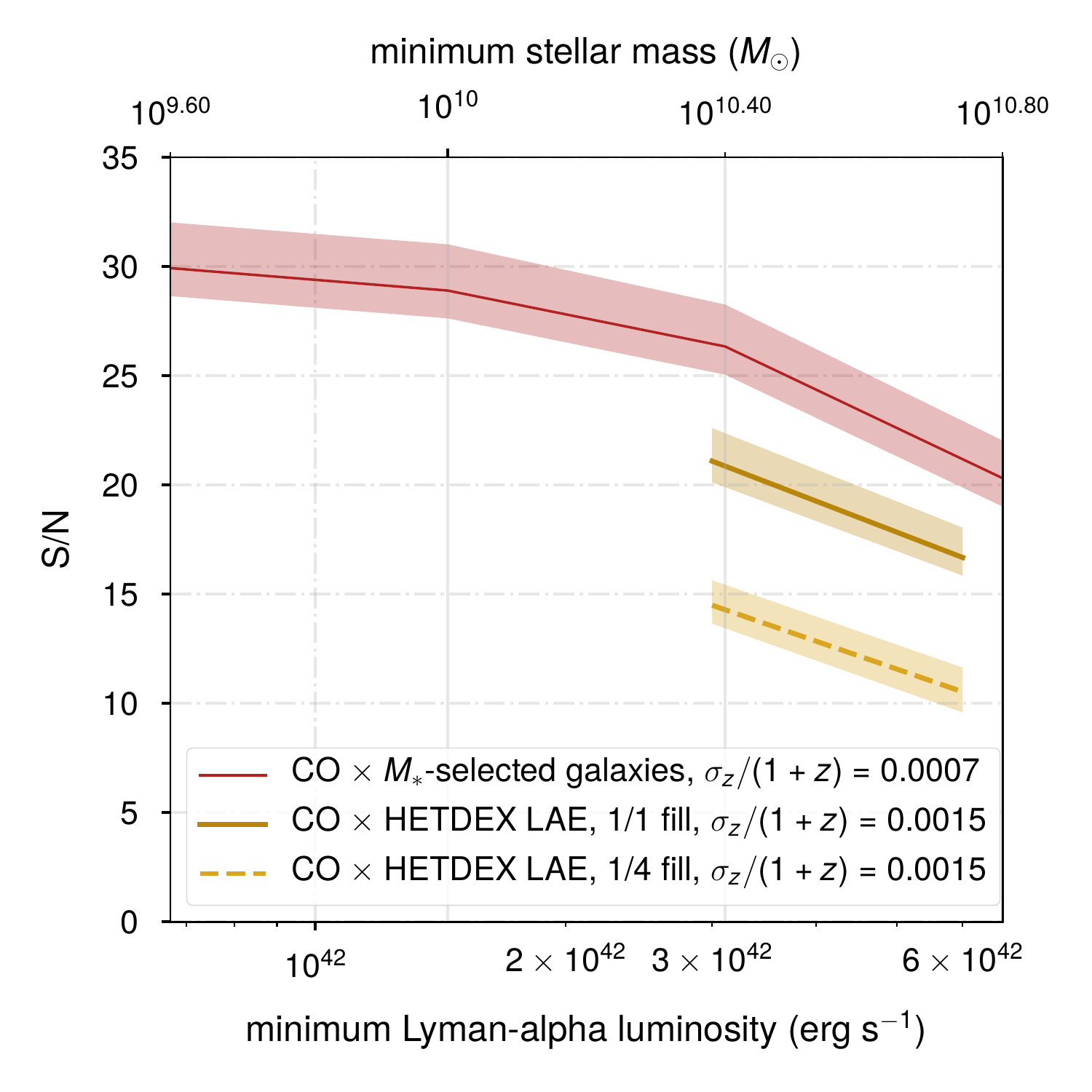}

\caption{Median (curves) and 95\% sample intervals (shaded areas surrounding curves) of cross spectrum signal-to-noise ratio for cross-correlation of CO temperature against mass-selected galaxies and LAEs (without and with sparse sampling), as a function of minimum stellar mass or Lyman-$\alpha$ luminosity. We loosely align the two survey limit metrics based on source abundances as reported in~\autoref{tab:COSMOS_SNR} and~\autoref{tab:HETDEX_SNR}---e.g.~cuts of $M_*>10^{10.4}\,M_\odot$ and $L_\mathrm{Ly\alpha}>3\times10^{42}$ erg s$^{-1}$ both result in $\sim10^4$ sources in a 2.5 deg$^2$ patch of the COMAP survey volume, or $\sim5\times10^3$ sources per deg$^2$ per $\Delta z=1$. Note, however, the different redshift resolutions assumed for the mass-selected galaxy sample and the LAE sample. All signal-to-noise ratios are quoted for a single 2.5 deg$^2$ patch observed for 1500 hours; we may expect up to a factor-of-$\sqrt{2}$ increase if two equivalent patches are observed for 1500 hours each.}
\label{fig:galSNR}
\end{figure}

We summarize the signal-to-noise ratios from COMAP--HETDEX cross-correlation (and expected LAE counts for applicable scenarios) in~\autoref{tab:HETDEX_SNR}, and compare the LAE cross-correlation signal-to-noise graphically against signal-to-noise from cross-correlation against a mass-selected galaxy sample in~\autoref{fig:galSNR}. Note that unlike the COSMOS field, the HETDEX survey footprint is partly well-matched with areas of relatively high observing efficiency for COMAP. Therefore, a CO observing campaign with sufficient data (in one patch of several) for a $\sim5\sigma$ CO auto detection could readily overlap with HETDEX to generate a $\sim15\sigma$ detection in cross-correlation, per the signal-to-noise ratios in~\autoref{tab:HETDEX_SNR}.

\subsection{Final Summary: Power Spectra and Sensitivities}
\label{sec:allres}
\begin{figure*}[t]
\centering\includegraphics[width=0.92\linewidth]{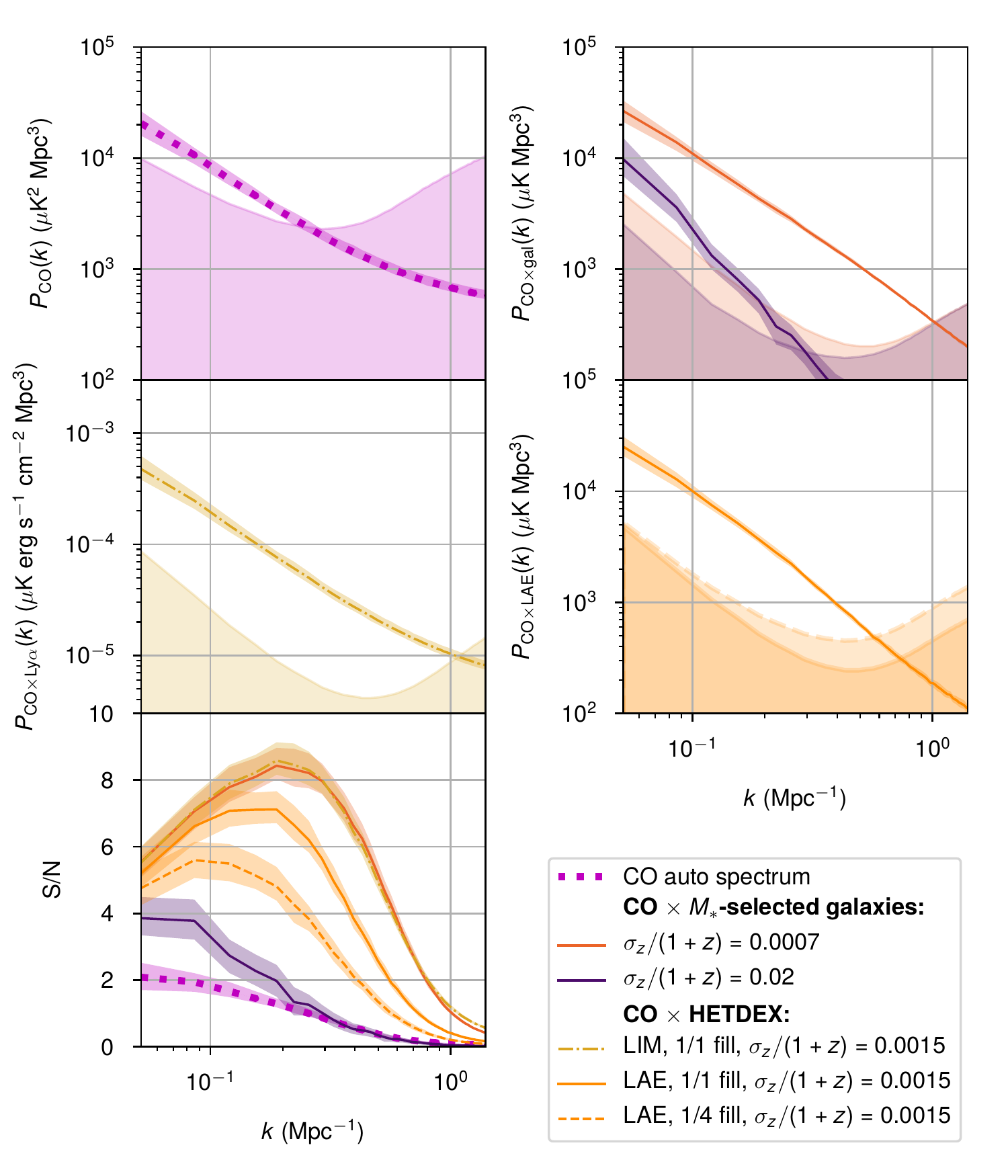}

\caption{\emph{Upper four panels:} Median (curves) and 95\% sample intervals (shaded areas surrounding curves) across 100 lightcones of indicated auto and cross spectra, with sensitivities indicated by the shaded areas of corresponding colour and line style extending to the lower limit of each panel. The $M_*$-selected galaxies exceed a minimum stellar mass of $10^{10}$ $M_\odot$, and the LAE samples exceed a minimum Ly-$\alpha$ luminosity of $3\times10^{42}$ erg s$^{-1}$. Unlike the CO $\times$ $M_*$-selected galaxy and CO $\times$ HETDEX LAE cross spectra, the CO $\times$ HETDEX LIM cross spectra do not account for the effect of redshift precision, which is folded instead into the sensitivity curve. \emph{Lower panel:} Median (curves) and 95\% sample intervals (shaded areas surrounding curves) of signal-to-noise ratio for all auto and cross spectra as a function of $k$, calculated in $k$-bins of width $\Delta k=0.035$. All $\mathrm{S/N}$ are quoted for a single patch of 2.5 deg$^2$ observed for 1500 hours; we may expect up to a factor-of-$\sqrt{2}$ increase if two equivalent patches are observed for 1500 hours each and a further roughly linear increase with more integration time.}
\label{fig:bigsensitivityplot}
\end{figure*}

To end this section, we show a plot of all auto and cross $P(k)$ with sensitivities in~\autoref{fig:bigsensitivityplot}. Note the shape of the CO auto $P(k)$, which flattens beyond $k\sim1$ Mpc$^{-1}$ as the shot-noise component of the power spectrum begins to dominate over the clustering component following the underlying matter distribution. Any such shot-noise component in the cross spectra is far less apparent, as expected from the random redshift errors wiping out smaller-scale correlations. (The exception is the CO $\times$ HETDEX LIM cross $P(k)$ plotted, which do not incorporate this effect, as it is folded into the accompanying sensitivity curve instead. Thus a shot-noise component is visible for this set of $P(k)$.) This matches what we also demonstrate in the $r(k)$ plots of~\autoref{fig:pzerr} and~\autoref{fig:Lyacuts}.

Note also that the shapes of the sensitivity curves clearly show the impact of cross-correlating against data with significantly higher angular resolution than the COMAP data (and thus reducing $\sigma_\perp$ in~\autoref{eq:Wbeam} by a factor of $\sqrt{2}$). We also plot the signal-to-noise ratio at each $k$ for all spectra considered, which shows the same.

Note finally that we find a different total signal-to-noise ratio if we consider uncertainties on the anisotropic power spectrum $P(k,\mu)$ (where $\mu=k_\parallel/k$ is the cosine of the $k$-space spherical polar angle, using $k_\parallel$ to describe the line-of-sight component of the vector $\mathbf{k}$) instead of the spherically-averaged $P(k)$, and may be higher for galaxy samples with $\sigma_z\geq0.01$. However, the enhancement is not enough to alter the fundamental conclusions of our work---see~\autoref{sec:PkmuSNR} for further discussion.

\section{Discussion}
\label{sec:discussion}
Broadly, the above results show that photometric redshift errors significantly attenuate the CO--galaxy cross-correlation at all scales, and overlapping with a spectroscopic dataset is the best path to a confident detection. We provide additional discussion of specific topics driving these results.

\subsection{Effects of Redshift Errors}
Our treatment of the effects of redshift errors is considerably simplified from what we may expect from real-world survey data. In particular, we simulate no bias and no catastrophic outliers for photometric redshifts, both of which could potentially further affect the cross-correlation signal.

Furthermore, photometric redshift algorithms usually compute a redshift probability density function (PDF) for each galaxy, which can be reduced to the point estimate that we consider. Indeed,~\cite{Asorey16} propose using the full redshift PDF for each galaxy in studies of angular galaxy clustering within bins of roughly twice the typical PDF widths.

However, the technique is of limited applicability when using line-intensity mapping to probe 3D clustering. In general, techniques using galaxy redshift PDFs will not (and are not meant to) recover the true redshift of the galaxy or the true galaxy density fluctuations in the survey volume. In fact, using the redshift PDFs for each galaxy instead of the estimated redshifts will merely convolve the PDF (around the estimated redshift) with the scatter distribution of the estimated redshift (versus the true redshift). This can only result in additional line-of-sight smearing and thus attenuation of the signal.

Furthermore, the detectability of the line-intensity signal relies in large part on the decrease in uncertainty from being able to access the line-of-sight fluctuation modes. In~\autoref{fig:chanSNR}, where we actually explore the idea of wider frequency channels (albeit not the idea of using full redshift PDFs), the signal-to-noise remains low despite recovery of the signal due to the loss of line-of-sight information.

Ultimately we must contend with some suppression of the cross spectrum signal from errors in galaxy redshifts, even for spectroscopic redshifts. In principle, we might consider calculating a transfer function to compensate for this suppression. However, the signal being suppressed in the first place diminishes our confidence in its detection, regardless of whether or not we can undo the suppression in analysis.

Furthermore, an accurate transfer function will require accurate and precise characterization of redshift errors, both of the mean offset or bias, and of the variance or scatter. For comparison, when binning populations of galaxies in a photometric survey to gauge cosmological parameters, both bias and scatter in redshift must be characterized within 0.003 to limit significant degradation in error, even using bins of $\Delta z\sim0.1$~\citep{Huterer06,Ma06}. In practice, line-intensity mapping can probe fluctuations across bins of $\Delta z$ as fine as $\sim0.001$, although the science is often more astrophysical than cosmological, especially in deep surveys with small sky fractions.

Such requirements in and of themselves should not render impossible studies of shot noise cross-correlation of the kind posited by~\cite{Wolz17} (which proposes cross-correlation against a spectroscopic galaxy survey, for which the errors may be more precisely characterized than for photometric redshifts) and~\cite{BreysseRahman17} (if extended to cross-correlations between line-intensity surveys). However, future studies may wish to take careful inventory of expected redshift errors. A thorough study of the feasibility of such techniques and the impact of these effects on shot noise detection significance is beyond the scope of this particular work, but still highly desirable as these cross-correlations become increasingly viable.

\subsection{The Slim Advantage of Lyman-$\alpha$ Line-Intensity Mapping over Individual Galaxy Detections}
\label{sec:LIMvsLAE}

Recall that we consider two possible outputs of HETDEX: the LAE overdensity $\delta_\mathrm{LAE,vox}$ for all emitters above a certain luminosity cut, and the total Lyman-$\alpha$ line-intensity quantified as $\nu_\mathrm{Ly\alpha}I_{\nu,\mathrm{Ly\alpha}}$. One might expect that since the latter includes Lyman-$\alpha$ emission from emitters below a realistic luminosity cut, the line-intensity cube should trace structures absent in the $\delta_\mathrm{LAE,vox}$ cube, and potentially significantly improve detectability in cross-correlation against $T_\mathrm{CO}$.

However, our simulated cross-correlation of $T_\text{CO}$ against $\nu_\mathrm{Ly\alpha}I_{\nu,\mathrm{Ly\alpha}}$ only provides a slim advantage in signal-to-noise---\replaced{10--25\% at best,}{40--60\%} based on the figures in~\autoref{tab:HETDEX_SNR}---over cross-correlation against LAEs detected with $L_\text{Ly$\alpha$}>3\times10^{42}$ erg s$^{-1}$\replaced{, especially after sparse sampling}{. This seems significant enough until considering that the signal-to-noise ratio is still about the same as a conventional spectroscopic galaxy survey with $M_{*,\text{min}}\sim10^{10}\,M_\odot$. Furthermore, in our model, LAEs with $L_\text{Ly$\alpha$}>3\times10^{42}$ erg s$^{-1}$ are typically in halos with $M_\text{vir}\gtrsim10^{12}\,M_\odot$ (looking at~\autoref{fig:hmll}), and the line-intensity data should trace twice as much signal or more by including emission from lower-mass halos (looking at~\autoref{fig:Lya_mmin} in~\autoref{sec:lyamodel_mmin})}\footnote{The auto power spectrum signal-to-noise is also quite similar for HETDEX as LIM survey and HETDEX as LAE survey\replaced{ for full-fill fields}, with a signal-to-noise ratio \replaced{of up to 81 possible in both scenarios}{of up to 113 for HETDEX LIM and 81 for HETDEX LAE} before sparse sampling, even across the limited sky area and range of $k$ considered in this work. With $1/4$-fill sparse sampling this ratio drops to \replaced{37}{104} for HETDEX LIM and 59 for HETDEX LAE.}. Note first that this ignores both the additional information that may be captured in line-intensity mapping (but is beyond the scope of our LAE-based simulations) and the additional challenges that would be inherent in line-intensity mapping, including contamination from zodiacal light and interloper emission in [\ion{O}{2}] \added{(the latter of which would be rejected in COMAP--HETDEX cross-correlation but would nonetheless contribute additional uncertainty about that cross signal\footnote{\added{See also~\citealt{Gong14} for a more direct masking-based rejection of [\ion{O}{2}] emission and a discussion of benefits and limitations of mitigation through cross-correlation. That work considers Lyman-$\alpha$ emission from $z\sim7$ but the broad conclusions should be extensible to $z\sim3$.}})}. With these caveats in mind, we should ask why overcoming such challenges and observing the full LAE population would appear to result in such little improvement in forecast $\mathrm{S/N}$.

One possible explanation may be the halo mass--line luminosity relations in~\autoref{fig:hmll}. The large step down in Lyman-$\alpha$ escape fraction around $\operatorname{SFR}\sim1\,M_\odot\,yr^{-1}$ or halo mass $M_\text{vir}\sim10^{11}\,M_\odot$ means that the slope of the halo mass--$L_\text{Ly$\alpha$}$ relation begins to decline above this mass, before the turnaround at $10^{12}\,M_\odot$ seen in both this relation and in the halo mass--$L_\text{CO}$ relation. This would lead to differences between the CO and Lyman-$\alpha$ signals in the relative contributions of halos with $M_\text{vir}\lesssim10^{11}\,M_\odot$ and those with $M_\text{vir}\gtrsim10^{12}\,M_\odot$, which may adversely affect the cross-correlation of the line-intensity signals. However, whether only the most massive halos behave similarly enough to co-correlate would certainly be a highly model-dependent effect. Furthermore, the Lyman-$\alpha$ relation is based solely on observed LAE densities, which may not result in a complete model of Lyman-$\alpha$ emission (as discussed briefly in~\autoref{sec:hmll}).

Another (not mutually exclusive) possibility is that LAE detection, and not Lyman-$\alpha$ intensity mapping, may actually be the optimal observation for HETDEX, which has low instrumental noise and high angular resolution and thus low source confusion. \replaced{Cheng et al.~(2018 in prep.)}{\cite{Cheng18}}~provide an overview of optimal observations in different noise and confusion regimes using an analytic source model. The work suggests that detecting individual sources rather than aggregate line-intensity mapping is the optimal observation for HETDEX, as well as for the higher-redshift Lyman-$\alpha$ observations that would be possible with the Cosmic Dawn Intensity Mapper~\citep[CDIM;][]{CDIM}, although not for the SPHEREx concept~\citep{SPHEREx}.

However, \replaced{Cheng et al.~(2018 in prep.)}{\cite{Cheng18}}~are also careful to note that the Lyman-$\alpha$ model used only incorporates point sources \added{(just as our own model relies only on LAE luminosity functions)} and does not take into account the expected extended emission from radiative transfer beyond galaxies. We have also discussed this as a limitation of our model, and repeat our caveat from~\autoref{sec:hmll} that a need exists for future work on implications of extended Lyman-$\alpha$ emission for HETDEX and COMAP--HETDEX cross-correlation.

\section{Conclusions}
\label{sec:conclusions}

We find that cross-correlation of COMAP Phase I data with galaxy surveys, both targeted galaxy surveys and blind Lyman-$\alpha$ surveys, could result in high signal-to-noise detections, but not unconditionally.
\begin{itemize}
\item With perfect or at least very precise redshift knowledge, the exercise could be done with as few as several thousand sources covering the COMAP survey volume, corresponding to a source abundance of $\gtrsim10^{-4}$ Mpc$^{-3}$ or $\gtrsim10^2$ per square degree per $\Delta z=0.1$.
\item However, to provide a significant advantage in cross-correlation alone over auto-correlation alone in signal-to-noise, the galaxy catalogue must achieve a redshift accuracy of $\sigma_z/(1+z)\lesssim0.003$, which is best obtained with low- to medium-resolution spectroscopy and will be challenging at best with photometry at high redshift.
\item If the redshift accuracy and source density satisfy the above, cross-correlations could result in a cross spectrum detection at a signal-to-noise of up to 15--30, compared to $\mathrm{S/N}\lesssim5$ for a CO auto spectrum detection (in a single patch). We expect this to be true in the case of cross-correlation with HETDEX, although this (and cross-correlation with Lyman-$\alpha$ surveys in general) requires further investigation with more faithful treatment of radiative processes.
\end{itemize}

We take care to note that targeting specific fields like COSMOS and targeting auto-correlation may not be mutually exclusive in general. Choosing one of the two is a necessity for noise-dominated surveys operating from sites where such fields cannot be mapped with high observing efficiency, but some line-intensity surveys may be in a position to observe a field like COSMOS with high observing efficiency. CONCERTO\footnote{\url{https://people.lam.fr/lagache.guilaine/CONCERTO.html}}, for instance, is to operate on the Atacama Pathfinder EXperiment (APEX) antenna located on Llano de Chajnantor, which is well-suited for observing the COSMOS field and other equatorial fields (being located at $23.01^\circ$ south latitude, versus the $37.23^\circ$ north latitude of the OVRO site). Catalogues in those fields may be able to provide priors for point sources and low-redshift CO emitters acting as foregrounds for the high-redshift [\ion{C}{2}] emission that CONCERTO targets. In particular, cross-correlation between $z\sim0.5$--2 galaxy surveys and CO emission captured in CONCERTO and other [\ion{C}{2}] experiments could enable novel science at both intermediate and high redshift, but we leave this possibility for others to investigate in future work. Overlapping with galaxy survey fields may also allow use of line-intensity maps as spectroscopic references (in addition to e.g.~quasars) for inferring clustering-redshift distributions of catalogued objects (see~\citealt{Menard13} for a description of the clustering redshift technique).

Furthermore, raw projected signal-to-noise is not an adequate singular basis for dismissing cross-correlating against less precise redshifts. Qualitative differences between the interpretation of an autocorrelation and that of a cross-correlation mean that even with the same signal-to-noise, a cross-correlation measurement may lend more confidence about the origin of the CO signal, and allow for better rejection of systematics and uncertainties beyond the fundamental sources of noise accounted for here. (HI intensity mapping provides a case study where auto spectra are biased by unknown systematics and a more robust constraint emerges from putting together the auto spectrum with the cross spectrum from cross-correlation of the HI data against a spectroscopic galaxy survey---see~\citealt{Switzer13}.) That said, we qualitatively expect and quantitatively confirm that the redshift accuracy typical of photometric surveys significantly affects our ability to detect 3D clustering in cross-correlation, and this gives us cause for concern in using photometric catalogues for cross-correlation against line-intensity surveys.

Existing photometric surveys should still inform cross-correlation prospects, as we may expect significant spectroscopic follow-up with instruments like PFS---and even surveys like HETDEX, which will use the COSMOS field for science verification and calibration~\citep{HETDEX2016}. However, we expect such surveys to operate beyond COSMOS---like in the SHELA patch, which already has deep multi-wavelength imagery in the optical and infrared through Spitzer/IRAC, NEWFIRM, and DECam, and will see full-fill HETDEX data in the next several years~\citep{SHELA,HETDEX2016} which could enable the high signal-to-noise cross-correlation detections discussed above. Cross-correlating against a photometric catalogue will not be a quick path to a detection for near-future line-intensity surveys like COMAP, but we find hope for future prospects as we wait for an influx of high-quality wide-field spectroscopic data in the coming years.

\acknowledgments{DTC, MPV, SEC, and RHW acknowledge support via NSF AST-1517598 and a seed grant from the Kavli Institute for Particle Astrophysics and Cosmology. KAC acknowledges funding from the NSF award AST-1518282 and the Keck Institute for Space Studies. HKE, MKF, HTI, and IKW acknowledge support from the Research Council of Norway through grant 251328. JOG acknowledges support from the Keck Institute for Space Studies, NSF AST-1517108, and the University of Miami. SEH acknowledges support from an STFC Consolidated Grant (ST/P000649/1). HP's research is supported by the Tomalla Foundation. We thank Tony Li for initial simulations and discussions that evolved into this work, and Clive Dickinson for crucial discussions and comments at the inception of this work. We also thank Llu\'{i}s Mas-Ribas for an enlightening discussion about Lyman-$\alpha$ blobs. Some of this work was presented and refined at the workshop `Cosmological Signals from Cosmic Dawn to the Present' held at the Aspen Center for Physics, which is supported by National Science Foundation grant PHY-1607611. We would like to acknowledge the organizers and participants of that workshop, including Yun-Ting Cheng for providing a draft version of work in preparation. We thank Matthew Becker for access to the Chinchilla cosmological simulation (\texttt{c400-2048}) used in this work.\added{ Finally, we thank an anonymous referee whose comments and suggestions greatly improved this manuscript.} This research made use of NASA's Astrophysics Data System Bibliographic Services. This work used computational resources at the SLAC National Accelerator Laboratory.}

\software{Astropy, a community-developed core Python package for astronomy~\citep{astropy}; Matplotlib~\citep{matplotlib}; \texttt{hmf}~\citep{hmf}.}
\bibliographystyle{aasjournal} 
\bibliography{correlate_references,Lya_refs}

\begin{thebibliography}{}
\expandafter\ifx\csname natexlab\endcsname\relax\def\natexlab#1{#1}\fi
\providecommand{\url}[1]{\href{#1}{#1}}

\bibitem[{{Adams} {et~al.}(2011){Adams}, {Blanc}, {Hill}, {Gebhardt}, {Drory},
  {Hao}, {Bender}, {Byun}, {Ciardullo}, {Cornell}, {Finkelstein}, {Fry},
  {Gawiser}, {Gronwall}, {Hopp}, {Jeong}, {Kelz}, {Kelzenberg}, {Komatsu},
  {MacQueen}, {Murphy}, {Odoms}, {Roth}, {Schneider}, {Tufts}, \&
  {Wilkinson}}]{HETDEXPilot}
{Adams}, J.~J., {Blanc}, G.~A., {Hill}, G.~J., {et~al.} 2011, \apjs, 192, 5

\bibitem[{{Anderson} {et~al.}(2018){Anderson}, {Luciw}, {Li}, {Kuo}, {Yadav},
  {Masui}, {Chang}, {Chen}, {Oppermann}, {Liao}, {Pen}, {Price},
  {Staveley-Smith}, {Switzer}, {Timbie}, \& {Wolz}}]{Anderson18}
{Anderson}, C.~J., {Luciw}, N.~J., {Li}, Y.-C., {et~al.} 2018, \mnras, 476,
  3382

\bibitem[{{Asorey} {et~al.}(2016){Asorey}, {Carrasco Kind}, {Sevilla-Noarbe},
  {Brunner}, \& {Thaler}}]{Asorey16}
{Asorey}, J., {Carrasco Kind}, M., {Sevilla-Noarbe}, I., {Brunner}, R.~J., \&
  {Thaler}, J. 2016, \mnras, 459, 1293

\bibitem[{{Astropy Collaboration} {et~al.}(2013){Astropy Collaboration},
  {Robitaille}, {Tollerud}, {Greenfield}, {Droettboom}, {Bray}, {Aldcroft},
  {Davis}, {Ginsburg}, {Price-Whelan}, {Kerzendorf}, {Conley}, {Crighton},
  {Barbary}, {Muna}, {Ferguson}, {Grollier}, {Parikh}, {Nair}, {Unther},
  {Deil}, {Woillez}, {Conseil}, {Kramer}, {Turner}, {Singer}, {Fox}, {Weaver},
  {Zabalza}, {Edwards}, {Azalee Bostroem}, {Burke}, {Casey}, {Crawford},
  {Dencheva}, {Ely}, {Jenness}, {Labrie}, {Lim}, {Pierfederici}, {Pontzen},
  {Ptak}, {Refsdal}, {Servillat}, \& {Streicher}}]{astropy}
{Astropy Collaboration}, {Robitaille}, T.~P., {Tollerud}, E.~J., {et~al.} 2013,
  \aap, 558, A33

\bibitem[{{Barger} {et~al.}(2012){Barger}, {Cowie}, \& {Wold}}]{Barger12}
{Barger}, A.~J., {Cowie}, L.~L., \& {Wold}, I.~G.~B. 2012, \apj, 749, 106

\bibitem[{{Behrens} {et~al.}(2018){Behrens}, {Byrohl}, {Saito}, \&
  {Niemeyer}}]{Behrens17}
{Behrens}, C., {Byrohl}, C., {Saito}, S., \& {Niemeyer}, J.~C. 2018, \aap, 614,
  A31

\bibitem[{{Behroozi} {et~al.}(2013{\natexlab{a}}){Behroozi}, {Wechsler}, \&
  {Conroy}}]{Behroozi13a}
{Behroozi}, P.~S., {Wechsler}, R.~H., \& {Conroy}, C. 2013{\natexlab{a}},
  \apjl, 762, L31

\bibitem[{{Behroozi} {et~al.}(2013{\natexlab{b}}){Behroozi}, {Wechsler}, \&
  {Conroy}}]{Behroozi13b}
---. 2013{\natexlab{b}}, \apj, 770, 57

\bibitem[{{Blanc} {et~al.}(2011){Blanc}, {Adams}, {Gebhardt}, {Hill}, {Drory},
  {Hao}, {Bender}, {Ciardullo}, {Finkelstein}, {Fry}, {Gawiser}, {Gronwall},
  {Hopp}, {Jeong}, {Kelzenberg}, {Komatsu}, {MacQueen}, {Murphy}, {Roth},
  {Schneider}, \& {Tufts}}]{Blanc11}
{Blanc}, G.~A., {Adams}, J.~J., {Gebhardt}, K., {et~al.} 2011, \apj, 736, 31

\bibitem[{{Breysse} {et~al.}(2014){Breysse}, {Kovetz}, \&
  {Kamionkowski}}]{Breysse14}
{Breysse}, P.~C., {Kovetz}, E.~D., \& {Kamionkowski}, M. 2014, \mnras, 443,
  3506

\bibitem[{{Breysse} \& {Rahman}(2017)}]{BreysseRahman17}
{Breysse}, P.~C., \& {Rahman}, M. 2017, \mnras, 468, 741

\bibitem[{{Bridge} {et~al.}(2018){Bridge}, {Hayes}, {Melinder}, {{\"O}stlin},
  {Gronwall}, {Ciardullo}, {Atek}, {Cannon}, {Gronke}, {Guaita}, {Hagen},
  {Herenz}, {Kunth}, {Laursen}, {Mas-Hesse}, \& {Pardy}}]{Bridge17}
{Bridge}, J.~S., {Hayes}, M., {Melinder}, J., {et~al.} 2018, \apj, 852, 9

\bibitem[{{Brocklehurst}(1971)}]{Brocklehurst71}
{Brocklehurst}, M. 1971, \mnras, 153, 471

\bibitem[{{Cai} {et~al.}(2014){Cai}, {Lapi}, {Bressan}, {De Zotti}, {Negrello},
  \& {Danese}}]{Cai14}
{Cai}, Z.-Y., {Lapi}, A., {Bressan}, A., {et~al.} 2014, \apj, 785, 65

\bibitem[{{Carilli} \& {Walter}(2013)}]{CW13}
{Carilli}, C.~L., \& {Walter}, F. 2013, \araa, 51, 105

\bibitem[{{Chabrier}(2003)}]{Chabrier03}
{Chabrier}, G. 2003, \pasp, 115, 763

\bibitem[{{Chang} {et~al.}(2010){Chang}, {Pen}, {Bandura}, \&
  {Peterson}}]{Chang10}
{Chang}, T.-C., {Pen}, U.-L., {Bandura}, K., \& {Peterson}, J.~B. 2010, \nat,
  466, 463

\bibitem[{{Cheng} {et~al.}(2016){Cheng}, {Chang}, {Bock}, {Bradford}, \&
  {Cooray}}]{Cheng16}
{Cheng}, Y.-T., {Chang}, T.-C., {Bock}, J., {Bradford}, C.~M., \& {Cooray}, A.
  2016, \apj, 832, 165

\bibitem[{{Cheng} {et~al.}(2018){Cheng}, {de Putter}, {Chang}, \&
  {Dore}}]{Cheng18}
{Cheng}, Y.-T., {de Putter}, R., {Chang}, T.-C., \& {Dore}, O. 2018, arXiv
  e-prints, arXiv:1809.06384

\bibitem[{{Chiang} {et~al.}(2013){Chiang}, {Wullstein}, {Jeong}, {Komatsu},
  {Blanc}, {Ciardullo}, {Drory}, {Fabricius}, {Finkelstein}, {Gebhardt},
  {Gronwall}, {Hagen}, {Hill}, {Jee}, {Jogee}, {Landriau}, {Mentuch Cooper},
  {Schneider}, \& {Tuttle}}]{Chiang13}
{Chiang}, C.-T., {Wullstein}, P., {Jeong}, D., {et~al.} 2013, \jcap, 12, 030

\bibitem[{{Chonis} {et~al.}(2013){Chonis}, {Blanc}, {Hill}, {Adams},
  {Finkelstein}, {Gebhardt}, {Kollmeier}, {Ciardullo}, {Drory}, {Gronwall},
  {Hagen}, {Overzier}, {Song}, \& {Zeimann}}]{Chonis13}
{Chonis}, T.~S., {Blanc}, G.~A., {Hill}, G.~J., {et~al.} 2013, \apj, 775, 99

\bibitem[{{Coil} {et~al.}(2011){Coil}, {Blanton}, {Burles}, {Cool},
  {Eisenstein}, {Moustakas}, {Wong}, {Zhu}, {Aird}, {Bernstein}, {Bolton}, \&
  {Hogg}}]{PRIMUS1}
{Coil}, A.~L., {Blanton}, M.~R., {Burles}, S.~M., {et~al.} 2011, \apj, 741, 8

\bibitem[{{Comaschi} \& {Ferrara}(2016)}]{ComaschiFerrara16}
{Comaschi}, P., \& {Ferrara}, A. 2016, \mnras, 455, 725

\bibitem[{{Cool} {et~al.}(2013){Cool}, {Moustakas}, {Blanton}, {Burles},
  {Coil}, {Eisenstein}, {Wong}, {Zhu}, {Aird}, {Bernstein}, {Bolton}, {Hogg},
  \& {Mendez}}]{PRIMUS2}
{Cool}, R.~J., {Moustakas}, J., {Blanton}, M.~R., {et~al.} 2013, \apj, 767, 118

\bibitem[{{Cooray} {et~al.}(2016){Cooray}, {Bock}, {Burgarella}, {Chary},
  {Chang}, {Dor{\'e}}, {Fazio}, {Ferrara}, {Gong}, {Santos}, {Silva}, \&
  {Zemcov}}]{CDIM}
{Cooray}, A., {Bock}, J., {Burgarella}, D., {et~al.} 2016, ArXiv e-prints,
  arXiv:1602.05178

\bibitem[{{Cowie} {et~al.}(2010){Cowie}, {Barger}, \& {Hu}}]{Cowie10}
{Cowie}, L.~L., {Barger}, A.~J., \& {Hu}, E.~M. 2010, \apj, 711, 928

\bibitem[{{Cowie} \& {Hu}(1998)}]{CowieHu98}
{Cowie}, L.~L., \& {Hu}, E.~M. 1998, \aj, 115, 1319

\bibitem[{{Croft} {et~al.}(2018){Croft}, {Miralda-Escud{\'e}}, {Zheng},
  {Blomqvist}, \& {Pieri}}]{Croft18}
{Croft}, R.~A.~C., {Miralda-Escud{\'e}}, J., {Zheng}, Z., {Blomqvist}, M., \&
  {Pieri}, M. 2018, ArXiv e-prints, arXiv:1806.06050

\bibitem[{{Croft} {et~al.}(2016){Croft}, {Miralda-Escud{\'e}}, {Zheng},
  {Bolton}, {Dawson}, {Peterson}, {York}, {Eisenstein}, {Brinkmann},
  {Brownstein}, {Cen}, {Delubac}, {Font-Ribera}, {Hamilton}, {Lee}, {Myers},
  {Palanque-Delabrouille}, {P{\^a}ris}, {Petitjean}, {Pieri}, {Ross}, {Rossi},
  {Schlegel}, {Schneider}, {Slosar}, {Vazquez}, {Viel}, {Weinberg}, \&
  {Y{\`e}che}}]{Croft16}
{Croft}, R.~A.~C., {Miralda-Escud{\'e}}, J., {Zheng}, Z., {et~al.} 2016,
  \mnras, 457, 3541

\bibitem[{{Davidzon} {et~al.}(2017){Davidzon}, {Ilbert}, {Laigle}, {Coupon},
  {McCracken}, {Delvecchio}, {Masters}, {Capak}, {Hsieh}, {Le F{\`e}vre},
  {Tresse}, {Bethermin}, {Chang}, {Faisst}, {Le Floc'h}, {Steinhardt}, {Toft},
  {Aussel}, {Dubois}, {Hasinger}, {Salvato}, {Sanders}, {Scoville}, \&
  {Silverman}}]{Davidzon17}
{Davidzon}, I., {Ilbert}, O., {Laigle}, C., {et~al.} 2017, \aap, 605, A70

\bibitem[{{Dawson} {et~al.}(2016){Dawson}, {Kneib}, {Percival}, {Alam},
  {Albareti}, {Anderson}, {Armengaud}, {Aubourg}, {Bailey}, {Bautista},
  {Berlind}, {Bershady}, {Beutler}, {Bizyaev}, {Blanton}, {Blomqvist},
  {Bolton}, {Bovy}, {Brandt}, {Brinkmann}, {Brownstein}, {Burtin}, {Busca},
  {Cai}, {Chuang}, {Clerc}, {Comparat}, {Cope}, {Croft}, {Cruz-Gonzalez}, {da
  Costa}, {Cousinou}, {Darling}, {de la Macorra}, {de la Torre}, {Delubac}, {du
  Mas des Bourboux}, {Dwelly}, {Ealet}, {Eisenstein}, {Eracleous}, {Escoffier},
  {Fan}, {Finoguenov}, {Font-Ribera}, {Frinchaboy}, {Gaulme}, {Georgakakis},
  {Green}, {Guo}, {Guy}, {Ho}, {Holder}, {Huehnerhoff}, {Hutchinson}, {Jing},
  {Jullo}, {Kamble}, {Kinemuchi}, {Kirkby}, {Kitaura}, {Klaene}, {Laher},
  {Lang}, {Laurent}, {Le Goff}, {Li}, {Liang}, {Lima}, {Lin}, {Lin}, {Lin},
  {Long}, {Lundgren}, {MacDonald}, {Geimba Maia}, {Malanushenko},
  {Malanushenko}, {Mariappan}, {McBride}, {McGreer}, {M{\'e}nard}, {Merloni},
  {Meza}, {Montero-Dorta}, {Muna}, {Myers}, {Nandra}, {Naugle}, {Newman},
  {Noterdaeme}, {Nugent}, {Ogando}, {Olmstead}, {Oravetz}, {Oravetz},
  {Padmanabhan}, {Palanque-Delabrouille}, {Pan}, {Parejko}, {P{\^a}ris},
  {Peacock}, {Petitjean}, {Pieri}, {Pisani}, {Prada}, {Prakash}, {Raichoor},
  {Reid}, {Rich}, {Ridl}, {Rodriguez-Torres}, {Carnero Rosell}, {Ross},
  {Rossi}, {Ruan}, {Salvato}, {Sayres}, {Schneider}, {Schlegel}, {Seljak},
  {Seo}, {Sesar}, {Shandera}, {Shu}, {Slosar}, {Sobreira}, {Streblyanska},
  {Suzuki}, {Taylor}, {Tao}, {Tinker}, {Tojeiro}, {Vargas-Maga{\~n}a}, {Wang},
  {Weaver}, {Weinberg}, {White}, {Wood-Vasey}, {Yeche}, {Zhai}, {Zhao}, {Zhao},
  {Zheng}, {Ben Zhu}, \& {Zou}}]{eBOSS}
{Dawson}, K.~S., {Kneib}, J.-P., {Percival}, W.~J., {et~al.} 2016, \aj, 151, 44

\bibitem[{{DESI Collaboration} {et~al.}(2016){DESI Collaboration}, {Aghamousa},
  {Aguilar}, {Ahlen}, {Alam}, {Allen}, {Allende Prieto}, {Annis}, {Bailey},
  {Balland}, \& et~al.}]{DESI}
{DESI Collaboration}, {Aghamousa}, A., {Aguilar}, J., {et~al.} 2016, ArXiv
  e-prints, arXiv:1611.00036

\bibitem[{{Dijkstra}(2017)}]{Dijkstra17}
{Dijkstra}, M. 2017, ArXiv e-prints, arXiv:1704.03416

\bibitem[{{Dijkstra} {et~al.}(2014){Dijkstra}, {Wyithe}, {Haiman}, {Mesinger},
  \& {Pentericci}}]{Dijkstra14}
{Dijkstra}, M., {Wyithe}, S., {Haiman}, Z., {Mesinger}, A., \& {Pentericci}, L.
  2014, \mnras, 440, 3309

\bibitem[{{Dopita} \& {Sutherland}(2003)}]{DopitaSutherland03}
{Dopita}, M.~A., \& {Sutherland}, R.~S. 2003, {Astrophysics of the diffuse
  universe}

\bibitem[{{Dor{\'e}} {et~al.}(2014){Dor{\'e}}, {Bock}, {Ashby}, {Capak},
  {Cooray}, {de Putter}, {Eifler}, {Flagey}, {Gong}, {Habib}, {Heitmann},
  {Hirata}, {Jeong}, {Katti}, {Korngut}, {Krause}, {Lee}, {Masters},
  {Mauskopf}, {Melnick}, {Mennesson}, {Nguyen}, {{\"O}berg}, {Pullen},
  {Raccanelli}, {Smith}, {Song}, {Tolls}, {Unwin}, {Venumadhav}, {Viero},
  {Werner}, \& {Zemcov}}]{SPHEREx}
{Dor{\'e}}, O., {Bock}, J., {Ashby}, M., {et~al.} 2014, ArXiv e-prints,
  arXiv:1412.4872

\bibitem[{{Drinkwater} {et~al.}(2018){Drinkwater}, {Byrne}, {Blake},
  {Glazebrook}, {Brough}, {Colless}, {Couch}, {Croton}, {Croom}, {Davis},
  {Forster}, {Gilbank}, {Hinton}, {Jelliffe}, {Jurek}, {Li}, {Martin},
  {Pimbblet}, {Poole}, {Pracy}, {Sharp}, {Smillie}, {Spolaor}, {Wisnioski},
  {Woods}, {Wyder}, \& {Yee}}]{WiggleZ}
{Drinkwater}, M.~J., {Byrne}, Z.~J., {Blake}, C., {et~al.} 2018, \mnras, 474,
  4151

\bibitem[{{Eriksen} \& {Gazta{\~n}aga}(2015)}]{Eriksen15}
{Eriksen}, M., \& {Gazta{\~n}aga}, E. 2015, \mnras, 452, 2168

\bibitem[{{Eriksen} {et~al.}(2018){Eriksen}, {Alarcon}, {Gaztanaga}, {Amara},
  {Cabayol}, {Carretero}, {Castander}, {Delfino}, {De Vicente}, {Fernandez},
  {Fosalba}, {Garcia-Bellido}, {Hildebrandt}, {Hoekstra}, {Joachimi},
  {Norberg}, {Miquel}, {Padilla}, {Refregier}, {Sanchez}, {Serrano},
  {Sevilla-Noarbe}, {Tallada}, {Tonello}, \& {Tortorelli}}]{Eriksen18}
{Eriksen}, M., {Alarcon}, A., {Gaztanaga}, E., {et~al.} 2018, ArXiv e-prints,
  arXiv:1809.04375

\bibitem[{{Fonseca} {et~al.}(2017){Fonseca}, {Silva}, {Santos}, \&
  {Cooray}}]{Fonseca17}
{Fonseca}, J., {Silva}, M.~B., {Santos}, M.~G., \& {Cooray}, A. 2017, \mnras,
  464, 1948

\bibitem[{{Fontana} {et~al.}(2014){Fontana}, {Dunlop}, {Paris}, {Targett},
  {Boutsia}, {Castellano}, {Galametz}, {Grazian}, {McLure}, {Merlin},
  {Pentericci}, {Wuyts}, {Almaini}, {Caputi}, {Chary}, {Cirasuolo},
  {Conselice}, {Cooray}, {Daddi}, {Dickinson}, {Faber}, {Fazio}, {Ferguson},
  {Giallongo}, {Giavalisco}, {Grogin}, {Hathi}, {Koekemoer}, {Koo}, {Lucas},
  {Nonino}, {Rix}, {Renzini}, {Rosario}, {Santini}, {Scarlata}, {Sommariva},
  {Stark}, {van der Wel}, {Vanzella}, {Wild}, {Yan}, \& {Zibetti}}]{HUGS}
{Fontana}, A., {Dunlop}, J.~S., {Paris}, D., {et~al.} 2014, \aap, 570, A11

\bibitem[{{Garel} {et~al.}(2012){Garel}, {Blaizot}, {Guiderdoni}, {Schaerer},
  {Verhamme}, \& {Hayes}}]{Garel12}
{Garel}, T., {Blaizot}, J., {Guiderdoni}, B., {et~al.} 2012, \mnras, 422, 310

\bibitem[{{Gazta{\~n}aga} {et~al.}(2012){Gazta{\~n}aga}, {Eriksen}, {Crocce},
  {Castander}, {Fosalba}, {Marti}, {Miquel}, \& {Cabr{\'e}}}]{Gaz12}
{Gazta{\~n}aga}, E., {Eriksen}, M., {Crocce}, M., {et~al.} 2012, \mnras, 422,
  2904

\bibitem[{{Giavalisco} {et~al.}(2004){Giavalisco}, {Ferguson}, {Koekemoer},
  {Dickinson}, {Alexander}, {Bauer}, {Bergeron}, {Biagetti}, {Brandt},
  {Casertano}, {Cesarsky}, {Chatzichristou}, {Conselice}, {Cristiani}, {Da
  Costa}, {Dahlen}, {de Mello}, {Eisenhardt}, {Erben}, {Fall}, {Fassnacht},
  {Fosbury}, {Fruchter}, {Gardner}, {Grogin}, {Hook}, {Hornschemeier}, {Idzi},
  {Jogee}, {Kretchmer}, {Laidler}, {Lee}, {Livio}, {Lucas}, {Madau},
  {Mobasher}, {Moustakas}, {Nonino}, {Padovani}, {Papovich}, {Park},
  {Ravindranath}, {Renzini}, {Richardson}, {Riess}, {Rosati}, {Schirmer},
  {Schreier}, {Somerville}, {Spinrad}, {Stern}, {Stiavelli}, {Strolger},
  {Urry}, {Vandame}, {Williams}, \& {Wolf}}]{GOODS}
{Giavalisco}, M., {Ferguson}, H.~C., {Koekemoer}, A.~M., {et~al.} 2004, \apjl,
  600, L93

\bibitem[{{Gong} {et~al.}(2014){Gong}, {Silva}, {Cooray}, \& {Santos}}]{Gong14}
{Gong}, Y., {Silva}, M., {Cooray}, A., \& {Santos}, M.~G. 2014, \apj, 785, 72

\bibitem[{{Grazian} {et~al.}(2015){Grazian}, {Fontana}, {Santini}, {Dunlop},
  {Ferguson}, {Castellano}, {Amorin}, {Ashby}, {Barro}, {Behroozi}, {Boutsia},
  {Caputi}, {Chary}, {Dekel}, {Dickinson}, {Faber}, {Fazio}, {Finkelstein},
  {Galametz}, {Giallongo}, {Giavalisco}, {Grogin}, {Guo}, {Kocevski},
  {Koekemoer}, {Koo}, {Lee}, {Lu}, {Merlin}, {Mobasher}, {Nonino}, {Papovich},
  {Paris}, {Pentericci}, {Reddy}, {Renzini}, {Salmon}, {Salvato}, {Sommariva},
  {Song}, \& {Vanzella}}]{Grazian15}
{Grazian}, A., {Fontana}, A., {Santini}, P., {et~al.} 2015, \aap, 575, A96

\bibitem[{{Gronwall} {et~al.}(2007){Gronwall}, {Ciardullo}, {Hickey},
  {Gawiser}, {Feldmeier}, {van Dokkum}, {Urry}, {Herrera}, {Lehmer}, {Infante},
  {Orsi}, {Marchesini}, {Blanc}, {Francke}, {Lira}, \& {Treister}}]{Gronwall07}
{Gronwall}, C., {Ciardullo}, R., {Hickey}, T., {et~al.} 2007, \apj, 667, 79

\bibitem[{{Hasinger} {et~al.}(2018){Hasinger}, {Capak}, {Salvato}, {Barger},
  {Cowie}, {Faisst}, {Hemmati}, {Kakazu}, {Kartaltepe}, {Masters}, {Mobasher},
  {Nayyeri}, {Sanders}, {Scoville}, {Suh}, {Steinhardt}, \& {Yang}}]{DEIMOS10k}
{Hasinger}, G., {Capak}, P., {Salvato}, M., {et~al.} 2018, \apj, 858, 77

\bibitem[{{Hayes}(2015)}]{Hayes2015}
{Hayes}, M. 2015, \pasa, 32, e027

\bibitem[{{Henry} {et~al.}(2015){Henry}, {Scarlata}, {Martin}, \&
  {Erb}}]{Henry15}
{Henry}, A., {Scarlata}, C., {Martin}, C.~L., \& {Erb}, D. 2015, \apj, 809, 19

\bibitem[{{Hill} \& {HETDEX Consortium}(2016)}]{HETDEX2016}
{Hill}, G.~J., \& {HETDEX Consortium}. 2016, in Astronomical Society of the
  Pacific Conference Series, Vol. 507, Multi-Object Spectroscopy in the Next
  Decade: Big Questions, Large Surveys, and Wide Fields, ed. I.~{Skillen},
  M.~{Balcells}, \& S.~{Trager}, 393

\bibitem[{{Hill} {et~al.}(2008){Hill}, {Gebhardt}, {Komatsu}, {Drory},
  {MacQueen}, {Adams}, {Blanc}, {Koehler}, {Rafal}, {Roth}, {Kelz}, {Gronwall},
  {Ciardullo}, \& {Schneider}}]{HETDEX}
{Hill}, G.~J., {Gebhardt}, K., {Komatsu}, E., {et~al.} 2008, in Astronomical
  Society of the Pacific Conference Series, Vol. 399, Panoramic Views of Galaxy
  Formation and Evolution, ed. T.~{Kodama}, T.~{Yamada}, \& K.~{Aoki}, 115

\bibitem[{{Hill} {et~al.}(2014){Hill}, {Tuttle}, {Drory}, {Lee}, {Vattiat},
  {DePoy}, {Marshall}, {Kelz}, {Haynes}, {Fabricius}, {Gebhardt}, {Allen},
  {Anwad}, {Bender}, {Blanc}, {Chonis}, {Cornell}, {Dalton}, {Good}, {Jahn},
  {Kriel}, {Landriau}, {MacQueen}, {Murphy}, {Peterson}, {Prochaska},
  {Nicklas}, {Ramsey}, {Roth}, {Savage}, \& {Snigula}}]{VIRUS}
{Hill}, G.~J., {Tuttle}, S.~E., {Drory}, N., {et~al.} 2014, in \procspie, Vol.
  9147, Ground-based and Airborne Instrumentation for Astronomy V, 91470Q

\bibitem[{{Hinshaw} {et~al.}(2013){Hinshaw}, {Larson}, {Komatsu}, {Spergel},
  {Bennett}, {Dunkley}, {Nolta}, {Halpern}, {Hill}, {Odegard}, {Page}, {Smith},
  {Weiland}, {Gold}, {Jarosik}, {Kogut}, {Limon}, {Meyer}, {Tucker}, {Wollack},
  \& {Wright}}]{WMAP9}
{Hinshaw}, G., {Larson}, D., {Komatsu}, E., {et~al.} 2013, \apjs, 208, 19

\bibitem[{{Hu} {et~al.}(1998){Hu}, {Cowie}, \& {McMahon}}]{Hu98}
{Hu}, E.~M., {Cowie}, L.~L., \& {McMahon}, R.~G. 1998, \apjl, 502, L99

\bibitem[{{Hummer} \& {Storey}(1987)}]{HS87}
{Hummer}, D.~G., \& {Storey}, P.~J. 1987, \mnras, 224, 801

\bibitem[{Hunter(2007)}]{matplotlib}
Hunter, J.~D. 2007, Computing In Science \& Engineering, 9, 90

\bibitem[{{Huterer} {et~al.}(2006){Huterer}, {Takada}, {Bernstein}, \&
  {Jain}}]{Huterer06}
{Huterer}, D., {Takada}, M., {Bernstein}, G., \& {Jain}, B. 2006, \mnras, 366,
  101

\bibitem[{{Juneau} {et~al.}(2005){Juneau}, {Glazebrook}, {Crampton},
  {McCarthy}, {Savaglio}, {Abraham}, {Carlberg}, {Chen}, {Le Borgne}, {Marzke},
  {Roth}, {J{\o}rgensen}, {Hook}, \& {Murowinski}}]{Juneau05}
{Juneau}, S., {Glazebrook}, K., {Crampton}, D., {et~al.} 2005, \apjl, 619, L135

\bibitem[{{Kennicutt} \& {Evans}(2012)}]{KE12}
{Kennicutt}, R.~C., \& {Evans}, N.~J. 2012, \araa, 50, 531

\bibitem[{{Kennicutt}(1998)}]{Kennicutt98}
{Kennicutt}, Jr., R.~C. 1998, \araa, 36, 189

\bibitem[{{Kennicutt} {et~al.}(1994){Kennicutt}, {Tamblyn}, \&
  {Congdon}}]{Kennicutt94}
{Kennicutt}, Jr., R.~C., {Tamblyn}, P., \& {Congdon}, C.~E. 1994, \apj, 435, 22

\bibitem[{{Knox}(1995)}]{Knox1995}
{Knox}, L. 1995, \prd, 52, 4307

\bibitem[{{Kovetz} {et~al.}(2017){Kovetz}, {Viero}, {Lidz}, {Newburgh},
  {Rahman}, {Switzer}, {Kamionkowski}, {Aguirre}, {Alvarez}, {Bock}, {Bond},
  {Bower}, {Bradford}, {Breysse}, {Bull}, {Chang}, {Cheng}, {Chung}, {Cleary},
  {Corray}, {Crites}, {Croft}, {Dor{\'e}}, {Eastwood}, {Ferrara}, {Fonseca},
  {Jacobs}, {Keating}, {Lagache}, {Lakhlani}, {Liu}, {Moodley}, {Murray},
  {P{\'e}nin}, {Popping}, {Pullen}, {Reichers}, {Saito}, {Saliwanchik},
  {Santos}, {Somerville}, {Stacey}, {Stein}, {Villaescusa-Navarro}, {Visbal},
  {Weltman}, {Wolz}, \& {Zemcov}}]{Kovetz17}
{Kovetz}, E.~D., {Viero}, M.~P., {Lidz}, A., {et~al.} 2017, ArXiv e-prints,
  arXiv:1709.09066

\bibitem[{{Laigle} {et~al.}(2016){Laigle}, {McCracken}, {Ilbert}, {Hsieh},
  {Davidzon}, {Capak}, {Hasinger}, {Silverman}, {Pichon}, {Coupon}, {Aussel},
  {Le Borgne}, {Caputi}, {Cassata}, {Chang}, {Civano}, {Dunlop}, {Fynbo},
  {Kartaltepe}, {Koekemoer}, {Le F{\`e}vre}, {Le Floc'h}, {Leauthaud}, {Lilly},
  {Lin}, {Marchesi}, {Milvang-Jensen}, {Salvato}, {Sanders}, {Scoville},
  {Smolcic}, {Stockmann}, {Taniguchi}, {Tasca}, {Toft}, {Vaccari}, \&
  {Zabl}}]{Laigle16}
{Laigle}, C., {McCracken}, H.~J., {Ilbert}, O., {et~al.} 2016, \apjs, 224, 24

\bibitem[{{Lawrence} {et~al.}(2007){Lawrence}, {Warren}, {Almaini}, {Edge},
  {Hambly}, {Jameson}, {Lucas}, {Casali}, {Adamson}, {Dye}, {Emerson},
  {Foucaud}, {Hewett}, {Hirst}, {Hodgkin}, {Irwin}, {Lodieu}, {McMahon},
  {Simpson}, {Smail}, {Mortlock}, \& {Folger}}]{UKIDSS}
{Lawrence}, A., {Warren}, S.~J., {Almaini}, O., {et~al.} 2007, \mnras, 379,
  1599

\bibitem[{{Le F{\`e}vre} {et~al.}(2015){Le F{\`e}vre}, {Tasca}, {Cassata},
  {Garilli}, {Le Brun}, {Maccagni}, {Pentericci}, {Thomas}, {Vanzella},
  {Zamorani}, {Zucca}, {Amorin}, {Bardelli}, {Capak}, {Cassar{\`a}},
  {Castellano}, {Cimatti}, {Cuby}, {Cucciati}, {de la Torre}, {Durkalec},
  {Fontana}, {Giavalisco}, {Grazian}, {Hathi}, {Ilbert}, {Lemaux}, {Moreau},
  {Paltani}, {Ribeiro}, {Salvato}, {Schaerer}, {Scodeggio}, {Sommariva},
  {Talia}, {Taniguchi}, {Tresse}, {Vergani}, {Wang}, {Charlot}, {Contini},
  {Fotopoulou}, {L{\'o}pez-Sanjuan}, {Mellier}, \& {Scoville}}]{VUDS}
{Le F{\`e}vre}, O., {Tasca}, L.~A.~M., {Cassata}, P., {et~al.} 2015, \aap, 576,
  A79

\bibitem[{{Leung} {et~al.}(2017){Leung}, {Acquaviva}, {Gawiser}, {Ciardullo},
  {Komatsu}, {Malz}, {Zeimann}, {Bridge}, {Drory}, {Feldmeier}, {Finkelstein},
  {Gebhardt}, {Gronwall}, {Hagen}, {Hill}, \& {Schneider}}]{Leung17}
{Leung}, A.~S., {Acquaviva}, V., {Gawiser}, E., {et~al.} 2017, \apj, 843, 130

\bibitem[{{Li} {et~al.}(2016){Li}, {Wechsler}, {Devaraj}, \& {Church}}]{Li16}
{Li}, T.~Y., {Wechsler}, R.~H., {Devaraj}, K., \& {Church}, S.~E. 2016, \apj,
  817, 169

\bibitem[{{Lidz} {et~al.}(2011){Lidz}, {Furlanetto}, {Oh}, {Aguirre}, {Chang},
  {Dor{\'e}}, \& {Pritchard}}]{Lidz11}
{Lidz}, A., {Furlanetto}, S.~R., {Oh}, S.~P., {et~al.} 2011, \apj, 741, 70

\bibitem[{{Lidz} \& {Taylor}(2016)}]{Lidz16}
{Lidz}, A., \& {Taylor}, J. 2016, \apj, 825, 143

\bibitem[{{Ma} {et~al.}(2006){Ma}, {Hu}, \& {Huterer}}]{Ma06}
{Ma}, Z., {Hu}, W., \& {Huterer}, D. 2006, \apj, 636, 21

\bibitem[{{Madau} \& {Dickinson}(2014)}]{MD14}
{Madau}, P., \& {Dickinson}, M. 2014, \araa, 52, 415

\bibitem[{{McCracken} {et~al.}(2012){McCracken}, {Milvang-Jensen}, {Dunlop},
  {Franx}, {Fynbo}, {Le F{\`e}vre}, {Holt}, {Caputi}, {Goranova}, {Buitrago},
  {Emerson}, {Freudling}, {Hudelot}, {L{\'o}pez-Sanjuan}, {Magnard}, {Mellier},
  {M{\o}ller}, {Nilsson}, {Sutherland}, {Tasca}, \& {Zabl}}]{UltraVISTA}
{McCracken}, H.~J., {Milvang-Jensen}, B., {Dunlop}, J., {et~al.} 2012, \aap,
  544, A156

\bibitem[{{M{\'e}nard} {et~al.}(2013){M{\'e}nard}, {Scranton}, {Schmidt},
  {Morrison}, {Jeong}, {Budavari}, \& {Rahman}}]{Menard13}
{M{\'e}nard}, B., {Scranton}, R., {Schmidt}, S., {et~al.} 2013, ArXiv e-prints,
  arXiv:1303.4722

\bibitem[{{Momcheva} {et~al.}(2016){Momcheva}, {Brammer}, {van Dokkum},
  {Skelton}, {Whitaker}, {Nelson}, {Fumagalli}, {Maseda}, {Leja}, {Franx},
  {Rix}, {Bezanson}, {Da Cunha}, {Dickey}, {F{\"o}rster Schreiber},
  {Illingworth}, {Kriek}, {Labb{\'e}}, {Ulf Lange}, {Lundgren}, {Magee},
  {Marchesini}, {Oesch}, {Pacifici}, {Patel}, {Price}, {Tal}, {Wake}, {van der
  Wel}, \& {Wuyts}}]{Momcheva16}
{Momcheva}, I.~G., {Brammer}, G.~B., {van Dokkum}, P.~G., {et~al.} 2016, \apjs,
  225, 27

\bibitem[{{Murphy} {et~al.}(2011){Murphy}, {Condon}, {Schinnerer}, {Kennicutt},
  {Calzetti}, {Armus}, {Helou}, {Turner}, {Aniano}, {Beir{\~a}o}, {Bolatto},
  {Brandl}, {Croxall}, {Dale}, {Donovan Meyer}, {Draine}, {Engelbracht},
  {Hunt}, {Hao}, {Koda}, {Roussel}, {Skibba}, \& {Smith}}]{Murphy11}
{Murphy}, E.~J., {Condon}, J.~J., {Schinnerer}, E., {et~al.} 2011, \apj, 737,
  67

\bibitem[{{Murray} {et~al.}(2013){Murray}, {Power}, \& {Robotham}}]{hmf}
{Murray}, S.~G., {Power}, C., \& {Robotham}, A.~S.~G. 2013, Astronomy and
  Computing, 3, 23

\bibitem[{{Osterbrock}(1989)}]{Osterbrock89}
{Osterbrock}, D.~E. 1989, {Astrophysics of gaseous nebulae and active galactic
  nuclei}

\bibitem[{{Padmanabhan}(2018)}]{Padmanabhan18}
{Padmanabhan}, H. 2018, \mnras, 475, 1477

\bibitem[{{Papovich} {et~al.}(2016){Papovich}, {Shipley}, {Mehrtens}, {Lanham},
  {Lacy}, {Ciardullo}, {Finkelstein}, {Bassett}, {Behroozi}, {Blanc}, {de
  Jong}, {DePoy}, {Drory}, {Gawiser}, {Gebhardt}, {Gronwall}, {Hill}, {Hopp},
  {Jogee}, {Kawinwanichakij}, {Marshall}, {McLinden}, {Mentuch Cooper},
  {Somerville}, {Steinmetz}, {Tran}, {Tuttle}, {Viero}, {Wechsler}, \&
  {Zeimann}}]{SHELA}
{Papovich}, C., {Shipley}, H.~V., {Mehrtens}, N., {et~al.} 2016, \apjs, 224, 28

\bibitem[{{Pengelly}(1964)}]{Pengelly64}
{Pengelly}, R.~M. 1964, \mnras, 127, 145

\bibitem[{{Pengelly} \& {Seaton}(1964)}]{PengellySeaton64}
{Pengelly}, R.~M., \& {Seaton}, M.~J. 1964, \mnras, 127, 165

\bibitem[{{Pullen} {et~al.}(2013){Pullen}, {Chang}, {Dor{\'e}}, \&
  {Lidz}}]{Pullen13}
{Pullen}, A.~R., {Chang}, T.-C., {Dor{\'e}}, O., \& {Lidz}, A. 2013, \apj, 768,
  15

\bibitem[{{Pullen} {et~al.}(2014){Pullen}, {Dor{\'e}}, \& {Bock}}]{Pullen14}
{Pullen}, A.~R., {Dor{\'e}}, O., \& {Bock}, J. 2014, \apj, 786, 111

\bibitem[{{Righi} {et~al.}(2008){Righi}, {Hern{\'a}ndez-Monteagudo}, \&
  {Sunyaev}}]{Righi08}
{Righi}, M., {Hern{\'a}ndez-Monteagudo}, C., \& {Sunyaev}, R.~A. 2008, \aap,
  489, 489

\bibitem[{{Salmon} {et~al.}(2015){Salmon}, {Papovich}, {Finkelstein}, {Tilvi},
  {Finlator}, {Behroozi}, {Dahlen}, {Dav{\'e}}, {Dekel}, {Dickinson},
  {Ferguson}, {Giavalisco}, {Long}, {Lu}, {Mobasher}, {Reddy}, {Somerville}, \&
  {Wechsler}}]{Salmon15}
{Salmon}, B., {Papovich}, C., {Finkelstein}, S.~L., {et~al.} 2015, \apj, 799,
  183

\bibitem[{{Santini} {et~al.}(2014){Santini}, {Maiolino}, {Magnelli}, {Lutz},
  {Lamastra}, {Li Causi}, {Eales}, {Andreani}, {Berta}, {Buat}, {Cooray},
  {Cresci}, {Daddi}, {Farrah}, {Fontana}, {Franceschini}, {Genzel}, {Granato},
  {Grazian}, {Le Floc'h}, {Magdis}, {Magliocchetti}, {Mannucci}, {Menci},
  {Nordon}, {Oliver}, {Popesso}, {Pozzi}, {Riguccini}, {Rodighiero}, {Rosario},
  {Salvato}, {Scott}, {Silva}, {Tacconi}, {Viero}, {Wang}, {Wuyts}, \&
  {Xu}}]{Santini14}
{Santini}, P., {Maiolino}, R., {Magnelli}, B., {et~al.} 2014, \aap, 562, A30

\bibitem[{{Schachter}(1991)}]{Schachter91}
{Schachter}, J. 1991, \pasp, 103, 457

\bibitem[{{Scoville} {et~al.}(2007){Scoville}, {Aussel}, {Brusa}, {Capak},
  {Carollo}, {Elvis}, {Giavalisco}, {Guzzo}, {Hasinger}, {Impey}, {Kneib},
  {LeFevre}, {Lilly}, {Mobasher}, {Renzini}, {Rich}, {Sanders}, {Schinnerer},
  {Schminovich}, {Shopbell}, {Taniguchi}, \& {Tyson}}]{COSMOS}
{Scoville}, N., {Aussel}, H., {Brusa}, M., {et~al.} 2007, \apjs, 172, 1

\bibitem[{{Seaton}(1964)}]{Seaton64}
{Seaton}, M.~J. 1964, \mnras, 127, 177

\bibitem[{{Silva} {et~al.}(2013){Silva}, {Santos}, {Gong}, {Cooray}, \&
  {Bock}}]{Silva13}
{Silva}, M.~B., {Santos}, M.~G., {Gong}, Y., {Cooray}, A., \& {Bock}, J. 2013,
  \apj, 763, 132

\bibitem[{{Sobral} {et~al.}(2014){Sobral}, {Best}, {Smail}, {Mobasher},
  {Stott}, \& {Nisbet}}]{Sobral14}
{Sobral}, D., {Best}, P.~N., {Smail}, I., {et~al.} 2014, \mnras, 437, 3516

\bibitem[{{Sobral} {et~al.}(2018){Sobral}, {Santos}, {Matthee},
  {Paulino-Afonso}, {Ribeiro}, {Calhau}, \& {Khostovan}}]{Sobral18}
{Sobral}, D., {Santos}, S., {Matthee}, J., {et~al.} 2018, \mnras, 476, 4725

\bibitem[{{Sobral} {et~al.}(2017){Sobral}, {Matthee}, {Best}, {Stroe},
  {R{\"o}ttgering}, {Oteo}, {Smail}, {Morabito}, \&
  {Paulino-Afonso}}]{Sobral17}
{Sobral}, D., {Matthee}, J., {Best}, P., {et~al.} 2017, \mnras, 466, 1242

\bibitem[{{Speagle} {et~al.}(2014){Speagle}, {Steinhardt}, {Capak}, \&
  {Silverman}}]{Speagle14}
{Speagle}, J.~S., {Steinhardt}, C.~L., {Capak}, P.~L., \& {Silverman}, J.~D.
  2014, \apjs, 214, 15

\bibitem[{{Steidel} {et~al.}(2011){Steidel}, {Bogosavljevi{\'c}}, {Shapley},
  {Kollmeier}, {Reddy}, {Erb}, \& {Pettini}}]{Steidel11}
{Steidel}, C.~C., {Bogosavljevi{\'c}}, M., {Shapley}, A.~E., {et~al.} 2011,
  \apj, 736, 160

\bibitem[{{Storey} \& {Hummer}(1995)}]{SH95}
{Storey}, P.~J., \& {Hummer}, D.~G. 1995, \mnras, 272, 41

\bibitem[{{Sun} {et~al.}(2018){Sun}, {Moncelsi}, {Viero}, {Silva}, {Bock},
  {Bradford}, {Chang}, {Cheng}, {Cooray}, {Crites}, {Hailey-Dunsheath},
  {Uzgil}, {Hunacek}, \& {Zemcov}}]{Sun18}
{Sun}, G., {Moncelsi}, L., {Viero}, M.~P., {et~al.} 2018, \apj, 856, 107

\bibitem[{{Switzer} {et~al.}(2013){Switzer}, {Masui}, {Bandura}, {Calin},
  {Chang}, {Chen}, {Li}, {Liao}, {Natarajan}, {Pen}, {Peterson}, {Shaw}, \&
  {Voytek}}]{Switzer13}
{Switzer}, E.~R., {Masui}, K.~W., {Bandura}, K., {et~al.} 2013, \mnras, 434,
  L46

\bibitem[{{Takada} {et~al.}(2014){Takada}, {Ellis}, {Chiba}, {Greene},
  {Aihara}, {Arimoto}, {Bundy}, {Cohen}, {Dor{\'e}}, {Graves}, {Gunn},
  {Heckman}, {Hirata}, {Ho}, {Kneib}, {Le F{\`e}vre}, {Lin}, {More},
  {Murayama}, {Nagao}, {Ouchi}, {Seiffert}, {Silverman}, {Sodr{\'e}},
  {Spergel}, {Strauss}, {Sugai}, {Suto}, {Takami}, \& {Wyse}}]{Takada14}
{Takada}, M., {Ellis}, R.~S., {Chiba}, M., {et~al.} 2014, \pasj, 66, R1

\bibitem[{{Tasca} {et~al.}(2017){Tasca}, {Le F{\`e}vre}, {Ribeiro}, {Thomas},
  {Moreau}, {Cassata}, {Garilli}, {Le Brun}, {Lemaux}, {Maccagni},
  {Pentericci}, {Schaerer}, {Vanzella}, {Zamorani}, {Zucca}, {Amorin},
  {Bardelli}, {Cassar{\`a}}, {Castellano}, {Cimatti}, {Cucciati}, {Durkalec},
  {Fontana}, {Giavalisco}, {Grazian}, {Hathi}, {Ilbert}, {Paltani}, {Pforr},
  {Scodeggio}, {Sommariva}, {Talia}, {Tresse}, {Vergani}, {Capak}, {Charlot},
  {Contini}, {de la Torre}, {Dunlop}, {Fotopoulou}, {Guaita}, {Koekemoer},
  {L{\'o}pez-Sanjuan}, {Mellier}, {Salvato}, {Scoville}, {Taniguchi}, \&
  {Wang}}]{VUDSDR1}
{Tasca}, L.~A.~M., {Le F{\`e}vre}, O., {Ribeiro}, B., {et~al.} 2017, \aap, 600,
  A110

\bibitem[{{Visbal} \& {Loeb}(2010)}]{VL10}
{Visbal}, E., \& {Loeb}, A. 2010, \jcap, 11, 016

\bibitem[{{Visbal} \& {McQuinn}(2018)}]{VisbalMcQuinn18}
{Visbal}, E., \& {McQuinn}, M. 2018, \apjl, 863, L6

\bibitem[{{Wolz} {et~al.}(2017){Wolz}, {Blake}, \& {Wyithe}}]{Wolz17}
{Wolz}, L., {Blake}, C., \& {Wyithe}, J.~S.~B. 2017, \mnras, 470, 3220

\bibitem[{{Yajima} {et~al.}(2014){Yajima}, {Li}, {Zhu}, {Abel}, {Gronwall}, \&
  {Ciardullo}}]{Yajima14}
{Yajima}, H., {Li}, Y., {Zhu}, Q., {et~al.} 2014, \mnras, 440, 776

\bibitem[{{Zheng} {et~al.}(2011){Zheng}, {Cen}, {Trac}, \&
  {Miralda-Escud{\'e}}}]{Zheng11}
{Zheng}, Z., {Cen}, R., {Trac}, H., \& {Miralda-Escud{\'e}}, J. 2011, \apj,
  726, 38

\end{thebibliography}

\appendix
\section{Lyman-$\alpha$ model details}
\label{sec:lyamodel}
We consider the SFR--$L_\text{Ly$\alpha$}$ relation from~\autoref{sec:hmll} in two parts: the `intrinsic' Lyman-$\alpha$ luminosity per unit SFR based purely on ionising emissivity (i.e.~the numeric coefficient in~\autoref{eq:Lyamodel}), and the escape fraction that modifies this luminosity (as given in~\autoref{eq:fescmodel}). We explain our rationale for each in~\autoref{sec:lyamodel_c} and~\autoref{sec:lyamodel_f} respectively, and compare simulated luminosity functions and power spectra to observations and previous work in~\autoref{sec:lyamodel_lf}.\added{ We also consider our choice of $10^{10}\,M_\odot$ as the minimum emitting halo mass in~\autoref{sec:lyamodel_mmin}.}
\subsection{Intrinsic Luminosity per SFR}
\label{sec:lyamodel_c}
The conversion is based on assuming a certain intrinsic H$\alpha$ luminosity per SFR and a Ly$\alpha$/H$\alpha$ line ratio of 8.7. Stellar synthesis modelling done in~\cite{Kennicutt94} (via~\citealt{Kennicutt98}) suggested that for a Salpeter IMF,
\begin{equation}\frac{L_{\rm H\alpha}}{\operatorname{SFR}}=\frac{1.26\times10^{41}\text{ erg s}^{-1}}{M_\odot\text{ yr}^{-1}}.\label{eq:SFR_Ha}\end{equation}
The calibration arises from models of stellar evolutionary tracks, ionising emissivity, and recombination rates for gas with $T=10^4$ K; \cite{Kennicutt94} cite~\cite{HS87} for the last item.

\cite{Murphy11} update this calibration with revised stellar synthesis models incorporating a Kroupa IMF, yielding a SFR per luminosity 0.68 times that of the old calibration, or a luminosity per SFR $(0.68)^{-1}=1.47$ times that of the old calibration. Any difference in this calibration due to using a Chabrier IMF is comparatively small, and we use the value from~\cite{Murphy11} unaltered\footnote{See the re-scalings of SFR--FUV conversion factors in~\cite{MD14} with different choices of IMF. These suggest that the right-hand side of~\autoref{eq:SFR_LIR} should be multiplied by $1.6$ if using the Salpeter IMF instead of the Chabrier IMF, and the right-hand side of~\autoref{eq:SFR_Ha} multiplied by 1.5 if using the Kroupa IMF instead of the Salpeter IMF. By contrast, the conversion factors assuming the Kroupa IMF or the Chabrier IMF are within 6\% of each other, which we are happy to deem sub-dominant to other modelling uncertainties.}.

The convention in previous literature is to convert the above $L_{\rm H\alpha}/\operatorname{SFR}$ ratio into a $L_{\text{Ly}\alpha}/\operatorname{SFR}$ ratio by assuming a Ly$\alpha$/H$\alpha$ ratio of 8.7. Common citations for this convention include
\begin{itemize}
\item \cite{Pengelly64} (communicated by Seaton), the first of three papers including~\cite{PengellySeaton64} and~\cite{Seaton64};
\item \cite{Brocklehurst71} (again communicated by Seaton, and in fact the content is very similar to~\citealt{Pengelly64});
\item \cite{HS87};
\item and~\cite{Hu98} (who cite~\cite{Brocklehurst71}, but are sometimes cited in isolation---in~\citealt{Hayes2015} and~\citealt{Bridge17}, for example).
\end{itemize}
Of these, only the last citation is strictly appropriate, as it is the only one explicitly stating a Ly$\alpha$/H$\alpha$ ratio of 8.7. The first three deal with hydrogen and helium recombination rates and line ratios---including the Balmer series and specifically the H$\alpha$/H$\beta$ ratio---for different possible gas densities and temperatures, and for different assumptions about whether the gas is optically thin (case A) or thick (case B) to the recombination lines. But as~\cite{Henry15} note, there needs to be additional information to link the Lyman-$\alpha$ line to the Balmer series.

\cite{Osterbrock89} is a possible source for the Ly$\alpha$/H$\alpha$ ratio. In the low-density limit, 68\% of recombinations lead to Lyman-$\alpha$ emission, and 45\% lead to H-$\alpha$ emission (see~\citealt{Dijkstra14} or~\citealt{Dijkstra17}). Combining this with the ratio of photon energies, we obtain a Ly$\alpha$/H$\alpha$ flux ratio of 8.2.

However, this is in the low-density limit, and collisional excitations at higher densities result in an enhanced ratio. Typical assumptions for the electron density fall within the range of $n_e=10^2$--$10^3$ cm$^{-3}$, and if we consult tables of line ratios as in~\cite{DopitaSutherland03} (which~\citealt{Henry15} consult for H$\alpha$/H$\beta$ and Ly$\alpha$/H$\beta$ tables, and are synthesised from~\citealt{SH95}), we find that 8.2--9.1 is a reasonable range for the line ratio given that density range at $T=10^4$ K (and assuming case B recombination). The conventional value of 8.7 appears to have been chosen (or at least kept) as a happy intermediate.

The resulting intrinsic conversion between star-formation rate and Lyman-$\alpha$ luminosity is
\begin{equation}\frac{L_{\text{Ly}\alpha}}{\operatorname{SFR}}=\frac{L_{\text{Ly}\alpha}}{L_{\text{H}\alpha}}\frac{\operatorname{SFR}_\text{K98}}{\operatorname{SFR}}\frac{L_{\rm H\alpha}}{\operatorname{SFR}_\text{K98}}=\frac{8.7}{0.68}\frac{1.26\times10^{41}\text{ erg s}^{-1}}{M_\odot\text{ yr}^{-1}}=\frac{1.6\times10^{42}\text{ erg s}^{-1}}{M_\odot\text{ yr}^{-1}}.\label{eq:lyamodel_C}\end{equation}
This is the origin of our value for the numeric coefficient in~\autoref{eq:Lyamodel}.
\subsection{The Escape Fraction}
\label{sec:lyamodel_f}
The above conversion operates under the assumption that recombination balances photoionisation within HII regions. However, this does not include the possibility of ionising radiation being absorbed by dust before it is able to trigger a photoionisation event, or the possibility of recombination line emission being absorbed by dust. There is also the possibility of ionising photons escaping the galaxy without a photoionisation event (in which case it will likely trigger an event in the intergalactic medium) or being absorbed in HI regions without triggering recombination line emission.

In this model, we ignore the last two possibilities for simplicity but model the first two, in an abbreviated version of the sort of model found in~\cite{Cai14}:
\begin{equation}L_{\text{Ly}\alpha}(\operatorname{SFR},z)=C\operatorname{SFR}f_\text{esc}^\text{ion}f_\text{esc}^{\text{Ly}\alpha},\end{equation}
where $C$ is the value obtained in~\autoref{eq:lyamodel_C}, and both escape fractions are functions of star-formation rate and redshift. For this work, as shown in~\autoref{eq:Lyamodel} and~\autoref{eq:fescmodel}, we lump the two escape fractions together into a single effective escape fraction relative to the intrinsic Ly$\alpha$ prediction.

We take $f_\text{esc}^\text{ion}\sim f_\text{esc}^{\text{Ly}\alpha}$, effectively squaring one escape fraction of UV photons against dust. Note that the escape/absorption mechanisms for ionising photons ($\lambda_\text{rest}\leq912$ \AA) and Lyman-$\alpha$ photons ($\lambda_\text{rest}=1216$ \AA) are in fact different, and any correlation between the two is subject to large amounts of scatter. The escape fractions are at least around the same order of magnitude, however---see the numbers obtained through simulations in~\cite{Yajima14} and the dust attenuation factors assumed in~\cite{Cai14} (the dust optical depth at 1216 \AA~is assumed to be 1.08 times the dust optical depth at 1350 \AA, which itself is assumed to be $1/\gamma\simeq1.18$ times the dust optical depth of ionising photons). 

The escape fraction is highly contrived to two ends:
\begin{itemize}
\item it increases monotonically with redshift (converging to 1 as $z\to\infty$),
\item and it decreases with higher star-formation rate.
\end{itemize}

The latter is easier to justify---it is natural to associated higher star-formation rate with more dust, and there is observational evidence for this correlation (see~\citealt{Santini14}). However, this in turn makes the former more difficult to justify: cosmic SFR density evolves non-monotonically with redshift---increasing up to $z\sim3$ before showing a clear decline after $z\sim2$~\citep{MD14}---suggesting that the redshift evolution of the escape fraction cannot be monotonic. However, (a) the scope of our modelling is limited to $z\gtrsim2$ where the behaviour may as well be monotonic, and (b) it may be possible for factors other than SFR to influence dust content in older/late-type galaxies, although there is considerable uncertainty around the latter.

Assigning specific numbers requires either a sophisticated simulation incorporating radiative transfer (like~\cite{Yajima14}) or observational constraints. Two simulation forecasts influence our choice of form and approximate parameter values, with subsequent fine-tuning based on observational constraints:
\begin{itemize}
\item Results in~\cite{Yajima14} show a median $f_\text{esc}^\text{ion}\sim0.2$ evolving very weakly with redshift and a non-monotonic evolution of $f_\text{esc}^{\text{Ly}\alpha}$, with median values ranging between $\sim0.3$ and $\sim0.9$. Scatter around the median is quite large, however.
\item Results in~\cite{Garel12} show a simulated escape fraction of near-unity for most simulated galaxies with SFR less than 1 $M_\odot$ yr$^{-1}$. The distribution of $f_\text{esc}(\operatorname{SFR})$ evolves strongly into a flat one with higher SFR, with an average of 21\% for simulated galaxies with SFR greater than 10 $M_\odot$ yr$^{-1}$.
\end{itemize}
\added{We model the total escape fraction $f_\text{esc}\equiv f_\text{esc}^\text{ion}f_\text{esc}^\mathrm{Ly\alpha}$ as the squared product of a generalised logistic function in redshift and an algebraic function in SFR:
\begin{equation}
f_\text{esc}(\mathrm{SFR},z)=\left[\left(1+e^{-\xi(z-z_0)}\right)^{-\zeta}\left(f_0+\frac{1-f_0}{1+(\mathrm{SFR}/\mathrm{SFR}_0)^\varsigma}\right)\right]^2.
\end{equation}
This form combines S-shaped curves in each variable, and shows the desired asymptotic behaviour discussed above. The function in redshift is an overall normalisation between 0 and 1, with a characteristic redshift $z_0$ acting as an inflection point of the redshift evolution, and $\xi$ and $\zeta$ controlling the shape. Meanwhile, the function in SFR changes between $f_0$ and $1$ around a characteristic $\mathrm{SFR}_0$, with $\varsigma$ again controlling the shape.
}
\subsection{Tuning and Comparison to Previous Work}
\label{sec:lyamodel_lf}
With the model above (including 0.3 dex log-normal scatter in SFR and in Lyman-$\alpha$ luminosity), we are able to translate the analytic halo mass function fit from~\cite{Behroozi13a,Behroozi13b} into simulated luminosity functions at different redshifts. We use a brute-force technique, randomly drawing from the halo mass function and applying the Lyman-$\alpha$ model to the masses drawn, and binning the resulting luminosities to obtain a luminosity function.

\begin{figure}[t]
\centering\includegraphics[width=0.64\linewidth]{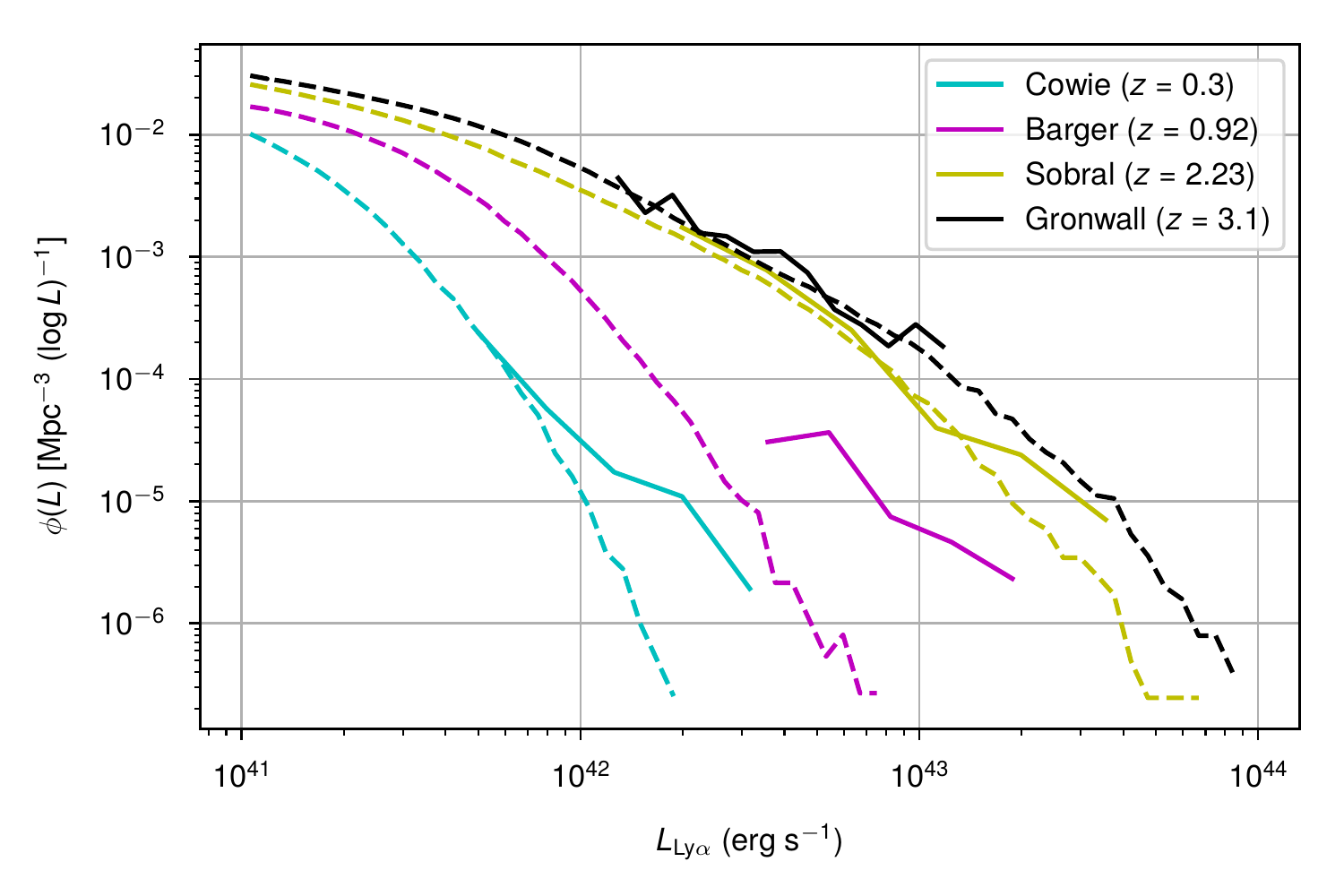}

\caption{Comparison of simulated Lyman-$\alpha$ luminosity functions (dashed curves) to observed LAE luminosity functions (solid curves) at four different redshifts, each from a different work: $z\sim0.3$ from~\cite{Cowie10}, $z\sim0.92$ from~\cite{Barger12}, $z\sim2.23$ from~\cite{Sobral17}, and $z\sim3.1$ from~\cite{Gronwall07}. We generate the simulated luminosity functions using $10^7$ random draws from the halo mass function, calculating a model Lyman-$\alpha$ luminosity for each mass, and binning these into log-luminosity bins. The model is specifically tuned to match the four observed luminosity functions as much as possible.}
\label{fig:Lya_lf_tune}
\end{figure}

We tune the escape fraction parameter values based on comparing these simulated data to observed luminosity functions (LF) at four different redshifts: $z\sim0.3$ from~\cite{Cowie10}, $z\sim0.92$ from~\cite{Barger12}, $z\sim2.23$ from~\cite{Sobral17}, and $z\sim3.1$ from~\cite{Gronwall07}. \added{The resulting parameters are $\xi=1.6$, $z_0=3.125$, $\zeta=1/4$, $f_0=0.18$, $\mathrm{SFR}_0\approx 1.29\,M_\odot$ yr$^{-1}$, and $\varsigma=0.875$.} We match the higher-redshift data better than the lower-redshift data, suggesting that our model cannot completely describe the strong evolution of the LAE LF from $z\sim0.3$ to $z\sim2$ (which is expected given the monotonic redshift evolution of the escape fraction, as discussed in the previous section). Note also, however, that there is support for a composite Schechter/power-law LF for low-to-intermediate redshift LAEs, and by and large we are trying to match only the Schechter part of this. We show a comparison of the simulated luminosity functions to these observed data in~\autoref{fig:Lya_lf_tune}.

\begin{figure}[t]
\centering\includegraphics[width=0.96\linewidth]{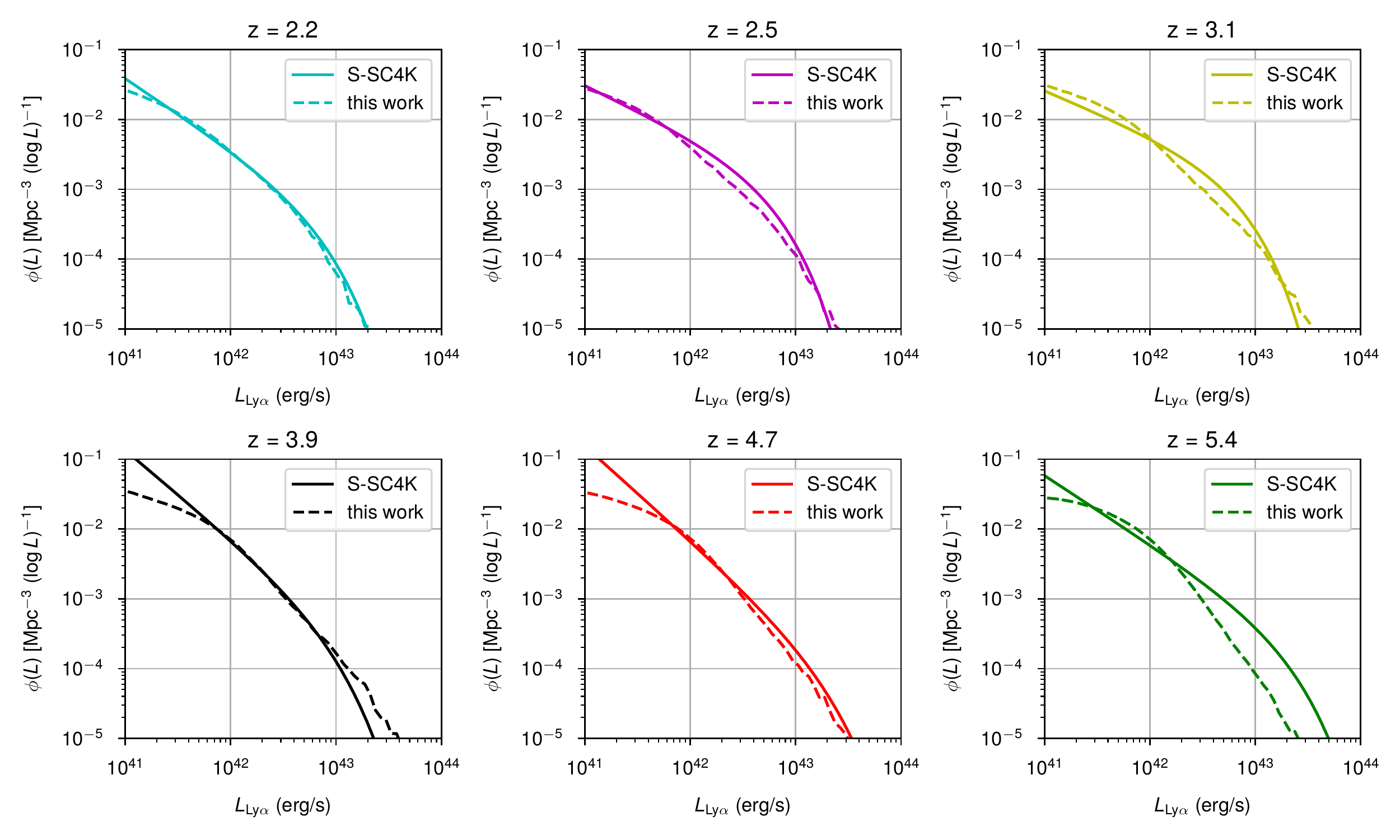}

\caption{Comparison of simulated Lyman-$\alpha$ luminosity functions (dashed curves) to S-SC4K LAE luminosity functions (solid curves) at six different redshifts from~\cite{Sobral18}. The latter are derived from a compilation of deep and wide LAE surveys. We generate the simulated luminosity functions using $5\times10^6$ random draws from the halo mass function, calculating a model Lyman-$\alpha$ luminosity for each mass, and binning these into log-luminosity bins.}
\label{fig:Lya_lf_compare}
\end{figure}

Without further tuning, we also compare against luminosity functions derived in~\cite{Sobral18} from a compilation of deep and wide LAE surveys (dubbed S-SC4K), and the plots in~\autoref{fig:Lya_lf_compare} show a reasonable match up to $z\sim5$.

\begin{figure}[t]
\centering\includegraphics[width=0.64\linewidth]{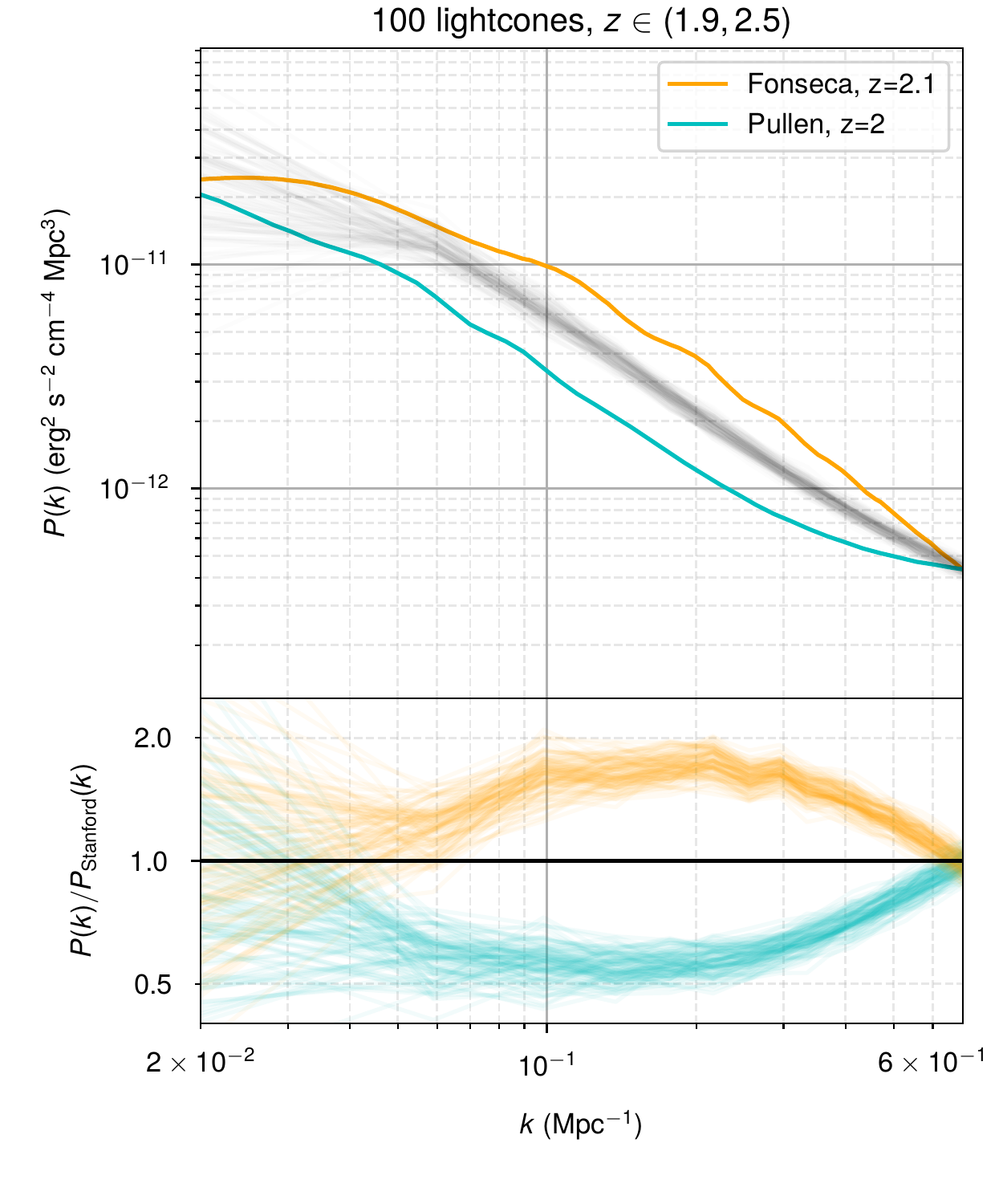}

\caption{Comparison of the simulated Lyman-$\alpha$ spherically-averaged 3D power spectrum $P(k)$ at redshift $1.9$--$2.5$ to simulated $P(k)$ at similar redshifts in previous works,~\cite{Fonseca17} and~\cite{Pullen14}. In the case of~\cite{Pullen14}, we use only the halo contribution, which is sub-dominant to IGM excitations in the model used in that work. \textbf{Upper panel:} $P(k)$ values from~\cite{Pullen14} (cyan),~\cite{Fonseca17} (orange), and this work (faint black lines, one for each lightcone used). \textbf{Lower panel:} $P(k)$ values from~\cite{Pullen14} (cyan),~\cite{Fonseca17} (orange) normalised by the $P(k)$ values from each lightcone in this work (dubbed $P_\text{Stanford}(k)$ in the plot).}
\label{fig:Lya_pspec_compare}
\end{figure}

Since the Chinchilla lightcones used in this work span $z=1.5$--3.5, we can use these to simulate Lyman-$\alpha$ fluctuations and power spectra at $z\sim2$ through the same methods used in the main work. We compare these in~\autoref{fig:Lya_pspec_compare} to power spectra in previous work in~\cite{Pullen14} (using only the halo contribution) and~\cite{Fonseca17}, and find our model yields predicted power spectra squarely in between the two previous works.

\added{\subsection{Minimum Halo Mass for Lyman-$\alpha$ Emission}
\label{sec:lyamodel_mmin}
As was the case for~\cite{Li16}, our choice to assign no line luminosities to halos below $10^{10}\,M_\odot$ in virial halo mass is partly pragmatic. From the point of view of simulation constraints, since we use a cosmological N-body box whose dark matter particle mass is only $5.9\times10^8h^{-1}\,M_\odot=8.4\times10^8\,M_\odot$, the halo population is severely incomplete for $M_\text{vir}\lesssim10^{10}\,M_\odot$. Unlike CO, however, Lyman-$\alpha$ emission does not require a particularly dusty or high-metallicity environment, so any physical mass cutoff for Lyman-$\alpha$ emission would likely be much lower than for CO emission.

Note, however, that this cutoff mainly affects our simulations of HETDEX line-intensity cubes, since the Lyman-$\alpha$ luminosity cutoffs used for our mock LAE catalogues correspond typically to halo masses well above a $10^{10}\,M_\odot$ cutoff. Furthermore, even considering line-intensity cubes, our model $L_\mathrm{Ly\alpha}(M_\text{vir})$ relation falls off quite sharply for $M_\text{vir}\lesssim10^{11}\,M_\odot$. Therefore, even if we assigned model luminosities to a well-represented population of halos with $M_\text{vir}\lesssim10^{10}\,M_\odot$, they would likely not contribute significantly to the signal.

We show this in~\autoref{fig:Lya_mmin} via analytic calculations of contributions to average line intensity from different ranges of halo masses. We calculate the line luminosity per volume $dL_\text{line}/dV = \int L_\text{line}(M)\,(dn/dM)\,dM$, where $dn/dM$ is the halo mass function fit in~\cite{Behroozi13b} at the appropriate redshift, and convert this into observer quantities of CO brightness temperature and Lyman-$\alpha$ $\nu I_\nu$. For both CO and Lyman-$\alpha$ emission, the contribution to mean line intensity falls off below $10^{11}\,M_\odot$ in halo mass for both lines, but especially rapidly for Lyman-$\alpha$ emission, and the analytic results suggest that we have captured a great majority of any expected signal using our cutoff halo mass of $10^{10}\,M_\odot$, at least for our assumed model.

\begin{figure}[t]
\centering\includegraphics[width=0.84\linewidth]{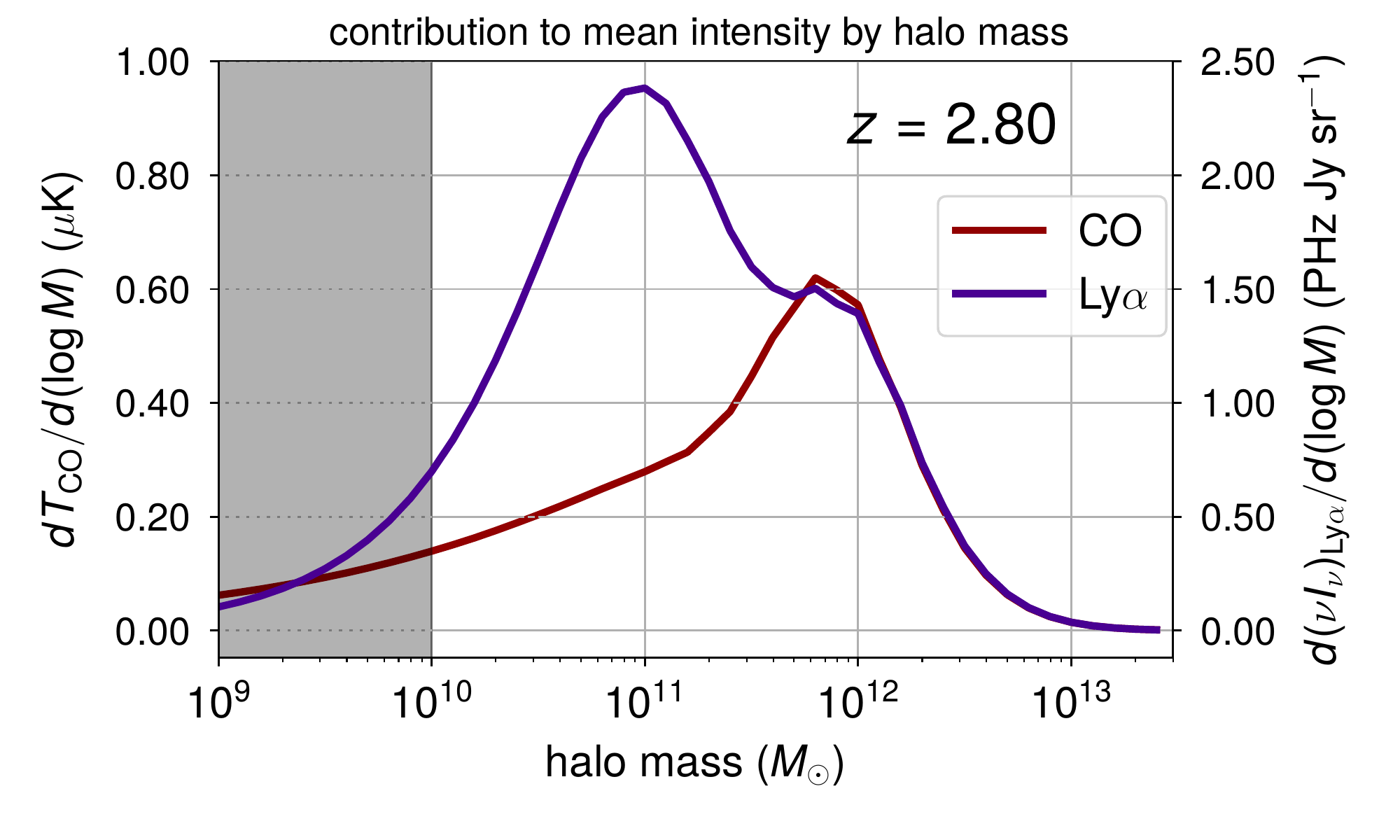}

\caption{Expected contributions by halo mass to mean intensity of CO and Lyman-$\alpha$ emission at $z=2.80$ (corresponding to the midpoint of the COMAP observing frequency band). We quantify the contribution from halo masses $M_\text{vir}\in[M,M+dM]$ to average CO temperature as $dT/d(\log{M})\propto L_\text{CO}(M)\,dn/d(\log{M})$, and to average Lyman-$\alpha$ $\nu I_\nu$ as $d(\nu I_\nu)/d(\log{M})\propto L_\text{Ly$\alpha$}(M)\,dn/d(\log{M})$. All calculations use the line-luminosity models in this work and the halo mass function fit from~\cite{Behroozi13b}. The grey shaded area indicates the range of halo masses below the cutoff mass of $10^{10}\,M_\odot$ used in our simulations.}
\label{fig:Lya_mmin}
\end{figure}
}
\added{
\section{Implementation of Log-scatter in Calculating Line Luminosities for Halos}
\label{sec:scatterorder}

As we describe in~\autoref{sec:hmll}, we use scaling relations to convert a halo's virial mass $M_\text{vir}$ and redshift $z$ to star-formation rate $\mathrm{SFR}$, and then to CO or Lyman-$\alpha$ line luminosity. We add log-normal scatter to halo properties at two points in the calculation:
\begin{itemize}
\item We add log-normal scatter to SFR while preserving the linear mean $\mathrm{SFR}(M_\text{vir},z)$. In practice, this means that for each halo, we calculate the expected mean SFR, then multiply this by a sample value from a log-normal distribution with a log-space standard deviation of $\sigma_\mathrm{SFR}=0.3$ (in units of dex) and a mean logarithm of $-\sigma_\mathrm{SFR}^2\ln{10}/2$. Thus, the mean logarithm is not equal to the logarithm of the linear mean SFR value, but rather $\log{[\avg{\mathrm{SFR}}/(M_\odot\text{ yr}^{-1})]}-\sigma_\mathrm{SFR}^2\ln{10}/2$, which is necessary for the linear mean of the distribution to be the desired $\avg{\mathrm{SFR}}$.
\item We then add log-normal scatter to $L_\text{CO}$ and $L_\mathrm{Ly\alpha}$ in the same manner, multiplying the mean line luminosity from the $L_\text{IR}$--$L_\text{line}$ relation by a sample value from a log-normal distribution with a log-space standard deviation of $\sigma_{L_\text{line}}$ (again in units of dex) but a mean logarithm of $-\sigma_{L_\text{line}}^2\ln{10}/2$.
\end{itemize}

The way in which we implement log-scatter thus preserves the linear mean SFR for a given halo mass and redshift, and preserves the linear mean line luminosity for a given SFR and redshift. This is justified, as we take at least the halo mass--SFR and SFR--$L_\text{CO}$ relations from literature. However, this is not always the same as preserving the linear mean line luminosity for a given halo mass and redshift, and we will focus on the CO luminosity to illustrate this point.

The log-normal distribution with natural log mean $\mu$ and natural log standard deviation $\sigma$ has linear mean $\exp{(\mu+\sigma^2/2)}$. However, in using our scaling relations, we want to preserve the linear mean of the dependent variable at each step. This means that if we have $y$ as a function of $x$, and there is a mean relation $\avg{y}(x)$ and desired log-scatter $\sigma_y$ in units of dex or $\sigma_y\ln{10}$ in natural log space, simply drawing from a log-normal distribution with natural log mean $\ln{[\avg{y}(x)]}$ and natural log standard deviation $\sigma_y\ln{10}$ will result in a linear mean of $[\avg{y}(x)]\exp{(\sigma_y^2\ln^2{10}/2)}$.

Therefore, what we need to do to preserve the linear mean of $\avg{y}(x)$ is draw from a log-normal distribution with natural log mean $\ln{[\avg{y}(x)]}-\sigma_y^2\ln^2{10}/2$ and natural log standard deviation $\sigma_y\ln{10}$. What we do in practice is equivalently multiply $\avg{y}(x)$ by a random variable $Z_y$ drawn from a log-normal distribution with natural log mean $-\sigma_y^2\ln^2{10}/2$ and natural log standard deviation $\sigma_y\ln{10}$. It is common notation to express that $Z_y$ is drawn from such a log-normal distribution---or equivalently, that $\ln{Z_y}$ is drawn from a normal distribution with the appropriate mean and standard deviation---by writing $\ln{Z_y}\sim\mathcal{N}(-\sigma_y^2\ln^2{10}/2,\sigma_y\ln{10})$.

Our fiducial model includes a $\operatorname{SFR}(M_\text{vir})$ relation (which also depends on redshift, but fix this for the time being) and a $L_\text{CO}(\operatorname{SFR})$ relation, both of which we scatter separately. Then for each halo $i$,
\begin{equation}\operatorname{SFR}_i = \avg{\operatorname{SFR}}(M_{\text{vir},i})\cdot\exp{X_\mathrm{SFR}},\text{ where }X_\mathrm{SFR}\sim\mathcal{N}(-\sigma_\mathrm{SFR}^2\ln^2{10}/2,\sigma_\mathrm{SFR}\ln{10}),\end{equation}
and
\begin{equation}L_{\text{CO},i} = \avg{L_{\text{CO}}}(\operatorname{SFR}_i)\cdot\exp{X_{L_\text{CO}}},\text{ where }X_{L_\text{CO}}\sim{\mathcal{N}(-\sigma_{L_\text{CO}}^2\ln^2{10}/2,\sigma_{L_\text{CO}}\ln{10})}.\end{equation}

The $\avg{L_\text{CO}}(\operatorname{SFR})$ relation specifically takes the form
\begin{equation}\log{\left(\frac{L_\text{CO}}{L_\odot}\right)} = \frac{1}{\alpha}\left[\log{\left(\frac{\operatorname{SFR}}{M_\odot\text{ yr}^{-1}}\right)}-\beta-10\right]-4.31,\end{equation}
once we have combined all the relations between SFR, IR luminosity, CO luminosity in observer units, and CO luminosity in intrinsic units.

\begin{align}
\log{L_{\text{CO},i}} &= \frac{1}{\alpha}\left[\log{\left(\frac{\operatorname{SFR}_i}{M_\odot\text{ yr}^{-1}}\right)}-\beta-10-4.31\alpha\right]+\log{\exp{X_{L_\text{CO}}}}
\\&=\frac{1}{\alpha}\left[\log{\left(\frac{\avg{\operatorname{SFR}}(M_{\text{vir},i})}{M_\odot\text{ yr}^{-1}}\cdot\exp{X_\mathrm{SFR}}\right)}-\beta-10-4.31\alpha\right]+\frac{X_{L_\text{CO}}}{\ln{10}}
\\&=\frac{1}{\alpha}\left[\log{\frac{\avg{\operatorname{SFR}}(M_{\text{vir},i})}{M_\odot\text{ yr}^{-1}}}-\beta-10-4.31\alpha\right]+\frac{X_\mathrm{SFR}}{\alpha\ln{10}}+\frac{X_{L_\text{CO}}}{\ln{10}}
\end{align}

Then the overall offset in the log mean versus na\"{i}vely combining the relations in log space comes out to be
\begin{equation}\avg{\log{L_{\text{CO},i}}} - \avg{\log{\avg{L_{\text{CO}}}[\avg{\operatorname{SFR}}(M_{\text{vir},i})]}} = \avg{\frac{X_\mathrm{SFR}}{\alpha\ln{10}}+\frac{X_{L_\text{CO}}}{\ln{10}}} = -\frac{\ln{10}}{2}\left(\frac{\sigma_\mathrm{SFR}^2}{\alpha}+\sigma_{L_\text{CO}}^2\right).\label{eq:logoffset_fid}\end{equation}
Note that this procedure, used for the work in our main text, should preserve the linear mean SFR for a given halo mass and the linear mean CO luminosity for a given SFR.

We now return to the idea of preserving the linear mean CO luminosity for a given halo mass and redshift, and how our fiducial model actually will not accomplish this. As we note in~\autoref{sec:hmll}, we may describe the total log-scatter in $L_\text{CO}$ with a total log-space standard deviation of $\sigma_\text{tot}=(\sigma_\mathrm{SFR}^2/\alpha^2+\sigma_{L_\text{CO}}^2)^{1/2}$---where the exponent of the SFR--$L_\text{CO}$ power law scales the originally applied log-scatter in SFR by $1/\alpha$. We may then consider combining the average $\mathrm{SFR}(M_\text{vir},z)$ and $L_\text{CO}(\mathrm{SFR})$ relations into a $L_\text{CO}(M_\text{vir},z)$ relation and simply applying a single log-scatter of $\sigma_\text{tot}$ (0.37 dex in our case) while preserving the linear mean $L_\text{CO}$ for fixed $M_\text{vir}$ and $z$. Then for each halo,
\begin{equation}L_{\text{CO},i} = \avg{L_{\text{CO}}}[\avg{\operatorname{SFR}}(M_{\text{vir},i})]\cdot\exp{X_{\text{tot}}},\text{ where }X_{\text{tot}}\sim{\mathcal{N}(-\sigma_{\text{tot}}^2\ln^2{10}/2,\sigma_{\text{tot}}\ln{10})},\end{equation}
from which we would obtain
\begin{equation}\avg{\log{L_{\text{CO},i}}} - \avg{\log{\avg{L_{\text{CO}}}[\avg{\operatorname{SFR}}(M_{\text{vir},i})]}} = \avg{\frac{X_{\text{tot}}}{\ln{10}}} = -\frac{\ln{10}}{2}\left(\frac{\sigma_\mathrm{SFR}^2}{\alpha^2}+\sigma_{L_\text{CO}}^2\right).\label{eq:logoffset_alt}\end{equation}

Thus, our fiducial log-mean $L_\text{CO}(M_\text{vir})$ offset of~\autoref{eq:logoffset_fid} differs from the log-mean offset required to preserve the linear mean $L_\text{CO}$ for a given halo mass, The difference in decimal log space between the right-hand sides of~\autoref{eq:logoffset_fid} and~\autoref{eq:logoffset_alt} is $-\sigma_\text{SFR}^2\ln{10}(1/\alpha-1/\alpha^2)/2$, corresponding to a multiplicative factor of $\exp{[-(\sigma_\text{SFR}\ln{10})^2(1/\alpha-1/\alpha^2)/2]}$. By separately preserving the linear mean SFR for a given halo mass and the linear mean CO luminosity for a given SFR, the linear mean CO luminosity for a halo mass is actually modified by this factor, relative to the expected value from combining the scaling relations with zero scatter.

For $\alpha=1.37$ and $\sigma_\text{SFR}=0.3$ (in units of dex), the effect is quite small---the linear mean $L_\text{CO}$ is 6\% below what might be expected from combining the mean scaling relations. However, the effect increases exponentially with $\sigma_\text{SFR}$, so for $\sigma_\text{SFR}$ of 1.0 dex, the linear mean $L_\text{CO}$ falls to half of what would be expected. This explains why, in Figure 5 of~\cite{Li16}, the $P(k)$ values at low $k$ fall with increasing $\sigma_\text{SFR}$ (although at the same time, increasing log-scatter in SFR also increases shot noise, which cushions the effect of not preserving the linear mean $L_\text{CO}$ for a given $M_\text{vir}$ value). Therefore, the details of the implementation of log-scatter become important if the scatter in SFR is high and the SFR--$L_\text{CO}$ power law is significantly sub- or super-linear.
}

\section{Lyman-$\alpha$ Modelling Beyond This Work: Overview of Radiative Processes}
\label{sec:Lyamodelbad}
The models used for CO and Lyman-$\alpha$ emission are both very simple models built on the galaxy--halo connection, assigning a luminosity to each halo identified in a dark matter simulation. This is already a significant simplification for CO emission, which depends on gas metallicity, AGN feedback, and other physical and environmental factors that a dark-matter-only simulation will not capture. The simplification is even more drastic in the case of Lyman-$\alpha$ emission, whose radiative transfer through the neutral gas of the circumgalactic and intergalactic media (CGM and IGM) alters observations beyond the simple escape fractions we posit.

\begin{itemize}
\item Scattering in the CGM results in diffuse Ly-$\alpha$ halos or blobs, significantly increasing the total flux over radii of $\sim10''$~\citep{Steidel11}. Since this diffuse surface brightness is extended and still relatively faint per solid angle, conventional targeted LAE surveys would not detect it, but line-intensity mappers like HETDEX may be able to.
\item Scattering in the IGM may result in anisotropic clustering observed in the Lyman-$\alpha$ intensity cube, as demonstrated in a simulation study from~\cite{Zheng11}. An analysis by~\cite{Croft16} of Lyman-$\alpha$ intensity in galaxy spectra from the Baryon Oscillation Spectroscopic Survey (BOSS), cross-correlated with BOSS quasars, reports this effect. However,~\cite{Croft18} have since reported a non-detection of any cross-correlation signal against the Lyman-$\alpha$ forest and a lower quasar cross-correlation signal than first reported, and no longer claim a quantitative measurement of anisotropic clustering. In addition, the results of another simulation study from~\cite{Behrens17} show a smaller anisotropy than was found in~\cite{Zheng11}. IGM scattering may have a greater effect by smoothing small-scale fluctuations, potentially leading to a strong dependence of the power spectrum log-slope on the mean IGM neutral fraction (see~\citealt{VisbalMcQuinn18}, showing this at $z\sim7$).
\item Emission from excitations in the IGM could be an additional factor, but while~\cite{Pullen14} found this to be a dominant contributor to the Lyman-$\alpha$ intensity signal,~\cite{Silva13} and~\cite{ComaschiFerrara16} did not.
\end{itemize}

Overall, radiative transfer significantly impacts the expected Lyman-$\alpha$ signal, and future forecasts should take into account the effects discussed above through sophisticated modelling of Lyman-$\alpha$ radiative processes.
\added{\section{An Analytic Check on the Effect of Redshift Errors on Power Spectra}
\label{sec:Wz_approx}
While we use $W^2(k)$ and $W^2_z(k)$ in the main text to describe attenuation of the auto and cross power spectra due to instrumental resolution, we may also use the same formalism to analytically calculate the expected attenuation of spectra due to redshift errors, by approximating the resulting effect on the galaxy density field as a simple convolution with a Gaussian profile. Since a discrete and relatively limited population of galaxies make up the density field, this is only an approximation, but sufficient at large scales.

Given the relevant comoving size $\sigma_\parallel$ of the Gaussian profile, we average the expected attenuation of $\exp{(-k_\parallel^2\sigma_\parallel^2)}$ within each $k$-shell to find $W_z(k)$. In the main text, we average across the discrete grid of $\mathbf{k}$ values that correspond to the discrete Fourier transform used to calculate the power spectra. However, in this section, we will obtain a closed-form expression for the attenuation with an analytic average, calculated across the full range of $\mu=k_\parallel/k$, which is the cosine of the spherical polar angle of $\mathbf{k}$. This ranges from $-1$ to $1$ but the quantity averaged is an even function of $\mu$, so
\[W^2_z(k) = \frac{\int d\mu\,\exp{(-k^2\mu^2\sigma_\parallel^2)}}{\int d\mu} = \int_0^1 d\mu\,\exp{(-k^2\mu^2\sigma_\parallel^2)}.\]
If we want to describe attenuation due to redshift errors that follow a Gaussian distribution with standard deviation $\sigma_z$ in redshift space, we would set $\sigma_\parallel\approx \sigma_{\parallel,\text{gal}}\equiv c\sigma_z/H(z)$, and the resulting $W^2_z(k)$ would describe attenuation of the galaxy auto spectrum due to redshift errors. As we discussed in~\autoref{sec:sensest}, if COMAP has much finer redshift resolution than the galaxy survey, then we would set $\sigma_\parallel=\sigma_{\parallel,\text{gal}}/\sqrt{2}$ to calculate the appropriate $W^2_z(k)$ for the CO--galaxy cross spectrum.

Thus, the expected attenuation of the galaxy density auto spectrum is
\begin{equation}W^2_{z,\text{gal}}(k) = \int_0^1 d\mu\,\exp{(-k^2\mu^2\sigma_{\parallel,\text{gal}}^2)} = \frac{\pi^{1/2}}{2k\sigma_{\parallel,\text{gal}}}\operatorname{erf}{(k\sigma_{\parallel,\text{gal}})},\label{eq:Wgal}\end{equation}
and the analogous $W^2_{z,\text{CO}\times\text{gal}}(k)$ for the CO--galaxy cross spectrum is
\begin{equation}W^2_{z,\text{CO}\times\text{gal}}(k) = \frac{\pi^{1/2}}{2^{1/2}k\sigma_{\parallel,\text{gal}}}\operatorname{erf}{\left(\frac{k\sigma_{\parallel,\text{gal}}}{\sqrt{2}}\right)}.\label{eq:WCOgal}\end{equation}
Since $\operatorname{erf}(x)/x\to2/\sqrt{\pi}$ as $x\to0$, both of the above should equal 1 for $\sigma_z=0$, but once $\sigma_{\parallel,\text{gal}}=c\sigma_{z}/H(z)\gtrsim k^{-1}$ the auto spectrum attenuates significantly at the given $k$, and the cross spectrum does the same once $\sigma_{\parallel,\text{gal}}\gtrsim 2^{1/2}k^{-1}$.

As we note in~\autoref{sec:sensest}, the range of $k$ represented in our simulations is $\sim0.02$ to $4$ Mpc$^{-1}$ (although we only plot $k$ above $\sim0.05$ Mpc$^{-1}$ as the lightcone-to-lightcone variance at $z\sim0.02$ Mpc$^{-1}$ is quite high), and $c/H(z=2.8)\sim10^3$ Mpc. So for the cross spectrum to decrease appreciably at $k\sim1$ Mpc$^{-1}$, at $z=2.8$ we only require $\sigma_z/(1+z)\gtrsim 0.0004$, and it will start decreasing appreciably at the lowest scales simulated once $\sigma_z/(1+z)\gtrsim0.02$ (and at the lowest scales plotted once $\sigma_z/(1+z)\gtrsim0.01$).

The attenuation of the cross spectrum is, of course, different from the attenuation of $r(k)$. Since the galaxy--galaxy auto spectrum is attenuated by $W^2_{z,\text{gal}}$ and the line--galaxy cross spectrum by $W^2_{z,\text{CO}\times\text{gal}}$ (and the line--line auto spectrum by a comparatively negligible amount), $r(k)$ is attenuated by a factor of
\begin{equation}\frac{W^2_{z,\text{CO}\times\text{gal}}(k)}{\sqrt{W^2_{z,\text{gal}}(k)}} = \frac{\pi^{1/4}}{(k\sigma_{\parallel})^{1/2}}\frac{\operatorname{erf}{(k\sigma_{\parallel}/\sqrt{2})}}{[\operatorname{erf}{(k\sigma_{\parallel})}]^{1/2}},\label{eq:Wrk}\end{equation}
which is approximately $1$ up to $k\sigma_{\parallel}\simeq1$, and $\approx\pi^{1/4}/(k\sigma_{\parallel})^{1/2}$ for $k\sigma_{\parallel}\gtrsim3$.

We note again that all of this assumes a Gaussian smoothing of the galaxy density field, while what really happens is Gaussian scattering of discrete redshifts. Since galaxies and very bright CO emitters (the dominant source of the shot noise in the CO auto spectrum) are discrete objects, and we have here considered only continuous CO temperature and galaxy density contrast fields, this analytic calculation is only an approximation, and breaks down particularly at high $k$. The power spectrum of the galaxy overdensity field goes to the inverse of the comoving galaxy density as $k\to\infty$ and Poisson noise dominates. Therefore, while redshift errors will attenuate the \emph{cross} shot noise (which does require coincidence of the CO peaks and the galaxies), the shot-noise component of the galaxy auto spectrum will remain unchanged. This means that at high $k$ the $r(k)$ attenuation is simply $W^2_{z,\text{CO}\times\text{gal}}(k)\sim 1/(k\sigma_{\parallel})$.

\begin{figure}[t]
\centering\includegraphics[width=0.92\linewidth]{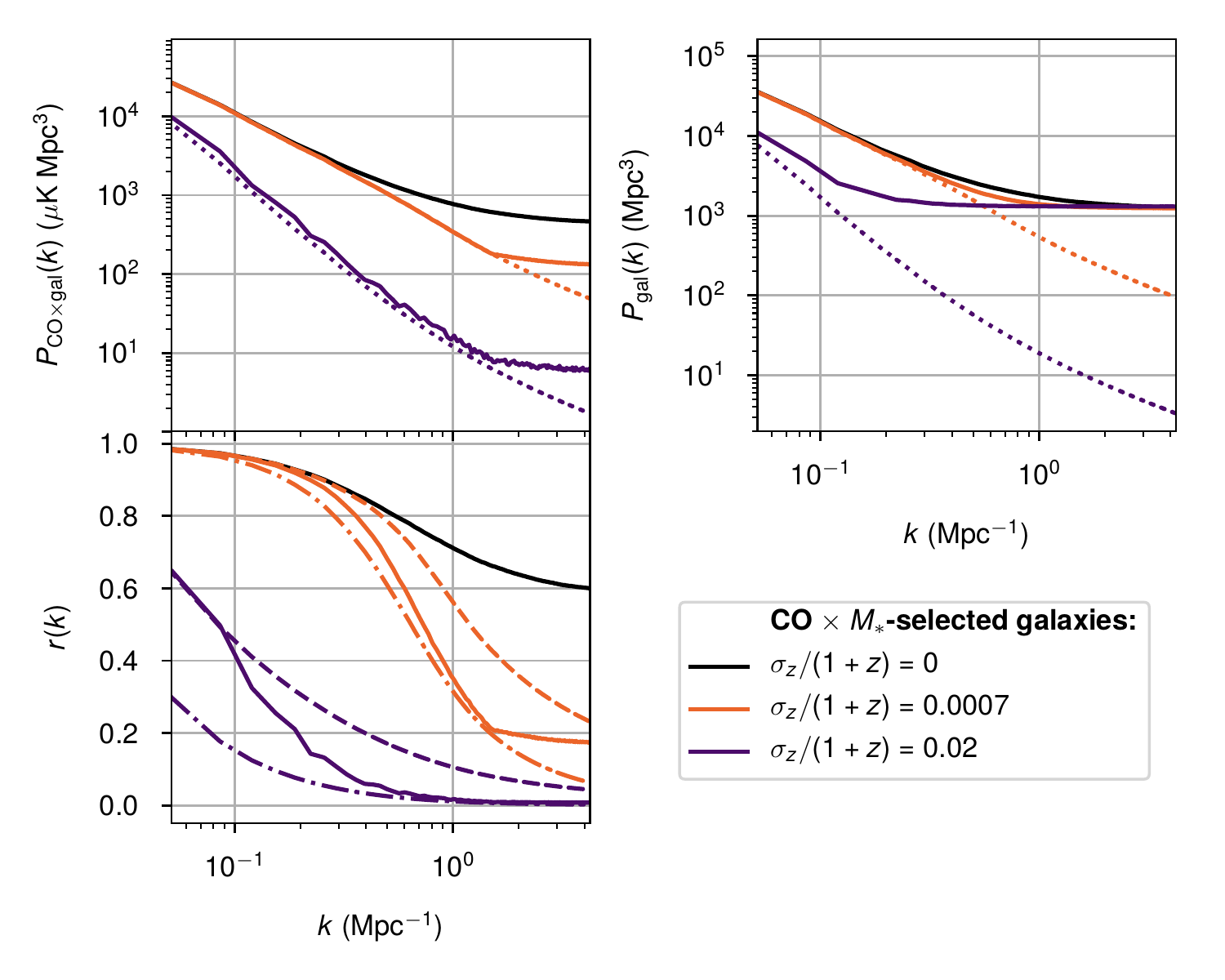}

\caption{A comparison of analytic expectations of attenuation of CO--galaxy cross-correlation and galaxy autocorrelation against simulations. The mock galaxy sample is selected based on a minimum $10^{10}\,M_\odot$ stellar mass. The solid curves show median quantities for different redshift errors; the dotted curves in the upper panels show the median spectrum for $\sigma_z/(1+z)=0$ multiplied by the analytically calculated attenuation of the CO--galaxy cross spectrum from~\autoref{eq:WCOgal} (\emph{upper left panel}) and  of the galaxy auto spectrum by~\autoref{eq:Wgal} (\emph{upper right panel}) for each nonzero $\sigma_z/(1+z)$. \emph{Lower panel:} analytic expectations of attenuation of $r(k)$ based on~\autoref{eq:Wrk} (dashed curves) and based only on~\autoref{eq:WCOgal} (dash-dotted curves) bracket the actual simulated results.}
\label{fig:rkcheck}
\end{figure}

We compare our analytic expectations to simulations in~\autoref{fig:rkcheck}. We find good agreement at low $k$, but with the shot-noise component of the galaxy auto spectrum unattenuated, the simulated attenuation of $r(k)$ is near the expression of~\autoref{eq:Wgal} at the lowest $k$ values considered but quickly approaches $W^2_{z,\text{CO}\times\text{gal}}(k)$ instead for higher $k$.
}
\section{Sensitivities for the Anisotropic Power Spectrum}
\label{sec:PkmuSNR}
A consideration of signal-to-noise for the anisotropic power spectrum, which is averaged in bins of $k$ and $\mu=k_\parallel/k$ (the cosine of the spherical polar angle of $\mathbf{k}$), is important for several reasons:
\begin{itemize}
\item The signal loss from galaxy redshift errors as considered in the main results of our paper should be highly anisotropic, disproportionately affecting line-of-sight modes (with large $\mu$).\added{ (This should be the only source of significant anisotropy in our $P(\mathbf{k})$, since we ignore the peculiar velocities of halos in our lightcones.)}
\item The beam attenuation calculated in~\autoref{eq:Wbeam} implicitly assumes a power spectrum that is isotropic in all three dimensions, which will not be the case if the previous point holds. We may then be mistaken in our estimate of signal loss due to the beam.
\end{itemize}
We expect the signal loss from redshift errors to be significant enough that the above points should not affect the basic conclusions of this work, but consider calculation of $P(k,\mu)$ and sensitivities for one realisation (for cross-correlation with $M_*$-selected galaxy samples only, using the fiducial $M_*>10^{10}\,M_\odot$ cut).

We can readily generalise the expressions in the main body of this work for $P(k,\mu)$ instead of $P(k)$\replaced{. In particular, w}{, working with two-dimensional bins in $k$ and $\mu$ (instead of binning only in $k$) to draw out line-of-sight anisotropies. If we consider the previous averaging of $P(\mathbf{k})$ into $P(k)$ to be in spherical shells of $k$, we now also divide $\mathbf{k}$-space into sectors based on values of $\mu$, and the averaging of $P(\mathbf{k})$ into $P(k,\mu)$ takes place within intersections of $k$-shells and $\mu$-sectors. W}e still assume that the power spectrum is isotropic in the two transverse dimensions, so that we can still separately average the beam attenuation as
\begin{equation}W^2(k,\mu) = {\bigl\langle\exp{(-k_\perp^2\sigma_\perp^2)}\bigr\rangle}_\mathbf{k},\end{equation}
\replaced{but the average is now}{which is similar to~\autoref{eq:Wbeam} in that it is an average of $\exp{(-k_\perp^2\sigma_\perp^2)}$ (the expected attenuation of $P(\mathbf{k})$ for any given $\mathbf{k}$) within a bin of discrete $\mathbf{k}$ corresponding to the discrete Fourier transform. However, this average is} over all discrete $\mathbf{k}$ that fall within the spherical shell centred at $k$ and the spherical sector corresponding to $\mu$, with the same bins defined by these shell-sector intersections used for averaging of $P(\mathbf{k})$ into $P(k,\mu)$.\added{ (Averaging the above across all $\mu$ would reproduce the $W^2(k)$ of~\autoref{eq:Wbeam}.)}

\subsection{An Approximate Analytic Example}

Note in particular that even with the same 3D $P(\mathbf{k})$, there is no reason to expect the total signal-to-noise to be the same for $P(k)$ and $P(k,\mu)$. For simplicity, take the CO auto spectrum (ignoring or absorbing $W^2(k)$) as an example. Suppose that we have $n_\mu$ $\mu$-bins in each $k$-shell such that
\begin{equation}N_\text{modes}(k) = \sum_i^{n_\mu} N_{m,i}(k)\end{equation}
and $N_{m,i}(k)$ represents the number of modes in the $\mu$-bin centred at $\mu_i$ falling within the $k$-shell centred at $k$. (From here on, summing over $i$ always implies a sum over all $n_\mu$ applicable values.)

In the main text, we calculate the spherically averaged spectrum $P(k)$ and then the signal-to-noise at each $k$ (which is then added in quadrature over all $k$ to obtain total signal-to-noise across all modes):
\begin{equation}\left[\mathrm{\frac{S}{N}}(k)\right]^2_\text{sph} = \frac{P(k)^2N_\text{modes}(k)}{[P_n+P(k)]^2},\end{equation}
where $P(k) = [\sum_i P(k,\mu_i)N_{m,i}(k)]/N_\text{modes}(k)$.
We substitute and simplify to obtain
\begin{equation}\left[\mathrm{\frac{S}{N}}(k)\right]^2_\text{sph} = \frac{[\sum_i P(k,\mu_i)N_{m,i}(k)]^2/N_\text{modes}(k)}{[P_n+\sum_i P(k,\mu_i)N_{m,i}(k)/N_\text{modes}(k)]^2}.\end{equation}
But if we took the signal-to-noise for each $\mu_i$ and then averaged, we would have
\begin{equation}\left[\mathrm{\frac{S}{N}}(k)\right]^2_\text{aniso} = \sum_i\left[\mathrm{\frac{S}{N}}(k,\mu_i)\right]^2 = \sum_i\frac{P(k,\mu_i)^2N_{m,i}(k)}{[P_n+P(k,\mu_i)]^2}.\end{equation}
It is difficult to see a way that the two can be generally equal. It is reasonable to approximate $N_{m,i}(k)\approx N_\text{modes}(k)/n_\mu$ (intervals in $\mu$ are roughly linear with intervals in the polar angle up to $\mu\sim0.5$, so the number of modes in each $\mu$-bin should be similar), in which case
\begin{equation}\left[\mathrm{\frac{S}{N}}(k)\right]^2_\text{sph} = \frac{[\sum_i P(k,\mu_i)]^2N_\text{modes}(k)/n_\mu^2}{[P_n+\sum_i P(k,\mu_i)/n_\mu]^2},\end{equation}
and
\begin{equation}\left[\mathrm{\frac{S}{N}}(k)\right]^2_\text{aniso} = \sum_i\frac{P(k,\mu_i)^2N_\text{modes}(k)/n_\mu}{[P_n+P(k,\mu_i)]^2}.\end{equation}
In the case that sample variance dominates our uncertainties, i.e.~$P_n\ll P(k,\mu_i)$,
\begin{equation}\left[\mathrm{\frac{S}{N}}(k)\right]^2_\text{aniso} = \sum_i\frac{N_\text{modes}(k)}{n_\mu} = N_\text{modes}(k) = \left[\mathrm{\frac{S}{N}}(k)\right]^2_\text{sph}.\end{equation}

However, we now demonstrate that in the extreme case where instrumental noise dominates the uncertainties and only one of $n_\mu$ bins has a nonzero value of $P(k,\mu_i)$, the signal-to-noise is significantly higher for $P(k,\mu)$ than for $P(k)$. (Note that it would be misleading to assume $P_n\ll P(k,\mu_i)$ for all $i$ when the right-hand side is zero for most $i$.) Take $\mu_0$ to be the bin with nonzero $P(k,\mu)$:
\begin{equation}\left[\mathrm{\frac{S}{N}}(k)\right]^2_\text{sph} = \frac{P(k,\mu_0)^2N_\text{modes}(k)/n_\mu^2}{[P_n+ P(k,\mu_0)/n_\mu]^2},\end{equation}
and
\begin{equation}\left[\mathrm{\frac{S}{N}}(k)\right]^2_\text{aniso} = \frac{P(k,\mu_0)^2N_\text{modes}(k)/n_\mu}{[P_n+P(k,\mu_0)]^2}.\end{equation}
If $P_n\gg P(k,\mu_0)$, as is typical in our simulated surveys,
\begin{equation}\left[\mathrm{\frac{S}{N}}(k)\right]^2_\text{sph} = \frac{P(k,\mu_0)^2N_\text{modes}(k)}{n_\mu^2P_n} = \frac{1}{n_\mu}\left[\mathrm{\frac{S}{N}}(k)\right]^2_\text{aniso},\end{equation}
resulting in a factor-of-$\sqrt{n_\mu}$ difference in signal-to-noise at this $k$.

\subsection{Results from One Realisation}

\begin{figure}[t]
\centering\includegraphics[width=0.84\linewidth]{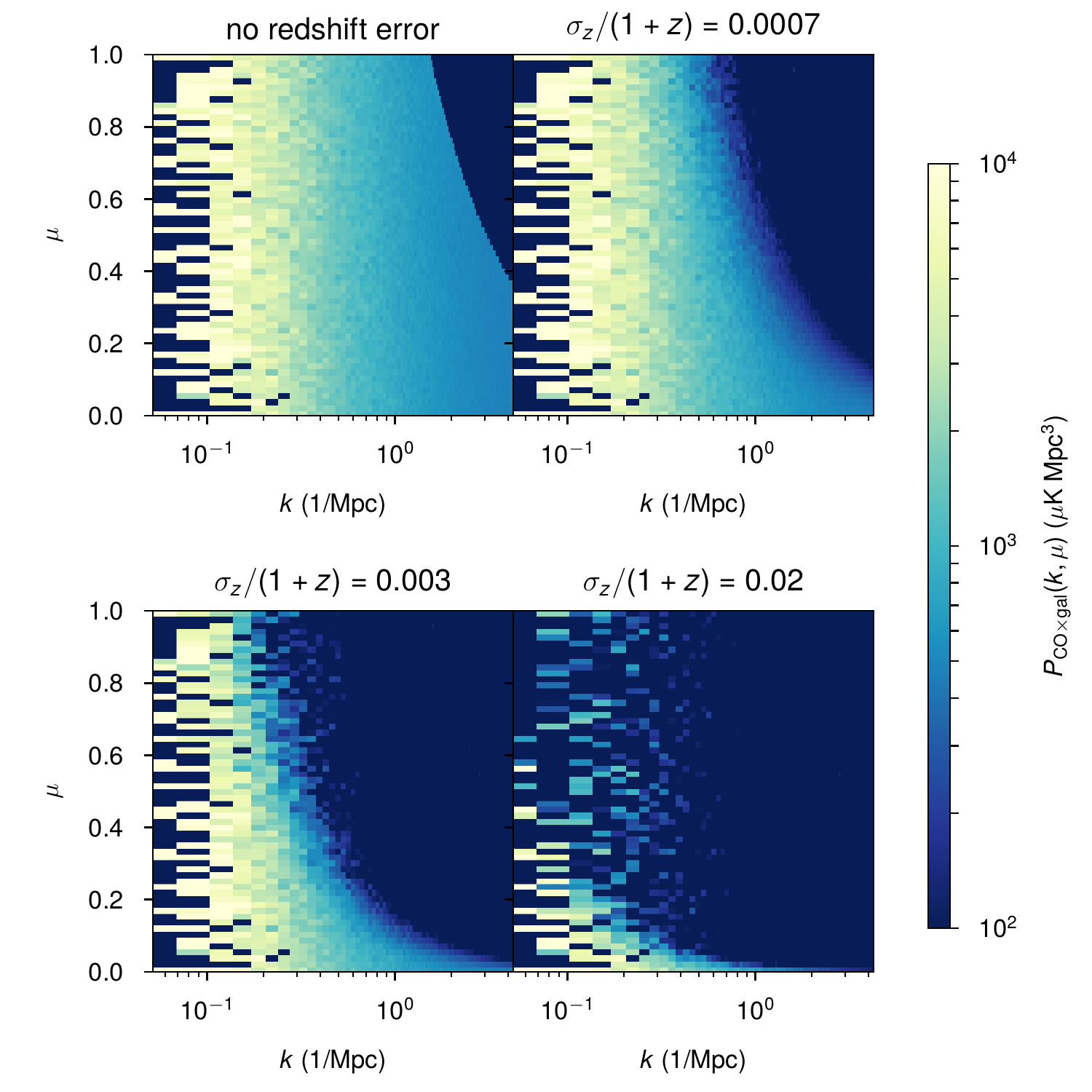}

\caption{The anisotropic CO-galaxy cross spectrum $P_{\text{CO}\times\text{gal}}(k,\mu)$ shown for one realisation of COMAP cross-correlation against a galaxy sample with $M_*>10^{10}\,M_\odot$. Each panel shows the cross spectrum for the galaxy $\sigma_z/(1+z)$ value indicated above it.}
\label{fig:Pkmu}
\end{figure}

For one lightcone out of our 100, we simulate cross-correlations between a CO cube and galaxy sample with $M_*>10^{10}\,M_\odot$ for all $\sigma_z/(1+z)$ values considered in the main text. However, we now obtain the anisotropic $P(k,\mu)$, which we show in~\autoref{fig:Pkmu}. As anticipated, the signal loss with increasing $\sigma_z/(1+z)$ is highly anisotropic and is greater for higher $\mu$ at any given $k$.

\begin{figure}[t]
\centering\includegraphics[width=0.54\linewidth]{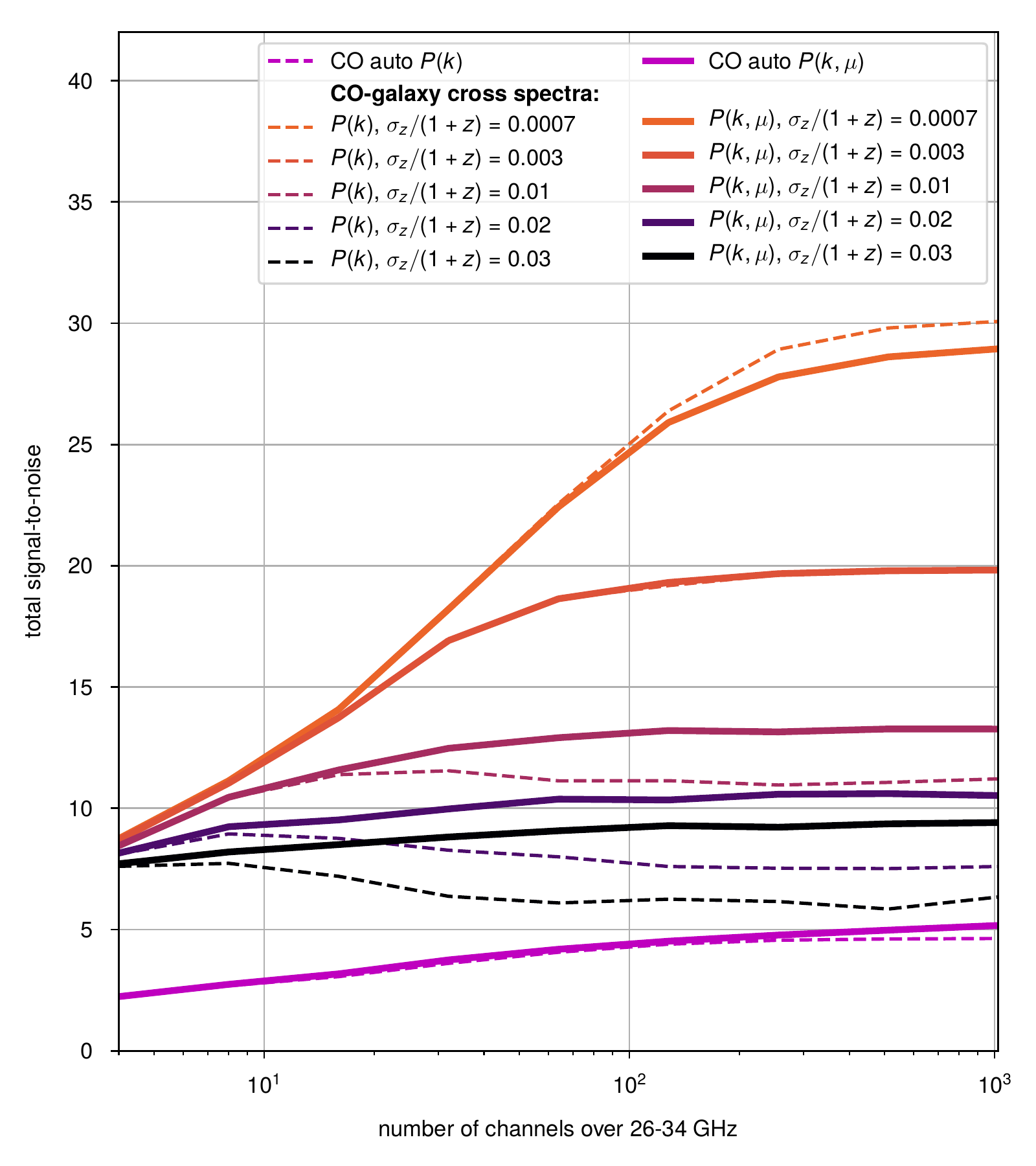}

\caption{A demonstration of the effect of COMAP line-of-sight resolution on the signal-to-noise ratio for auto and cross spectra, and specifically how it differs for the anisotropic $P(k,\mu)$ from what we show for $P(k)$ in~\autoref{fig:chanSNR}. We express frequency resolution here as number of channels across the spectrometer bandwidth, and show how it affects total signal-to-noise over all scales $\text{S}/\text{N}=[\sum_k(\text{S}/\text{N})_k^2]^{1/2}$ in one realisation for simulated CO auto spectra and CO--galaxy cross spectra---both spherically averaged $P(k)$ (dashed curves) and anisotropic $P(k,\mu)$ (solid curves)---for different galaxy $\sigma_z/(1+z)$ values. The simulated galaxy sample is selected with a minimum stellar mass of $\log{(M_\mathrm{*,min}/M_\odot)}=10.0$. All signal-to-noise ratios are quoted for a single patch of 2.5 deg$^2$ observed for 1500 hours; we may expect up to a factor-of-$\sqrt{2}$ improvement if two equivalent patches are observed for 1500 hours each.}
\label{fig:Pkmu_SNR}
\end{figure}

We also show in~\autoref{tab:Pkmu_SNR} the total signal-to-noise ratios across all modes for this realisation, comparing between $P(k)$ and $P(k,\mu)$. The difference is notable for high values of $\sigma_z/(1+z)$, but not as great as the $\sqrt{n_\mu}$ in the simple calculation above (which would have been a factor of almost 8). Furthermore, as we show in~\autoref{fig:Pkmu_SNR}, the change in signal-to-noise with frequency resolution differs from what we show in~\autoref{fig:chanSNR} for $P(k)$. Working with $P(k,\mu)$ by definition separates the line-of-sight modes from the transverse modes, and thus the different degrees of attenuation experienced due to redshift error. Then the only effect of decreasing the number of frequency channels in the survey volume is to decrease the number of modes averaged and thus to increase uncertainties, and the slight gain in $P(k)$ signal-to-noise shown in~\autoref{fig:chanSNR} for $\sigma_z/(1+z)\geq0.01$ is absent in the $P(k,\mu)$ signal-to-noise curves in~\autoref{fig:Pkmu_SNR}. We thus consider our basic conclusion---that photometric errors significantly reduce any advantage in detection significance from cross-correlation---to be unchanged.

\begin{deluxetable}{lccc}
\tabletypesize{\footnotesize}
\tablewidth{0.9\linewidth}
\tablecaption{\label{tab:Pkmu_SNR}
Total signal-to-noise ratio over all modes for spherically-averaged power spectra ($\mathrm{S/N}_\text{sph}$) and anisotropic power spectra ($\mathrm{S/N}_\text{aniso}$) in one realisation.}
\tablehead{\\[-1em] Power spectrum & \added{$\sigma_z/(1+z)$ for galaxies} & $\mathrm{S/N}_\text{sph}$ & $\mathrm{S/N}_\text{aniso}$}
\startdata
\replaced{$P_{\text{CO}\times\text{gal}}$}{CO--galaxy cross} & $0.$    &33.2&37.7\\
\replaced{$P_{\text{CO}\times\text{gal}}$}{CO--galaxy cross} & $0.0007$&29.9&28.6\\
\replaced{$P_{\text{CO}\times\text{gal}}$}{CO--galaxy cross} & $0.003 $&19.9&19.9\\
\replaced{$P_{\text{CO}\times\text{gal}}$}{CO--galaxy cross} & $0.01  $&11.3&13.3\\
\replaced{$P_{\text{CO}\times\text{gal}}$}{CO--galaxy cross} & $0.02  $& 7.7&10.6\\
\replaced{$P_{\text{CO}\times\text{gal}}$}{CO--galaxy cross} & $0.03  $& 6.6&9.5\\
\replaced{$P_\text{CO}$ (auto spectrum)}{CO auto} & \dots & 4.7 & 5.0\\
\hline
\enddata
\tablecomments{For the galaxy sample, $\log{(M_\mathrm{*,min}/M_\odot)}=10$, with a galaxy count of $2.9\times10^4$ without redshift errors. All signal-to-noise ratios are still quoted for a single patch of 2.5 deg$^2$ observed for 1500 hours; we may expect up to a factor-of-$\sqrt{2}$ increase if two equivalent patches are observed for 1500 hours each and a further roughly linear increase with more integration time.}
\end{deluxetable}

\end{document}